\documentclass[12pt,preprint,flushrt]{aastex}
\usepackage{graphicx,natbib}
\citeindextrue
\newcommand{\lowomega}{\ensuremath{22\pm2\% }}
\newcommand{\highomega}{\ensuremath{7\pm3\% }}
\newcommand{\totalomega}{\ensuremath{29\pm4\% }}
\newcommand{\honeno}{\ion{H}{1}}
\newcommand{\hone}{\honeno\ }
\newcommand{\WG}{{\bf W}\ensuremath{_G}}
\newcommand{\PKS}{PKS\ 2155-304}
\newcommand{\Hnaught}{\ensuremath{H_0}}
\newcommand{\z}{\ensuremath{z}}
\newcommand{\zzerono}{\ensuremath{z=0}}
\newcommand{\zzero}{\zzerono\ }

\newcommand{\bb}{\ensuremath{b}}
\newcommand{\ang}{~\mbox{\AA}}
\newcommand{\nopercmtwono}{{\rm cm}\ensuremath{^{-2}}}

\newcommand{\persecondno}{\ {\rm s}\ensuremath{^{-1}}}
\newcommand{\percmtwo}{\ \nopercmtwono\ }
\newcommand{\percmtwono}{\ \nopercmtwono}

\newcommand{\perhzno}{\ {\rm Hz}\ensuremath{^{-1}}}

\newcommand{\perMpcno}{~{\rm Mpc}\ensuremath{^{-1}}}

\newcommand{\persrno}{~{\rm sr}\ensuremath{^{-1}}}
\newcommand{\nomang}{{\rm m}\mbox{\AA}}
\newcommand{\noang}{\mbox{\AA}}
\newcommand{\mang}{~{\rm m}\mbox{\AA}}

\newcommand{\Ang}{\ang\ }
\newcommand{\Mang}{\mang\ }
\newcommand{\Nno}{\ensuremath{\mathcal{N}}}

\newcommand{\Wno}{\ensuremath{\mathcal{W}}}
\newcommand{\W}{\Wno~}
\newcommand{\Ws}{\Wno s}
\newcommand{\Wi}{\ensuremath{\Wno_i}}
\newcommand{\nW}{\ensuremath{n(\Wno)}}
\newcommand{\nWi}{\ensuremath{n(\Wi)}}
\newcommand{\subH}{\ensuremath{_{\rm HI}}}
\newcommand{\fHI}{\ensuremath{f}\subH}
\newcommand{\THI}{\ensuremath{T}\subH}

\newcommand{\etno}{et~al.}
\newcommand{\et}{\etno\ }
\newcommand{\eti}{\etno}
\newcommand{\etl}{\et}

\newcommand{\taueff}{\ensuremath{\tau_{\rm eff}}}
\newcommand{\dtaudz}{d\taueff/\dz}
\newcommand{\hsfi}{\ensuremath{h^{-1}_{70}}}

\newcommand{\hohi}{\ensuremath{h^{-1}_{100}}}
\newcommand{\hhi}{\ensuremath{h_{100}}}

\newcommand{\Jno}{\ensuremath{J_0}}
\newcommand{\Jm}{\ensuremath{J_{-23}}}

\newcommand{\Jnuv}{\ensuremath{J_{\nu}}}
\newcommand{\sig}{\ensuremath{\sigma}\ }
\newcommand{\signo}{\ensuremath{\sigma}}

\newcommand{\onesig}{1\signo\ }
\newcommand{\onesigno}{1\signo}
\newcommand{\twosig}{2\signo\ }
\newcommand{\threesig}{3\signo\ }

\newcommand{\foursig}{4\signo\ }
\newcommand{\about}{\ensuremath{\sim}}
\newcommand{\zabout}[1]{\ensuremath{\z\sim#1}}
\newcommand{\nokmsno}{{\rm km~s}\ensuremath{^{-1}}}
\newcommand{\kmsno}{~\nokmsno}
\newcommand{\nommsno}{{\rm Mm}\persecondno}
\newcommand{\mmsno}{~\nommsno}
\newcommand{\mms}{\mmsno\ }

\newcommand{\kms}{\kmsno\ }

\newcommand{\lya}{Ly\ensuremath{\alpha} }
\newcommand{\lyano}{Ly\ensuremath{\alpha}}
\newcommand{\lyb}{Ly\ensuremath{\beta} }

\newcommand{\Nhno}{\ensuremath{N\subH}}
\newcommand{\fulllognh}{\ensuremath{\log{\left[\Nhno(\nopercmtwono)\right]}}}
\newcommand{\lognh}{\ensuremath{\log{\left[\Nhno\right]}}}
\newcommand{\logNh}{\lognh\ }
\newcommand{\Nh}{\Nhno\ }
\newcommand{\lowrange}{\ensuremath{12.3\leq\logNh\leq14.5}}

\newcommand{\Nmin}{\ensuremath{N_{\rm min}}}
\newcommand{\nNh}{\ensuremath{n(\Nhno)}}

\newcommand{\nNhno}{\ensuremath{n(\Nhno)}}

\newcommand{\nz}{\ensuremath{n(\z)}}

\newcommand{\myb}{25\kmsno}

\newcommand{\cgs}{ {\rm ergs}~{\rm cm}\ensuremath{^{-2}}\persecondno\perhzno\persrno}
\newcommand{\bobs}{\bb}

\newcommand{\bmsd}{\ensuremath{b_{\rm obs}}}
\newcommand{\SL}{\ensuremath{SL}}
\newcommand{\SLs}{\ensuremath{SL}s}
\newcommand{\definite}{\ensuremath{SL\ge4\signo}}
\newcommand{\expanded}{\ensuremath{SL\ge3\signo}}
\newcommand{\tent}{\ensuremath{3\signo\le SL\lt4\signo}}
\newcommand{\real}{\ensuremath{SL\ge4\signo}}

\newcommand{\cDz}{\ensuremath{c\Delta}\z}

\newcommand{\Dz}{\ensuremath{\Delta}\z}
\newcommand{\delz}{\ensuremath{\delta}\z}
\newcommand{\DW}{\ensuremath{\Delta}\Wno}
\newcommand{\DNh}{\ensuremath{\Delta}\Nhno}
\newcommand{\Dv}{\ensuremath{\Delta v}}

\newcommand{\dz}{{\rm d}\z}
\newcommand{\pz}{\partial\z}
\newcommand{\Nobs}{\ensuremath{N_{\rm obs}}}
\newcommand{\Nran}{\ensuremath{N_{\rm ran}}}
\newcommand{\Nsym}{C}

\newcommand{\NofNhno}{\ensuremath{\Nsym\subH}}
\newcommand{\NofNh}{\ensuremath{\Nsym\subH}~}

\newcommand{\betaNh}{\ensuremath{\NofNhno \Nhno^{-\beta}}}
\newcommand{\highbeta}{1.33}
\newcommand{\Ehighbeta}{0.30}
\newcommand{\highconstant}{\ensuremath{5.2\pm4.9}}
\newcommand{\lowbeta}{1.65}
\newcommand{\Elowbeta}{0.07}
\newcommand{\lowconstant}{\ensuremath{10.3\pm1.0}}
\newcommand{\gt}{\ensuremath{>}}
\newcommand{\lt}{\ensuremath{<}}

\newcommand{\zrange}{\ensuremath{0.002\lt\z\lt0.069}}

\newcommand{\zem}{\ensuremath{z_{\rm em}}}
\newcommand{\prox}{c\zem -- 1200\kms}
\newcommand{\proxno}{c\zem -- 1200\kmsno}
\newcommand{\lowz}{low-\z\ }
\newcommand{\lowzno}{low-\z }
\newcommand{\highz}{high-\z\ }
\newcommand{\highzno}{high-\z }
\newcommand{\lownhno}{low-\Nhno}
\newcommand{\highnhno}{high-\Nhno}
\newcommand{\lownh}{\lownhno\ }
\newcommand{\highnh}{\highnhno\ }
\newcommand{\lowzlya}{low-\z\ \lya}
\newcommand{\dn}{{\rm d}\Nno}
\newcommand{\dNh}{{\rm d}\Nh}

\newcommand{\pNh}{\partial\Nh}

\newcommand{\dW}{{\rm d}\Wno}
\newcommand{\pNhno}{\partial\Nhno}
\newcommand{\pW}{\partial\Wno}
\newcommand{\dndz}{\dn/\dz\ }
\newcommand{\dndzz}{\dn/\dz(\z)}

\newcommand{\dndzno}{\dn/\dz}
\newcommand{\dndzzero}{\ensuremath{(\dndzno)_{\zzero}}}
\newcommand{\dndzzerono}{\ensuremath{(\dndzno)_{\zzerono}}}
\newcommand{\zonelog}{\ensuremath{\log\left[1+\z\right]}}
\newcommand{\dndzlog}{\ensuremath{\log\left[\dndzno\right]}}
\newcommand{\logdndzzero}{\ensuremath{\log\left[\dndzzerono\right]}}

\newcommand{\dtwondnhdzover}{\ensuremath{{\partial^2 \Nno\over \pz~\pNh}}}
\newcommand{\dtwondnhdz}{\ensuremath{\partial^2\Nno/\pz\pNh}}
\newcommand{\dtwondnhdzno}{\ensuremath{\partial^2\Nno/\pz\pNhno}}
\newcommand{\dtwondzdnhno}{\dtwondnhdzno}

\newcommand{\dtwondzdnh}{\dtwondnhdz}
\newcommand{\dtwondWdz}{\ensuremath{{\partial^2\Nno/\pz\pW}}}
\newcommand{\dtwondWdzover}{\ensuremath{{\partial^2 \Nno \over \pz~\pW}}}

\newcommand{\cz}{\ensuremath{cz}}

\newcommand{\kimrange}{\ensuremath{13.1\lt\lognh\lt14.0}}

\newcommand{\bvalue}{\bb-value\ }
\newcommand{\bvalues}{\bb-values\ }
\newcommand{\bvalueno}{\bb-value}
\newcommand{\bvaluesno}{\bb-values}
\newcommand{\lam}{\ensuremath{\lambda}}

\newcommand{\Dlam}{\ensuremath{\Delta\lam}}
\newcommand{\Plam}{\ensuremath{P(\lam)}}

\newcommand{\NV}[1]{\ion{N}{5}~\lam#1}
\newcommand{\NI}[1]{\ion{N}{1}~\lam#1}

\newcommand{\CI}[1]{\ion{C}{1}~\lam#1}
\newcommand{\CIII}[1]{\ion{C}{3}~\lam#1}

\newcommand{\FeII}[1]{\ion{Fe}{2}~\lam#1}
\newcommand{\SII}[1]{\ion{S}{2}~\lam#1}
\newcommand{\SiII}[1]{\ion{Si}{2}~\lam#1}
\newcommand{\Sithree}{\ion{Si}{3}~\lam1206.5}
\newcommand{\OVI}[1]{O~VI~\lam#1}
\newcommand{\SiIII}[1]{\ion{Si}{3}~\lam#1}
\newcommand{\SFsixty}{\SiII{1260.4}\ +\ \FeII{1260.5}}
\newcommand{\NVdoublet}{\ion{N}{5}~\ensuremath{\lambda\lambda}1238.8, 1242.8}

\newcommand{\MGdoublet}{\ion{Mg}{2}~\ensuremath{\lambda\lambda}1239.9, 1240.4}
\newcommand{\OVIdoublet}{\ion{O}{6}~\ensuremath{\lambda\lambda}1031.9, 1037.6}
\newcommand{\NItriplet}{\ion{N}{1}~\ensuremath{\lambda\lambda}1199.5, 1200.2, 1200.7}
\newcommand{\SIItriplet}{\ion{S}{2}~\ensuremath{\lambda\lambda}1250.6, 1253.8, 1259.5}

 \newcommand{\omegab}{\ensuremath{\Omega_{\rm b}}}

\newcommand{\omegalya}{\ensuremath{\Omega_{\lyano}}}
\newcommand{\vlsrno}{\ensuremath{V_{\rm lsr}}}
\newcommand{\vlsr}{\vlsrno~}
\newcommand{\vobsno}{\ensuremath{V_{\rm obs}}}

\newcommand{\Nabs}{109}

\newcommand{\comboNabs}{187}

\newcommand{\nstis}{15}

\shorttitle{Statistics of the \lowzlya Forest}
\shortauthors{Penton, Stocke, \& Shull}
\received{2003 August 15}
\begin{document}
\title{The Local \lya Forest IV: STIS G140M Spectra and Results on the Distribution and Baryon Content of \ion{H}{1} Absorbers
\footnote{Based on observations with the NASA/ESA Hubble Space Telescope, obtained at the Space
Telescope Science Institute, which is operated by the Association of Universities for Research in
Astronomy, Inc. under NASA contract No. NAS5-26555.}}
\author{Steven V. Penton, John T. Stocke, and J. Michael Shull\altaffilmark{2}}
\affil{Center for Astrophysics and Space Astronomy, Department of Astrophysical and Planetary Sciences,
University of Colorado, Boulder CO, 80309}
\altaffiltext{2}{also at JILA, University of Colorado and National Institute of Standards and Technology.}
\email{spenton@casa.colorado.edu, stocke@casa.colorado.edu, mshull@casa.colorado.edu}
\begin{abstract}
We present HST~STIS/G140M spectra of \nstis\ extragalactic targets, which we combine with GHRS/G160M 
data to examine the statistical properties of the \lowzlya forest.
With STIS, we detect \Nabs\ \lya absorbers at significance level (\SL) $\geq 4\signo$
over \zrange, with a total redshift pathlength $\Dz=0.770$. 
Our combined sample consists of \comboNabs\ \lya absorbers with \real\ over $\Dz = 1.157$. 
We evaluate the physical
properties of these \lya absorbers and compare them to their \highz counterparts. 
Using two different models for \lya forest absorbers, we determine that the
warm, photoionized IGM contains
\totalomega\ of the total baryon inventory at \zzero (assuming $\Jno =1.3 \times 10^{-23}$\cgs).
We derive the distribution in column density,
\Nhno$^{-\lowbeta\pm \Elowbeta}$ for $12.5\leq\fulllognh\le14.5$, breaking to a flatter 
slope above $\lognh \approx 14.5$.
As with the high equivalent width ($\Wno\gt240\mang$) absorbers, the number density of
low-\W absorbers at \zzero is well above the extrapolation of \dndz from $\z\gt2$. 
However, $\logdndzzero = 1.40 \pm 0.08$ for $\Wno\gt240\mang$ is $25\%$ below the value obtained by the HST QSO Key Project,
a difference that may arise from line blending. The slowing of the number density evolution of high-\W
\lya clouds is not as great as previously measured, and the 
break to slower evolution may occur later than previously  suggested (\zabout{1.0} rather than 1.6). 
We find a $7.2\sigma$ excess in the two-point correlation function (TPCF)
of \lya absorbers for velocity separations $\Dv\leq260$\kmsno,  which is exclusively due to the higher column density clouds.
From our previous result that higher column density
\lya clouds cluster more strongly with galaxies, this TPCF suggests a physical
difference between the higher and lower column density clouds in our sample. 
The systematic error
produced by cosmic variance on these results increases the total errors
on derived quantities by $\sim\sqrt{2}$.
\end{abstract}
\section{Introduction}\label{sec:obs}
Since the discovery of the high-redshift \lya forest over
30 years ago, these abundant absorption features in the spectra
of QSOs have been used as evolutionary probes of the intergalactic
medium (IGM), galactic halos, large-scale structure, and chemical
evolution.
In the past few years,
these discrete \lya lines have been interpreted in the context of
N-body hydrodynamical models \citep{Cen94,Hernquist96,Zhang97,Dave99}
as arising from baryon density
fluctuations associated with gravitational instability during structure formation.
However, the detailed physical processes governing the recycling of metal-enriched gas,
from galaxy disks into extended but still gravitationally-bound galaxy halos or
into gravitationally-unbound winds, have not been included in these simulations to any
precision. Therefore, the physical and causal relationship between the \lya
forest absorbers and galaxies is still uncertain and controversial \citep[see][]{Mulchaey02}.
{\it Hubble Space Telescope} (HST) UV spectroscopy with the Faint Object Spectrograph (FOS) and Goddard High Resolution
Spectrograph (GHRS) in the past, with the Space Telescope Imaging Spectrograph (STIS) at
present, and with the Cosmic Origins Spectrograph (COS) in the future, can make a
significant contribution to this problem. This is due to the low redshift ($\z\lt0.1$) of the absorbers 
discovered with HST, allowing a detailed scrutiny of the nearby galaxy distribution not possible at \highzno.

At high redshift, the \lya absorption lines evolve rapidly with redshift, 
$d{\cal N}/dz \propto (1+z)^{\gamma}$, where $\gamma\approx 2.2$ 
for $1.5\leq\z\leq4$ \citep{Kim01}.
A major surprise came when HST discovered \lya absorption lines toward the quasar 3C~273 at $z_{\rm em}=0.158$, using
both FOS \citep{Bahcall91} and GHRS \citep{Morris91,Morris93}.
The number of these absorbers was far in excess of their expected number
based upon an extrapolation from \highz \citep{Weymann98}.
Current evidence suggests that the evolution of the \lya forest slowed dramatically at $\z\lt1.6$,
probably as a result of the collapse and assembly of baryonic structures in the IGM
together with the decline in the intensity of the ionizing radiation field \citep{Theuns98,Dave99,Haardt96,Shull99b}.
Detailed measurements of the \lya forest evolution in the 
interval $\z\lt1.5$, for equivalent widths $\Wno\gt240\mang$, are described in the FOS Key Project papers:
the three catalog papers \citep{Bahcall93,Bahcall96,Jannuzi98}
and the evolutionary analysis \citep{Weymann98}.

In previous papers in our series \citep[][ Papers~I, II, III, respectively]{PaperI, PaperII, PaperIII},
we used the HST/GHRS and the G160M first-order grating with \about19\kms resolution
to study the very low-redshift ($\z\lt0.07$) \lya forest at lower column densities than
was possible with the Key Project data 
(our limiting \W is \about15\mang\, or $\fulllognh\geq12.5$; hereafter, \Nh is implied to be in units of \percmtwono). 
Paper~II showed that the number density evolution of \lownh \lya forest absorbers 
also exhibits a rapid decline from higher redshift, but perhaps with only a minimal slowing in that
evolution at \zabout{1}. These low \Nhno\, absorbers show a small excess power in
the cloud-cloud two-point correlation function (TPCF) amplitude and only at $\cDz\leq150$\kms (Paper~II). 
Unlike results based on ground-based galaxy surveys near the \highnh FOS absorbers \citep{Lanzetta95, Chen98}, 
the low column density absorbers do not correlate closely with galaxies (Paper~III). 
In our GHRS sample, we found that $22\pm8$\% of the absorbers ``reside'' in galaxy voids. 
Even the remaining 78\% of the absorbers are not close to galaxies,
but may align with large-scale structures of galaxies (Paper~III) as predicted by recent simulations
\citep{Cen94, Dave99}. Thus, just as at \highzno, there appears to be a 
physical distinction between
the higher column density absorbers ($\lognh\geq14$) discovered in the
Key Project work and the lower column density absorbers investigated in Papers~I--III.
Earlier results from our study have appeared in various research papers
and reviews \citep{Stocke95, Shull97,Shull99a, Stocke02, Stocke03, Shull02, Shull03}.

In this paper we more than double our \lya sample using \nstis\ STIS targets. Combining these STIS/G140M results with 
the GHRS sample analyzed in Paper~II, we confirm and extend our previous results and
improve the statistics of our conclusions concerning the local \lya forest.
However, despite the much improved statistics, cosmic variance in absorber numbers and properties
may still be an important, although diminished, factor in the error analysis.
We also obtain several new results: 
(1) an inconclusive search for very broad, shallow absorbers in these spectra (see \S~\ref{sec:broad}); 
(2) a more accurate determination (\S~\ref{sec:dndzz2}) of the evolution in number density of high and low
column density absorbers, which differs from the Key Project
 in suggesting that the fast evolution of higher column density absorbers
may persist to \zabout{1}; 
(3) a more accurate accounting of the baryon density in the local \lya absorbers using
two different formulations (\S~\ref{sec:omega_b});
and (4) a separate cloud-cloud TPCF for higher and lower column density absorbers in our sample, which
shows a difference in clustering (\S~\ref{sec:TPCF}).

This paper is organized as follows. In \S~\ref{sec:spec}, we 
present the target sample and describe the basic data reduction and
analysis process. We also discuss the limitations of these data for
obtaining \bvaluesno. In 
\S~\ref{sec:REW} and \S~\ref{sec:Onh}, we discuss the basic properties of
our measured rest-frame equivalent width (\Wno) and \hone column 
density (\Nhno) distributions for our
\lya absorbers and compare them to higher-\z\ distributions.
In \S~\ref{sec:Z}, we discuss the \z\ distribution of the \lowzlya
forest within the small redshift range (\zrange) of our spectra, 
as well as the cumulative Lyman continuum opacity of
these absorbers and the \z\ evolution in the number density of lines, \dndzno.
In \S~\ref{sec:omega_b} we present a new accounting of the local IGM baryon
density which finds \totalomega\ of the baryons in the photoionized \lya absorbers.
In \S~\ref{sec:TPCF}, we analyze the cloud-cloud TPCF for \lowzlya clouds.
Section~\ref{sec:conc} summarizes the important conclusions of this  investigation.
The spectra and the detailed line list are presented in an Appendix. As in our previous
work, we assume a Hubble constant of \Hnaught $= 70 h_{70}$\kmsno\perMpcno.
\section{The HST/STIS+G140M Sample and Spectral Processing}\label{sec:spec}
In Table~\ref{stis_objects} we present the basic physical data for the 
\nstis\ STIS sightlines observed and analyzed in this program.
This table summarizes the J2000 positions in celestial and Galactic 
coordinate frames (columns 2-7) and emission-line redshifts (\z; column 8)
of our HST targets. All redshifts are the optical, narrow emission line 
redshifts as reported from
the NASA/IPAC Extragalactic Database (NED).\footnote{The NASA/IPAC 
Extragalactic Database (NED) is operated by the Jet
Propulsion Laboratory, California Institute of Technology, under 
contract with NASA.}
Also included in Table~\ref{stis_objects}, but discussed in detail later, are by column:
(9) the adjustment required to convert the observed wavelength scale to 
the Local Standard of Rest (LSR), where $\Delta \vlsr=\vlsr - \vobsno$,
is determined from the location of the available Galactic absorption 
lines (see Paper~I);
(10) the order ({$O$}) of the polynomial used to normalize the 
continuum of each target,
and (11) the mean signal-to-noise ratio (SNR) of each spectrum per 3.22 pixel (41.2\kms full
width at half maximum, FWHM) resolution element (RE). 
The Galactic \hone LSR velocity measurements are taken from the 
Leiden/Dwingeloo surveys \citep[LDS,][]{Dwing} for all targets except for PKS~2005-489,
 which we obtained from the Parkes HIPASS \citep{HIPASS} survey. The LDS
 velocities have an estimated uncertainty of $\pm2$\kmsno, while the HIPASS
 data have $\pm5$\kmsno.
\setlength{\tabcolsep}{1mm}
\begin{deluxetable}{lrrcccccrrrr}
\tablecolumns{11}
\tabletypesize{\footnotesize}
\tablecaption{Our 15 HST/STIS/G140M Targets \label{stis_objects}}
\tablewidth{0pt}
\tablehead{
\colhead{Target} &
\colhead{RA\tablenotemark{a}} &
\colhead{DEC\tablenotemark{a}} &
\colhead{RA\tablenotemark{a}} &
\colhead{DEC\tablenotemark{a}}&
\colhead{l} &
\colhead{b} &
\colhead{\z} &
\colhead{$\Delta \vlsr$\tablenotemark{b}} &
\colhead{$O$\tablenotemark{c}} &
\colhead{SNR\tablenotemark{d}} \\
\colhead{}  &
\colhead{(hh:mm:ss)}&
\colhead{(dd:mm:ss)}&
\colhead{($^{\circ}$)} &
\colhead{($^{\circ}$)} &
\colhead{($^{\circ}$)} &
\colhead{($^{\circ}$)} &
\colhead{ }&
\colhead{(\nokmsno)}&
\colhead{ }&
\colhead{ }  \\
\colhead{(1)}&\colhead{(2)}&
\colhead{(3)}&\colhead{(4)}&
\colhead{(5)}&\colhead{(6)}&
\colhead{(7)}&\colhead{(8)}&
\colhead{(9)}&\colhead{(10)}&
\colhead{(11)}
}
\startdata
HE1029-140&10 31 54.3&-14 16 52.4& 157.976& -14.281&-100.67&  36.51&  0.0860&23.0&          19&19.6\\
IIZW136&21 32 27.8&+10 08 19.5& 323.116&  10.139&  63.67& -29.07&  0.0630&26.9&          22&21.0\\
MR2251-178&22 54 05.9&-17 34 55.3& 343.524& -17.582&  46.20& -61.33&  0.0644& 8.9&          13&27.5\\
MRK478&14 42 07.5&+35 26 22.5& 220.531&  35.440&  59.24&  65.03&  0.0791&17.6&          21&24.8\\
MRK926&23 04 43.5&-08 41 08.4& 346.181&  -8.686&  64.09& -58.76&  0.0473&22.0&          29& 8.0\\
MRK1383&14 29 06.4&+01 17 05.1& 217.277&   1.285& -10.78&  55.13&  0.0865&32.4&          11&24.2\\
NGC985&02 34 37.7&-08 47 15.7&  38.657&  -8.788&-179.16& -59.49&  0.0431& 4.0&          23&21.1\\
PG0804+761&08 10 58.6&+76 02 42.6& 122.744&  76.045& 138.28&  31.03&  0.1000&15.4&          19&26.6\\
PG1116+215&11 19 08.7&+21 19 17.8& 169.786&  21.322&-136.64&  68.21&  0.1763&22.8&           3&19.1\\
PG1211+143&12 14 17.7&+14 03 12.0& 183.574&  14.053& -92.45&  74.31&  0.0809&12.9&          17&20.2\\
PG1351+640&13 53 15.8&+63 45 45.1& 208.316&  63.763& 111.89&  52.02&  0.0882&10.3&          24&15.3\\
PKS2005-489&20 09 25.5&-48 49 54.3& 302.356& -48.832&  -9.63& -32.60&  0.0710&-3.2&           8&37.5\\
TON-S180&00 57 20.0&-22 22 56.3&  14.333& -22.382& 139.00& -85.07&  0.0620&-5.8&          22&20.8\\
TON-1542&12 32 03.6&+20 09 29.3& 188.015&  20.158& -90.56&  81.74&  0.0630&17.5&          29&16.8\\
VIIZW118&07 07 13.1&+64 35 58.3& 106.805&  64.600& 151.36&  25.99&  0.0797&20.5&           8&19.9\\
\enddata

\tablenotetext{a}{J2000 coordinates.}
\tablenotetext{b}{$\Delta \vlsr=\vlsr - \vobsno$, as determined from the location of the Galactic absorption lines.}
\tablenotetext{c}{The order ($O$) of the polynomial used to normalize the spectrum.}
\tablenotetext{d}{Average signal-to-noise ratio (SNR) per resolution element of the target spectrum where extragalactic \lya may be detected.}

\end{deluxetable}
\normalsize

Our STIS spectra were recalibrated on September 21, 2002 using version V3.0 of STSDAS 
and V2.13b of CALSTIS via on-the-fly recalibration (OTFR) from the HST archive. 
Our observations were performed with the $52\arcsec \times 0.2\arcsec$ aperture, in either the
1195--1248\Ang or 1245--1299\Ang settings.
For all objects except \objectname[]{VII~ZW~118} both settings were used for each object. 
In general, the post-processing of our STIS spectra was identical to the procedures
used for our GHRS sample (Paper~I). 
The differences in the procedures for handling STIS data compared to the GHRS are:
\begin{itemize}
\item{The 1195\Ang cutoff in the lower STIS setting allows us to identify  Galactic
\NItriplet, and \SiIII{1206.5} absorption lines. The \ion{N}{1} lines contribute to the determination and
adjustment of our spectra to the local standard of rest (LSR).}
\item{Owing to concerns regarding the RE (see below), all spectral fits were performed on
the raw data. In our GHRS sample, spectral fits were performed on data
smoothed to the approximate spectral resolution.}
\item{Continuum normalization was performed by constructing a polynomial
of the order given in
Table~\ref{stis_objects}, combined with Gaussian components for emission
intrinsic to the target QSO (\lyano, \NI{1134.9}, and \CIII{1175.7}, when present)
and a Galactic \lya absorption component. 
The specific emission components included for each target are given in the Appendix.}
\item{Based upon the \threesig difference between the \Ws\ of features measured prior 
to and after our continuum normalization, we added a 4.2\% uncertainty in \W
in quadrature. In our GHRS sample we added a 3.4\% continuum-level uncertainty.}
\end{itemize}

Owing to the fact that the STIS 0.2\arcsec\ slit corresponds to a velocity shift of $\pm40$\kms 
(from a centered target), we have chosen not to use the heliocentric velocities
provided by the standard reduction software. Instead, we use the strong,
low ionization Galactic absorption lines of N~I, S~II, and Si~II present
in our spectra to provide a velocity zero point. 
We assume that these lines occur at the same LSR velocity as the dominant \ion{H}{1} emission in that direction.
As with our GHRS reduction, we expect that the wavelengths and recession velocities
quoted in Table~\ref{linelist_nh} and in the Appendix have accuracies of at least $\pm5$\kms.
The limitation is the accuracy with which \vlsrno(\honeno) can be determined from 
the 21 cm \hone emission line profile and the assumed correspondence between the \hone 
and the gas that gives rise to the Galactic metal absorption lines listed above.

\begin{figure}[htb] \epsscale{0.67}
\plotone{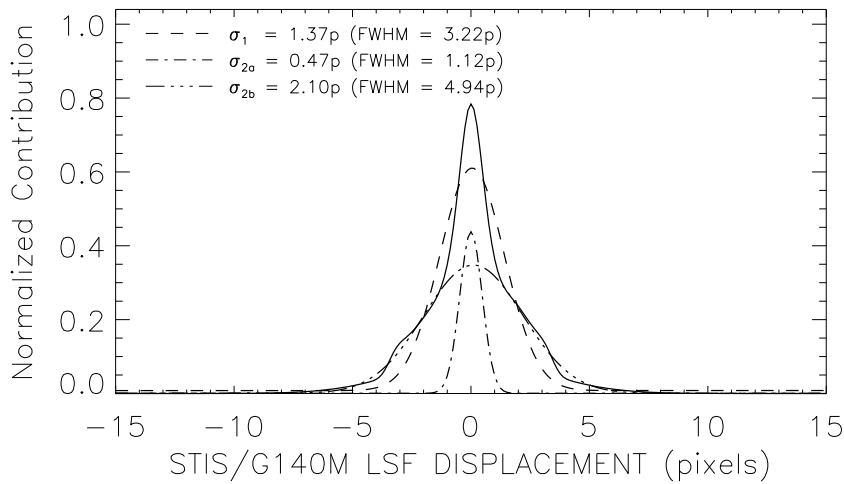}
\caption{\label{STIS_RE} The solid line displays the LSF at 1200\Ang for STIS/G140M 
 and the $52\arcsec\times0.2\arcsec$ aperture.
The LSF is best modeled as the combination of two Gaussian components of approximately equal 
strengths with FWHMs of 1.12 and 4.94 pixels ($\sigma_{2a}$=0.47 pixels, $\sigma_{2b}$=2.10 pixels),
shown as the dot-dashed Gaussians. 
The best-fit single Gaussian component, shown as the dashed line, has a FWHM of 3.22 pixels
($\sigma_{1}$=1.37 pixels). 
We use this latter value for the RE when calculating the significances and \bvalues of our 
absorption features.}
\end{figure}

For our bandpass, grating, and aperture, the STIS data handbook reports that the spectral 
RE, defined as the FWHM of the line spread function (LSF), is \about1.7  (0.05\ang) pixels at 1200\Ang 
($\about21$\kmsno). 
As shown in Figure~\ref{STIS_RE}, the actual LSF has considerable wings (solid line), and is best
modeled as a combination of two Gaussian components of approximately equal strengths.
The two components have FWHMs of 1.12 and 4.94 pixels, corresponding to Gaussian widths, $\sigma_{2a}$ and
$\sigma_{2b}$ in Figure~\ref{STIS_RE}, of 0.47 and 2.10 pixels respectively. 
The best-fit single Gaussian component, also shown in Figure~\ref{STIS_RE}, has a
FWHM of 3.22 pixels (\about $41.2$\kmsno),  corresponding to a Gaussian width, $\sigma_{1}$, of 1.37 pixels (17.5\kmsno). 
We believe this to be the best value for approximating the STIS spectral RE
using a single-Gaussian fit, not the 1.7-pixel value obtained from the STIS instrument handbook.
This RE is used to determine the significance levels (\SLs) and
\bvalues of our absorption features
(see Paper~I), but it does not affect the measured \Ws.
Choosing to use this single Gaussian approximation slightly
underestimates the \SLs\ of narrow absorption features in our sample,
but it does not affect the \Ws.

Doppler widths and \bvalues are estimated from the velocity widths
(\WG=\bmsd/$\sqrt{2}$) of our fitted Gaussian components to the unsmoothed spectra. 
As such, they are not true measurements of the actual \bvaluesno, as when fitting Voigt profiles
or determining a curve of growth,
but rather velocity dispersions assuming that the \lya absorption lines may contain multiple
velocity components \citep{Shull00} and are not heavily saturated. 
The latter is a particularly good assumption for the large number of low-\W lines 
($\Wno\lt75$\mang), but it becomes increasingly suspect for the higher \W lines.
Unlike our GHRS analysis, owing to the uncertainty in accurately characterizing
the STIS LSF, we elected not to smooth our data with the LSF prior to
fitting the components. Thus, our observed \bvalue (\bmsd) is simply the
convolution of the instrumental profile ($\bobs_{LSF}$) and the actual \bvalue (\bobs) of the absorber. 
The \bvalues add in quadrature, $\bobs^2=\bmsd^2-\bobs_{LSF}^2=2(\WG^2-\sigma_{1}^2)$.
Therefore, we are hampered in detecting \lya absorbers with observed 
\bvalues near or below $\bmsd=\sqrt{2}~\sigma_{1} \approx 25$\kmsno.

Motion of the target in the STIS aperture during our lengthy
exposures further broadens the LSF. In an attempt to correct for this, 
the jitter files from the HST observations were used to degrade the 
LSF in a manner unique for each observation. 
For our data, the removal of the jitter appears to be imperfect due to the potentially 
large amplitude (the 0.2\arcsec\ slit width corresponds to \about80\kms in our 
bandpass) and the incomplete details of the temporal extent of large jitter excursions. 
Nevertheless, for each observation, the degradation of the LSF was simulated, measured, and  
removed in quadrature from our observed \bvaluesno. 
Tables~\ref{linelist_nh} and \ref{linelist_all} present these corrected \bvaluesno. 
A comparison between the STIS/G140M \bvalues and the GHRS/G160M \bvalues in Paper~I 
reveals that the STIS values are significantly
larger (factor \about1.5 in both the median and the mean) than the GHRS values. 
Since some of these same targets were observed with the STIS E140M, direct
comparisons of \bvalues for \about $40$ \lya absorbers are currently available to us.
While the dispersion is substantial, the \bvalues reported here are \about2 times larger
than those measured from the higher resolution HST/STIS echelle spectra.
We attribute this difference to spacecraft drift and residual jitter that we were unable 
to remove. Additionally, at the time of our observations, the HST drift rate 
for a typical exposure was $\ge 0.02\arcsec$, or $10$\kms per hour for the G140M.
Therefore, we consider these \bvalues suspect and do not use them in any analysis.

Making an accurate measurement of \bvalues is important in determining the
actual \hone column densities (\Nhno) of the saturated
\lya absorbers, since each \bvalue produces a different curve of growth
for the upper range of column densities in our sample.
It has been shown that the only reliable method of deriving \bvalues 
for partially saturated \lya forest lines is by a curve-of-growth using
higher Lyman series lines \citep{Hurwitz98,Shull00}.
Significantly lower \bvalues (factor of 2 in the median) are found from the curve-of-growth
technique \citep{Shull00}. 
For example, the 1586\kms absorber in the \objectname[]{3C~273} sightline has $\bb=72$\kms
from the \lya profile \citep{Weymann95}, while the curve of growth value is 16\kms \citep{Sembach01}. 
The column density estimates based upon these \bvalues differ by a factor of 40.
This disagreement in estimating \bvalues can be understood if the \lya absorption profiles 
include non-thermal broadening from cosmological expansion and infall, arising from velocity
shear and  a few  unresolved velocity components. 
\citet{Hu95} reach the same conclusion, based upon \zabout{3} Keck HIRES QSO spectra.
Therefore, given the difficulties in measuring the LSFs for individual or co-added spectra 
and in interpreting the \bvalues derived from them, we elect to assume a constant \bvalue of 
25\kms as in Paper~II.
This \bvalue is similar to the median value found in a higher resolution study of 
\zzero \lya absorbers by \citet{Dave01}.
We avoid all analyses using the individual \bvalues in Tables~\ref{linelist_nh} and \ref{linelist_all}, and 
we strongly suggest that others do the same. In Table~\ref{linelist_nh} we 
list column densities for the \lya absorbers with \bvalues of 20, 25, and 30\kms to illustrate 
the uncertainties in the inferred column densities.

In Appendix~\ref{sec:ApA}, we present the STIS spectra, including  sensitivity and available 
pathlength estimates and descriptions of relevant Galactic and extragalactic
data for each spectrum, as well as a list of all absorption lines. 
We quote scientific results for the \real\ sample. We have verified that 
the noise characteristics of our spectra are Gaussian distributed around the continuum 
fits shown on the spectra in the Appendix.
Given the number of independent REs in our spectra, we expect that 10\% or \about 5 of the \tent\ 
absorption lines are likely to be spurious, while  $\lesssim0.1\%$ spurious absorptions would be expected
in our \real\ sample.
To ensure that our study examines only the IGM, as explained in Paper~I, we 
exclude all Galactic metal-line absorbers and absorbers ``intrinsic'' to the AGN, 
with $(cz_{AGN}-cz_{abs})\lt1200$\kmsno, 
to create the STIS \lya \real\ sample, presented in Table~\ref{linelist_nh}.
Owing to the substantial breadth of the \lya absorptions intrinsic 
to the AGN, metal lines in these systems can be at redshifts
significantly displaced ($\pm 100$\kmsno) from the best-fit redshifts for \lyano. 
This is particularly true when the \lya absorptions are significantly
blended, making precise wavelength determination of components problematical. 
Using a relaxed criterion for association of \lya and metal line absorbers ($\pm 100$\kmsno), we have
searched for metal lines  associated with intrinsic absorbers. 
Usually, these lines fall outside our observed waveband 
or the intrinsic absorber is weak or absent. However, we now identify 
three weak absorbers at 1241.62, 1242.25 and 1244.38\Ang in the \objectname[]{MRK~279} sightline as intrinsic \Sithree. Higher resolution STIS echelle observations
of this target confirm these revised identifications (Arav, private communication).
Similarly, two weak, intrinsic \ion{Si}{3} absorptions are present in the \objectname[]{MR~2251-178} sightline.
While these identifications are slightly more uncertain than the others in Table~\ref{linelist_all} 
(due to the relaxed wavelength criterion),
there are so few of them that they do not alter the statistical results presented here.

By column, Table~\ref{linelist_nh} lists:
(1) target name; 
(2) LSR-adjusted absorber wavelength and error in\ang; these errors include
the Gaussian centroid measurement error added in quadrature with the estimated
$\pm5$\kms error in setting the wavelength scale zero point (see above);
(3) absorber recession velocity, \cz~in\kmsno;
(4) observed \bvalue (\bmsd) in\kmsno;
(5) resolution-corrected \bvalue in\kmsno;
(6) rest-frame equivalent width in\Mang with the total uncertainty as previously described;
(7-9) estimated column densities in\percmtwo assuming \bvalues of 20, 25, 30\kmsno;
and (10) significance levels (\SL) of each absorber in our STIS sample. A similar table
for our GHRS sample of \lya absorbers can be found in Paper~II, Table~1. 
As described in Paper~I, the uncertainties, \DW, for the \W values 
are from the Gaussian fit to each feature and are not the significance level, \SL, of the 
absorption feature; typically $\Wno/\DW\lt\SL$.
\begin{deluxetable}{lccccclllr}
\tableheadfrac{0.02}
\tablecolumns{10}
\tabletypesize{\footnotesize}
\tablecaption{H~I Column Densities (\Nhno) of \real\ \lya features\label{linelist_nh}}
\tablewidth{0pt}
\tablehead{
\colhead{Target} &
\colhead{Wavelength\tablenotemark{a}} &
\colhead{Velocity\tablenotemark{b}} &
\colhead{\bmsd} &
\colhead{\bobs\tablenotemark{c}} &
\colhead{\Wno\tablenotemark{d}} &
\multicolumn{3}{c}{\Nhno\tablenotemark{e}} &
\colhead{SL\tablenotemark{f}} \\
\colhead{Name} &
\colhead{(\AA)} &
\colhead{(\nokmsno)} &
\colhead{(\nokmsno)} &
\colhead{(\nokmsno)} &
\colhead{(m\AA)} &
\colhead{\bb=20} &
\colhead{\bb=25} &
\colhead{\bb=30} &
\colhead{}
}
\startdata

HE1029-140& 1223.66 $\pm$ 0.04& 1971 $\pm$ 11& 59 $\pm$ 9& 53 $\pm$ 10& 110 $\pm$ 39&13.46&13.42&13.40& 10.1\\
HE1029-140& 1224.60 $\pm$ 0.04& 2202 $\pm$ 10& 43 $\pm$ 14& 35 $\pm$ 17& 45 $\pm$ 31&12.97&12.96&12.95& 4.5\\
HE1029-140& 1225.50 $\pm$ 0.04& 2423 $\pm$ 9& 49 $\pm$ 4& 42 $\pm$ 5& 183 $\pm$ 32&13.84&13.75&13.70& 17.9\\
HE1029-140& 1234.01 $\pm$ 0.06& 4523 $\pm$ 15& 58 $\pm$ 17& 52 $\pm$ 19& 59 $\pm$ 37&13.11&13.09&13.08& 6.4\\
HE1029-140& 1277.99 $\pm$ 0.04& 15369 $\pm$ 10& 51 $\pm$ 23& 44 $\pm$ 26& 30 $\pm$ 32&12.78&12.78&12.77& 4.2\\
HE1029-140& 1278.38 $\pm$ 0.04& 15464 $\pm$ 9& 48 $\pm$ 3& 42 $\pm$ 3& 278 $\pm$ 26&14.50&14.17&14.03& 38.0\\
HE1029-140& 1292.47 $\pm$ 0.05& 18941 $\pm$ 12& 43 $\pm$ 11& 36 $\pm$ 13& 38 $\pm$ 20&12.88&12.88&12.87& 5.5\\
HE1029-140& 1293.35 $\pm$ 0.04& 19157 $\pm$ 10& 38 $\pm$ 5& 30 $\pm$ 7& 62 $\pm$ 18&13.13&13.11&13.10& 9.4\\
IIZW136& 1249.41 $\pm$ 0.03& 8320 $\pm$ 7& 41 $\pm$ 3& 33 $\pm$ 4& 192 $\pm$ 22&13.90&13.79&13.73& 27.2\\
IIZW136& 1264.67 $\pm$ 0.04& 12084 $\pm$ 9& 38 $\pm$ 5& 29 $\pm$ 7& 71 $\pm$ 21&13.21&13.19&13.17& 9.6\\
IIZW136& 1265.52 $\pm$ 0.04& 12294 $\pm$ 9& 50 $\pm$ 12& 44 $\pm$ 14& 52 $\pm$ 27&13.05&13.03&13.02& 7.0\\
IIZW136& 1272.55 $\pm$ 0.03& 14026 $\pm$ 8& 37 $\pm$ 4& 27 $\pm$ 6& 72 $\pm$ 19&13.21&13.19&13.18& 10.4\\
IIZW136& 1285.80 $\pm$ 0.03& 17293 $\pm$ 7& 83 $\pm$ 3& 79 $\pm$ 3& 486 $\pm$ 23&16.74&15.77&15.11& 105.3\\
MR2251-178& 1224.74 $\pm$ 0.04& 2237 $\pm$ 11& 21 $\pm$ 7& $\lt 20$ & 39 $\pm$ 34&12.90&12.89&12.89& 4.1\\
MR2251-178& 1224.96 $\pm$ 0.06& 2291 $\pm$ 14& 36 $\pm$ 14& 26 $\pm$ 20& 52 $\pm$ 46&13.04&13.03&13.02& 5.6\\
MR2251-178& 1227.98 $\pm$ 0.05& 3035 $\pm$ 13& 52 $\pm$ 13& 46 $\pm$ 15& 60 $\pm$ 32&13.11&13.10&13.08& 6.5\\
MR2251-178& 1228.67 $\pm$ 0.04& 3205 $\pm$ 10& 73 $\pm$ 4& 69 $\pm$ 4& 349 $\pm$ 37&15.21&14.61&14.32& 38.7\\
MR2251-178& 1233.38 $\pm$ 0.04& 4368 $\pm$ 11& 55 $\pm$ 17& 49 $\pm$ 19& 40 $\pm$ 28&12.92&12.91&12.90& 4.8\\
MR2251-178& 1252.25 $\pm$ 0.04& 9021 $\pm$ 11& 40 $\pm$ 10& 32 $\pm$ 13& 51 $\pm$ 28&13.04&13.02&13.01& 6.6\\
MR2251-178& 1255.14 $\pm$ 0.04& 9735 $\pm$ 10& 50 $\pm$ 3& 43 $\pm$ 3& 181 $\pm$ 23&13.83&13.74&13.69& 24.0\\
MR2251-178& 1257.74 $\pm$ 0.04& 10375 $\pm$ 11& 69 $\pm$ 24& 65 $\pm$ 25& 38 $\pm$ 30&12.89&12.88&12.87& 5.5\\
MR2251-178& 1272.86 $\pm$ 0.04& 14103 $\pm$ 11& 29 $\pm$ 7& $\lt 20$ & 20 $\pm$ 10&12.58&12.58&12.58& 4.4\\
MR2251-178& 1278.80 $\pm$ 0.04& 15569 $\pm$ 10& 21 $\pm$ 5& $\lt 20$ & 17 $\pm$ 8&12.51&12.50&12.50& 4.2\\
MR2251-178& 1280.48 $\pm$ 0.04& 15982 $\pm$ 10& 36 $\pm$ 5& 26 $\pm$ 7& 30 $\pm$ 9&12.77&12.77&12.76& 8.2\\
MRK478& 1222.09 $\pm$ 0.04& 1582 $\pm$ 10& 52 $\pm$ 4& 45 $\pm$ 4& 194 $\pm$ 31&13.90&13.79&13.74& 22.5\\
MRK478& 1251.31 $\pm$ 0.04& 8788 $\pm$ 10& 63 $\pm$ 3& 58 $\pm$ 3& 290 $\pm$ 30&14.61&14.24&14.08& 39.3\\
MRK478& 1295.12 $\pm$ 0.04& 19593 $\pm$ 10& 49 $\pm$ 5& 43 $\pm$ 6& 84 $\pm$ 18&13.30&13.27&13.26& 16.2\\
MRK926& 1245.05 $\pm$ 0.05& 7245 $\pm$ 12& 70 $\pm$ 14& 66 $\pm$ 14& 179 $\pm$ 75&13.83&13.73&13.69& 9.7\\
MRK926& 1246.07 $\pm$ 0.06& 7496 $\pm$ 15& 51 $\pm$ 17& 45 $\pm$ 20& 72 $\pm$ 53&13.21&13.19&13.18& 4.4\\
MRK926& 1255.11 $\pm$ 0.05& 9726 $\pm$ 12& 42 $\pm$ 11& 33 $\pm$ 14& 77 $\pm$ 44&13.25&13.23&13.21& 4.8\\
MRK926& 1256.28 $\pm$ 0.10& 10015 $\pm$ 23& 77 $\pm$ 32& 73 $\pm$ 34& 66 $\pm$ 61&13.17&13.15&13.14& 4.4\\
MRK926& 1263.19 $\pm$ 0.04& 11720 $\pm$ 9& 33 $\pm$ 3& 22 $\pm$ 5& 117 $\pm$ 25&13.50&13.46&13.43& 11.4\\
MRK1383& 1244.44 $\pm$ 0.03& 7094 $\pm$ 8& 36 $\pm$ 6& 26 $\pm$ 8& 54 $\pm$ 19&13.06&13.05&13.04& 7.3\\
MRK1383& 1245.51 $\pm$ 0.04& 7359 $\pm$ 9& 39 $\pm$ 7& 31 $\pm$ 10& 35 $\pm$ 15&12.85&12.84&12.84& 6.8\\
MRK1383& 1250.10 $\pm$ 0.03& 8490 $\pm$ 8& 69 $\pm$ 4& 65 $\pm$ 5& 218 $\pm$ 30&14.05&13.90&13.82& 30.7\\
MRK1383& 1251.97 $\pm$ 0.04& 8951 $\pm$ 10& 55 $\pm$ 9& 49 $\pm$ 10& 66 $\pm$ 22&13.16&13.14&13.13& 9.5\\
MRK1383& 1257.20 $\pm$ 0.05& 10242 $\pm$ 12& 59 $\pm$ 14& 53 $\pm$ 15& 54 $\pm$ 27&13.06&13.05&13.04& 7.7\\
MRK1383& 1261.28 $\pm$ 0.04& 11247 $\pm$ 9& 49 $\pm$ 9& 42 $\pm$ 10& 61 $\pm$ 24&13.13&13.11&13.10& 8.5\\
MRK1383& 1278.79 $\pm$ 0.03& 15566 $\pm$ 8& 52 $\pm$ 3& 46 $\pm$ 3& 282 $\pm$ 20&14.53&14.19&14.05& 45.9\\
MRK1383& 1280.08 $\pm$ 0.07& 15883 $\pm$ 18& 36 $\pm$ 26& 27 $\pm$ 34& 26 $\pm$ 31&12.71&12.70&12.70& 4.2\\
MRK1383& 1283.15 $\pm$ 0.06& 16640 $\pm$ 15& 50 $\pm$ 18& 44 $\pm$ 20& 25 $\pm$ 20&12.70&12.69&12.69& 4.1\\
NGC985& 1224.41 $\pm$ 0.08& 2156 $\pm$ 19& 66 $\pm$ 25& 61 $\pm$ 27& 51 $\pm$ 42&13.03&13.02&13.01& 4.5\\
PG0804+761& 1220.32 $\pm$ 0.05& 1147 $\pm$ 11& 46 $\pm$ 4& 38 $\pm$ 4& 165 $\pm$ 29&13.75&13.67&13.63& 18.9\\
PG0804+761& 1221.87 $\pm$ 0.05& 1530 $\pm$ 12& 51 $\pm$ 9& 45 $\pm$ 10& 78 $\pm$ 28&13.25&13.23&13.22& 10.5\\
PG0804+761& 1222.24 $\pm$ 0.05& 1621 $\pm$ 11& 40 $\pm$ 11& 32 $\pm$ 15& 41 $\pm$ 27&12.92&12.91&12.91& 5.6\\
PG0804+761& 1238.18 $\pm$ 0.04& 5552 $\pm$ 11& 56 $\pm$ 3& 50 $\pm$ 4& 324 $\pm$ 44&14.95&14.44&14.21& 58.5\\
PG0804+761& 1238.61 $\pm$ 0.05& 5658 $\pm$ 11& 48 $\pm$ 43& 41 $\pm$ 51& 28 $\pm$ 33&12.74&12.73&12.73& 5.0\\
PG0804+761& 1247.62 $\pm$ 0.04& 7880 $\pm$ 11& 22 $\pm$ 5& $\lt 20$ & 18 $\pm$ 9&12.53&12.53&12.53& 4.2\\
PG0804+761& 1287.02 $\pm$ 0.05& 17597 $\pm$ 12& 53 $\pm$ 7& 47 $\pm$ 8& 72 $\pm$ 21&13.21&13.19&13.18& 12.0\\
PG0804+761& 1287.68 $\pm$ 0.05& 17758 $\pm$ 13& 43 $\pm$ 10& 36 $\pm$ 11& 38 $\pm$ 18&12.88&12.87&12.87& 6.3\\
PG0804+761& 1289.98 $\pm$ 0.05& 18326 $\pm$ 11& 26 $\pm$ 5& $\lt 20$ & 37 $\pm$ 14&12.87&12.86&12.85& 6.1\\
PG0804+761& 1292.38 $\pm$ 0.05& 18918 $\pm$ 12& 40 $\pm$ 9& 32 $\pm$ 11& 34 $\pm$ 17&12.83&12.82&12.82& 5.8\\
PG1116+215& 1221.75 $\pm$ 0.04& 1499 $\pm$ 9& 58 $\pm$ 11& 52 $\pm$ 12& 82 $\pm$ 33&13.28&13.26&13.24& 8.6\\
PG1116+215& 1235.59 $\pm$ 0.03& 4913 $\pm$ 8& 39 $\pm$ 7& 30 $\pm$ 9& 90 $\pm$ 32&13.33&13.31&13.29& 9.4\\
PG1116+215& 1250.21 $\pm$ 0.03& 8518 $\pm$ 8& 51 $\pm$ 3& 45 $\pm$ 4& 227 $\pm$ 32&14.11&13.94&13.85& 27.2\\
PG1116+215& 1254.99 $\pm$ 0.03& 9696 $\pm$ 8& 35 $\pm$ 8& 25 $\pm$ 11& 68 $\pm$ 34&13.18&13.16&13.15& 8.8\\
PG1116+215& 1265.78 $\pm$ 0.12& 12357 $\pm$ 30& 92 $\pm$ 37& 89 $\pm$ 39& 88 $\pm$ 82&13.32&13.29&13.28& 10.6\\
PG1116+215& 1266.47 $\pm$ 0.21& 12529 $\pm$ 51& 83 $\pm$ 53& 80 $\pm$ 55& 44 $\pm$ 53&12.96&12.95&12.94& 5.4\\
PG1116+215& 1269.61 $\pm$ 0.06& 13301 $\pm$ 16& 87 $\pm$ 21& 83 $\pm$ 22& 65 $\pm$ 34&13.16&13.14&13.13& 8.3\\
PG1116+215& 1287.44 $\pm$ 0.03& 17698 $\pm$ 7& 56 $\pm$ 5& 51 $\pm$ 6& 167 $\pm$ 31&13.76&13.68&13.64& 20.0\\
PG1116+215& 1289.58 $\pm$ 0.04& 18227 $\pm$ 9& 45 $\pm$ 8& 38 $\pm$ 9& 77 $\pm$ 28&13.25&13.22&13.21& 8.9\\
PG1116+215& 1291.75 $\pm$ 0.07& 18763 $\pm$ 16& 62 $\pm$ 21& 57 $\pm$ 23& 42 $\pm$ 31&12.94&12.93&12.92& 4.8\\
PG1211+143& 1224.31 $\pm$ 0.06& 2130 $\pm$ 18& 100 $\pm$ 3& 97 $\pm$ 3& 186 $\pm$ 19&13.86&13.76&13.71& 19.1\\
PG1211+143& 1235.72 $\pm$ 0.06& 4944 $\pm$ 17& 63 $\pm$ 7& 58 $\pm$ 7& 189 $\pm$ 46&13.88&13.77&13.72& 21.1\\
PG1211+143& 1236.09 $\pm$ 0.06& 5036 $\pm$ 17& 42 $\pm$ 5& 34 $\pm$ 6& 154 $\pm$ 40&13.69&13.63&13.59& 17.6\\
PG1211+143& 1242.49 $\pm$ 0.06& 6615 $\pm$ 18& 41 $\pm$ 10& 32 $\pm$ 13& 89 $\pm$ 54&13.33&13.30&13.29& 10.9\\
PG1211+143& 1244.06 $\pm$ 0.06& 7002 $\pm$ 17& 59 $\pm$ 5& 54 $\pm$ 6& 150 $\pm$ 30&13.67&13.61&13.57& 17.7\\
PG1211+143& 1247.11 $\pm$ 0.06& 7752 $\pm$ 17& 79 $\pm$ 6& 75 $\pm$ 6& 159 $\pm$ 26&13.72&13.65&13.61& 25.7\\
PG1211+143& 1268.43 $\pm$ 0.06& 13010 $\pm$ 17& 60 $\pm$ 7& 54 $\pm$ 8& 216 $\pm$ 56&14.03&13.89&13.81& 29.3\\
PG1211+143& 1268.66 $\pm$ 0.06& 13068 $\pm$ 17& 40 $\pm$ 5& 31 $\pm$ 7& 95 $\pm$ 30&13.37&13.34&13.32& 13.0\\
PG1211+143& 1270.63 $\pm$ 0.06& 13552 $\pm$ 17& 28 $\pm$ 5& $\lt 20$ & 44 $\pm$ 17&12.95&12.94&12.94& 5.9\\
PG1211+143& 1277.72 $\pm$ 0.06& 15301 $\pm$ 17& 60 $\pm$ 4& 55 $\pm$ 5& 308 $\pm$ 54&14.78&14.34&14.15& 40.5\\
PG1211+143& 1278.22 $\pm$ 0.06& 15426 $\pm$ 17& 100 $\pm$ 3& 97 $\pm$ 3& 691 $\pm$ 25&17.71&17.38&16.73& 90.6\\
PG1211+143& 1278.95 $\pm$ 0.06& 15605 $\pm$ 17& 49 $\pm$ 5& 42 $\pm$ 5& 132 $\pm$ 28&13.58&13.53&13.50& 17.2\\
PG1211+143& 1281.84 $\pm$ 0.06& 16318 $\pm$ 17& 59 $\pm$ 26& 54 $\pm$ 28& 51 $\pm$ 49&13.03&13.02&13.01& 7.2\\
PG1211+143& 1294.05 $\pm$ 0.06& 19328 $\pm$ 18& 78 $\pm$ 3& 74 $\pm$ 3& 564 $\pm$ 31&17.27&16.52&15.70& 92.4\\
PG1211+143& 1294.61 $\pm$ 0.06& 19468 $\pm$ 18& 61 $\pm$ 3& 56 $\pm$ 3& 249 $\pm$ 26&14.27&14.04&13.93& 41.7\\
PG1211+143& 1295.17 $\pm$ 0.06& 19604 $\pm$ 18& 65 $\pm$ 11& 60 $\pm$ 12& 67 $\pm$ 25&13.17&13.15&13.14& 11.4\\
PG1351+640& 1246.68 $\pm$ 0.04& 7647 $\pm$ 13& 20 $\pm$ 4& $\lt 20$ & 34 $\pm$ 15&12.84&12.83&12.83& 4.3\\
PG1351+640& 1297.29 $\pm$ 0.04& 20127 $\pm$ 14& 26 $\pm$ 5& $\lt 20$ & 42 $\pm$ 18&12.93&12.92&12.92& 5.1\\
PKS2005-489& 1226.83 $\pm$ 0.06& 2752 $\pm$ 15& 40 $\pm$ 12& 31 $\pm$ 15& 24 $\pm$ 15&12.68&12.67&12.67& 5.1\\
PKS2005-489& 1235.73 $\pm$ 0.05& 4947 $\pm$ 13& 64 $\pm$ 3& 59 $\pm$ 3& 299 $\pm$ 26&14.70&14.29&14.11& 68.1\\
PKS2005-489& 1236.19 $\pm$ 0.05& 5061 $\pm$ 12& 47 $\pm$ 3& 40 $\pm$ 3& 281 $\pm$ 21&14.53&14.19&14.04& 64.1\\
PKS2005-489& 1246.45 $\pm$ 0.05& 7589 $\pm$ 13& 53 $\pm$ 20& 47 $\pm$ 23& 14 $\pm$ 11&12.42&12.42&12.41& 4.5\\
PKS2005-489& 1266.74 $\pm$ 0.06& 12594 $\pm$ 16& 50 $\pm$ 14& 44 $\pm$ 16& 22 $\pm$ 13&12.63&12.62&12.62& 5.3\\
PKS2005-489& 1273.76 $\pm$ 0.05& 14326 $\pm$ 13& 46 $\pm$ 6& 39 $\pm$ 7& 49 $\pm$ 13&13.01&13.00&12.99& 11.8\\
PKS2005-489& 1277.57 $\pm$ 0.06& 15265 $\pm$ 14& 35 $\pm$ 7& 25 $\pm$ 10& 27 $\pm$ 12&12.72&12.72&12.71& 6.6\\
PKS2005-489& 1285.84 $\pm$ 0.05& 17306 $\pm$ 13& 52 $\pm$ 3& 47 $\pm$ 3& 294 $\pm$ 15&14.65&14.26&14.10& 72.7\\
PKS2005-489& 1294.65 $\pm$ 0.05& 19478 $\pm$ 13& 44 $\pm$ 3& 37 $\pm$ 3& 249 $\pm$ 15&14.26&14.03&13.93& 63.3\\
TON-S180& 1223.45 $\pm$ 0.05& 1919 $\pm$ 12& 43 $\pm$ 9& 35 $\pm$ 11& 66 $\pm$ 28&13.16&13.15&13.13& 6.5\\
TON-S180& 1226.92 $\pm$ 0.05& 2774 $\pm$ 12& 38 $\pm$ 10& 28 $\pm$ 13& 49 $\pm$ 26&13.01&13.00&12.99& 4.9\\
TON-S180& 1227.77 $\pm$ 0.05& 2985 $\pm$ 12& 33 $\pm$ 9& 21 $\pm$ 13& 41 $\pm$ 23&12.92&12.91&12.91& 4.3\\
TON-S180& 1237.98 $\pm$ 0.05& 5502 $\pm$ 11& 57 $\pm$ 4& 51 $\pm$ 5& 268 $\pm$ 54&14.42&14.12&14.00& 31.2\\
TON-S180& 1241.18 $\pm$ 0.05& 6290 $\pm$ 12& 38 $\pm$ 8& 28 $\pm$ 10& 54 $\pm$ 23&13.06&13.04&13.03& 6.2\\
TON-S180& 1244.13 $\pm$ 0.04& 7017 $\pm$ 11& 61 $\pm$ 4& 56 $\pm$ 4& 222 $\pm$ 29&14.07&13.91&13.84& 30.0\\
TON-S180& 1257.90 $\pm$ 0.05& 10415 $\pm$ 13& 38 $\pm$ 10& 30 $\pm$ 13& 41 $\pm$ 23&12.93&12.92&12.91& 5.0\\
TON-S180& 1268.03 $\pm$ 0.05& 12912 $\pm$ 12& 85 $\pm$ 14& 81 $\pm$ 15& 107 $\pm$ 39&13.44&13.40&13.38& 14.3\\
TON-S180& 1268.66 $\pm$ 0.05& 13068 $\pm$ 12& 57 $\pm$ 5& 52 $\pm$ 5& 140 $\pm$ 27&13.62&13.56&13.53& 18.7\\
TON-S180& 1270.47 $\pm$ 0.05& 13515 $\pm$ 12& 63 $\pm$ 6& 58 $\pm$ 7& 140 $\pm$ 29&13.62&13.56&13.53& 19.0\\
TON-S180& 1271.15 $\pm$ 0.05& 13681 $\pm$ 11& 59 $\pm$ 4& 54 $\pm$ 4& 212 $\pm$ 29&14.01&13.87&13.80& 28.1\\
TON-1542& 1220.48 $\pm$ 0.03& 1186 $\pm$ 8& 53 $\pm$ 5& 47 $\pm$ 5& 294 $\pm$ 56&14.65&14.26&14.09& 19.7\\
TON-1542& 1223.36 $\pm$ 0.03& 1895 $\pm$ 8& 43 $\pm$ 4& 35 $\pm$ 5& 216 $\pm$ 42&14.04&13.89&13.82& 16.9\\
TON-1542& 1226.06 $\pm$ 0.03& 2563 $\pm$ 8& 55 $\pm$ 4& 49 $\pm$ 5& 248 $\pm$ 41&14.26&14.03&13.93& 19.6\\
TON-1542& 1282.41 $\pm$ 0.03& 16458 $\pm$ 8& 39 $\pm$ 3& 31 $\pm$ 4& 105 $\pm$ 20&13.43&13.39&13.37& 15.0\\
VIIZW118& 1222.65 $\pm$ 0.04& 1721 $\pm$ 10& 50 $\pm$ 12& 43 $\pm$ 14& 54 $\pm$ 29&13.06&13.04&13.04& 5.9\\
VIIZW118& 1225.33 $\pm$ 0.03& 2382 $\pm$ 7& 56 $\pm$ 14& 50 $\pm$ 16& 68 $\pm$ 38&13.18&13.16&13.15& 7.8\\
VIIZW118& 1225.65 $\pm$ 0.02& 2460 $\pm$ 6& 51 $\pm$ 3& 44 $\pm$ 4& 267 $\pm$ 35&14.41&14.12&14.00& 31.1\\
VIIZW118& 1234.30 $\pm$ 0.03& 4595 $\pm$ 7& 32 $\pm$ 8& 20 $\pm$ 13& 35 $\pm$ 20&12.85&12.84&12.83& 4.6\\
VIIZW118& 1234.70 $\pm$ 0.03& 4693 $\pm$ 7& 61 $\pm$ 19& 55 $\pm$ 21& 45 $\pm$ 31&12.98&12.96&12.96& 6.0\enddata
\tablenotetext{a}{Corrected to LSR using Galactic absorption features.}
\tablenotetext{b}{Non-relativistic velocity (v=$cz$) relative to the Galactic LSR.}
\tablenotetext{c}{\bvalue after STIS resolution element correction.}
\tablenotetext{d}{Rest-frame equivalent width of \lyano.}
\tablenotetext{e}{H~I column density (\Nhno) assuming $\bobs= 20, 25, 30\kmsno$.}
\tablenotetext{f}{Significance Level (SL) of the \lya absorption feature, in $\sigma$.}
\end{deluxetable}

\subsection{Broad, Shallow, \lya Absorption Lines}\label{sec:broad}
In keeping with the reduction and analysis procedures developed in Paper~I for the GHRS
spectra, we arbitrarily limit the width of any intervening absorption to $\bobs\leq100$\kmsno. 
If the broadening were entirely thermal, the corresponding gas temperature appropriate for photoionized gas in the IGM
would be $\THI=(m_H b^2 / 2 k)=(37,800~{\rm K})(b/25\kmsno)^2$.
If this limit is exceeded, the absorption is subdivided until all subcomponents have 
$\bobs\leq100$\kmsno. Higher resolution spectroscopy (e.g., STIS echelle spectra of some of our
targets) generally supports this procedure, as does the median observed difference between
\bvalues obtained from \lya absorption line widths and \bvalues derived from a
curve-of-growth analysis \citep{Shull00}. These broad lines often consist of two or
more components when observed at higher spectral resolution. On the other hand, some of these
broad features are quite shallow and may not be real absorption, but rather
undulations in the underlying continuum of the target AGN. Because the 100\kms limit is
arbitrary, it is important to scrutinize the impact this assumption has on the line lists.

The generally higher SNR of the STIS spectra (20-40 per RE compared to 10-20 for our GHRS
spectra) allows a more accurate continuum fitting as well as the ability to
detect more subtle absorptions. We found seven possible broad absorbers in our full GHRS+STIS sample,
five coming from the STIS spectra presented here (see Table~\ref{broad_lya}). 
Table~\ref{broad_lya} lists by column: 
(1) target name; 
(2) wavelength range of the broad absorber; 
(3) wavelength centroid of the absorbers identified by our analysis procedures in the wavelength regions of column (2); 
(4) \bvalueno(s) of these absorbers; and 
(5) significance level of these absorbers. 
The possible broad absorbers do not appear as single absorbers in our line lists, either because
they have been arbitrarily subdivided to obtain $b\lt100$\kmsno, or they do not achieve
a \real.
Visual scrutiny of the wavelength regions in Table~\ref{broad_lya} give the strong
impression that most of these possible broad lines do subdivide into narrower components close
to their fitted values. 
Others (e.g., the \objectname[]{NGC~985} entry) may not be real. 
The \objectname[]{Fairall~9} entries are particularly uncertain because they occur on the blue wing of the \lya emission line
of the target. To be conservative, we have listed all of the possible broad absorbers in Table~\ref{broad_lya}, 
so that their maximal impact can be estimated. 
The net effect of relaxing the arbitrary upper limit on \bvalue would 
be to reduce the line list by five entries, if all of the broad absorbers are real.
Therefore, the number of these possible broad absorbers is small and their statistical effect on the sample as
a whole is negligible.

On the other hand, these absorptions could be real detections of the warm-hot IGM ($T = 10^{5-7}~{\rm K}$)
and so should be investigated further. 
A thermally broadened \lya line with \bb = 100\kms would have $\THI=6\times 10^5~{\rm K}$, at which temperature
the neutral fraction in the collisionally ionized case, $\fHI\approx5\times 10^{-7}$, would be extremely small.
Such components would be difficult to detect in \lyano, unless the total hydrogen column density approached $10^{20}\percmtwono$.
The more likely scenario of multiple velocity components, with dispersion exceeding 100\kmsno, complicates the
search for a broad shallow component.
High-SNR STIS echelle spectra or COS medium resolution spectra are required to
determine if these absorptions have multiple subcomponents. 
However, higher resolution data
will not determine unambiguously whether these absorptions are real or AGN continuum features.
We note that no very broad absorption line candidates come from the spectra of the four BL~Lac objects
in our sample; BL~Lac objects could facilitate better searches for broad \lya absorbers
given the power-law appearance of their continua.
Detections of the \ion{O}{6} doublet associated with these broad features would be compelling,
but most of these \lya absorptions are so weak that current far-UV spectrographs have insufficient 
sensitivity to detect \ion{O}{6} associated with the entries in Table~\ref{broad_lya}. 
Therefore, our observations show no strong evidence for very broad ($\bobs\gt100$\kmsno) 
absorptions in the local \lya forest.
\begin{deluxetable}{lcccc}
\tablecaption{Possible Broad \lya features \label{broad_lya}}
\tabletypesize{\footnotesize}
\tablecolumns{5}
\tablewidth{285pt}
\tablehead{
\colhead{Target}&
\colhead{Wavelength\tablenotemark{a}}&
\colhead{Absorber\tablenotemark{b}}&
\colhead{\bvalueno}&
\colhead{SL}\\
\colhead{Name}&
\colhead{Range (\noang)}&
\colhead{Wavelength (\noang)}&
\colhead{(\nokmsno)}&
\colhead{}
}
\startdata

MRK~1383      &1279.6--1280.8 &1280.1 &27 &4.2\\
              &               &1280.3 &30 &3.6\\
              &               &1280.6 &35 &3.8\\ 
NGC~985       &1223.0--1224.0 &1223.6 &35 &3.1\\
PG0804+761    &1261.0--1262.0 &1261.3 &33 &3.8\\
PG1116+216    &1265.0--1266.8 &1265.8 &88 &10.6\\
              &               &1266.5 &44 &5.4\\
PKS2005-489   &1226.7--1227.4 &1226.9 &24 &5.1\\
Fairall~9     &1263.5--1264.8 &1264.0 &42 &7.0\\
              &               &1264.7 &44 &10.2\\
Fairall~9     &1265.0--1266.0 &1265.1 &23 &9.5\\
              &               &1265.4 &19 &3.8\\
              &               &1266.0 &32 &7.0\enddata
\tablenotetext{a}{Wavelength range of possible, single broad \lya absorption line.}
\tablenotetext{b}{Wavelength centroids of absorption lines found by our automated procedure
by requiring $\bb \lt 100\kmsno$.}
\end{deluxetable}

\section{Rest-Frame Equivalent Width Distribution}\label{sec:REW}
\begin{figure}[htb]\epsscale{0.67}
\plotone{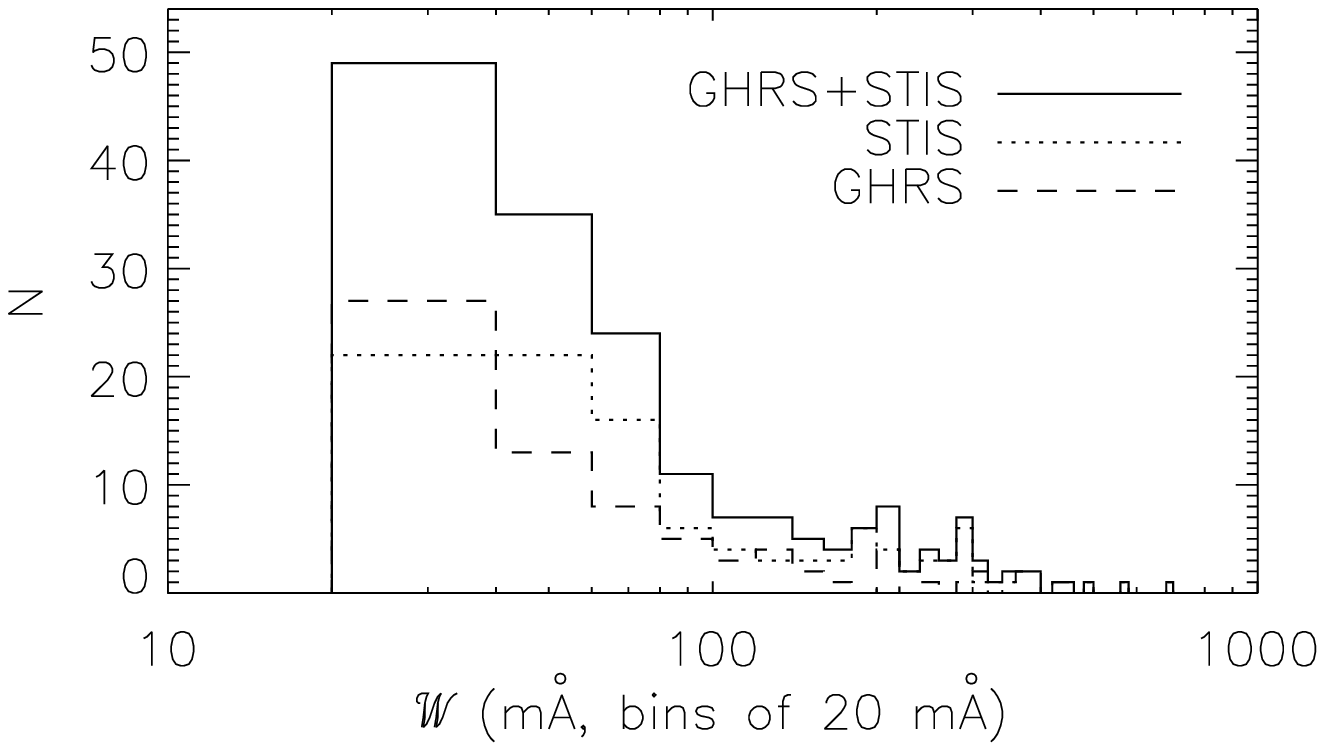}
\caption{\label{ew_dist} Observed rest-frame equivalent
width (\Wno) distribution of all \comboNabs\ intergalactic \real\ \lya absorbers in the current (STIS, dotted),
previous (GHRS, dashed) and cumulative (STIS+GHRS, solid) samples.
This distribution has not been corrected
for the non-uniform wavelength and sensitivity coverage of our
observations. No \lya absorptions classified as intrinsic are included in this 
distribution.}
\end{figure}

In Figure~\ref{ew_dist} we display the rest-frame equivalent width (\Wno) 
distribution, $\Nno(\Wi)$, for all of our (\real) \lya features.
As expected, we detect an increasing number of absorbers at decreasing 
\Wno, down to our detection limit. 
Because our spectra are of varying sensitivity and  wavelength coverage, 
this observed \W distribution is not the true \W distribution.
As in Paper~II, we must account for incompleteness due to SNR variations across each 
spectrum as well as between target spectra.
To determine the true \W distribution, we normalize the \W line density 
by the available pathlength, \Dz(\Wno), which is a function of \W and \z, since our 
spectra have varying \foursig detection limits across the observed
waveband and each spectrum covers a different waveband (See Paper~II for details).

\begin{figure}[htb] \epsscale{0.67} 
\plotone{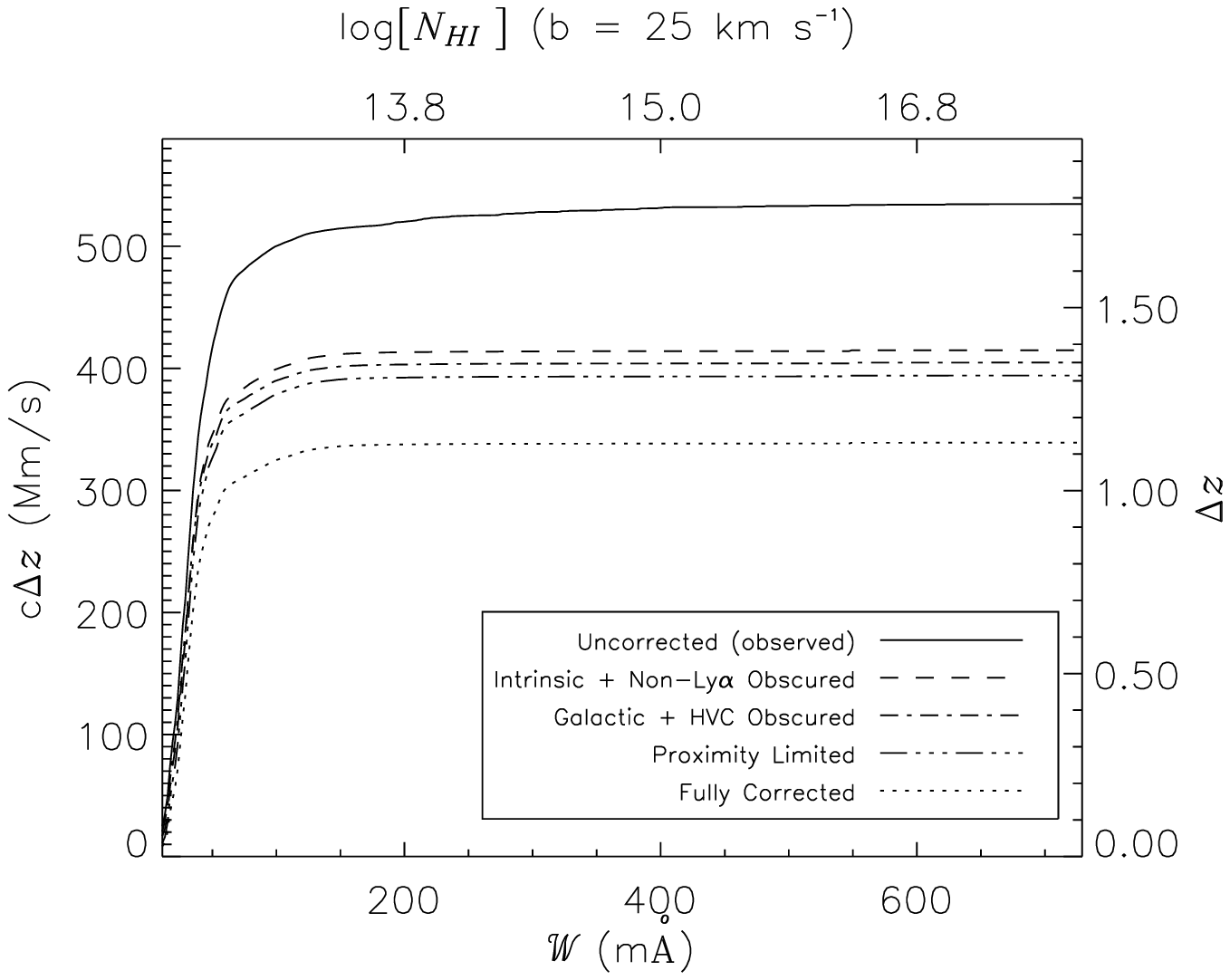}
\caption{\label{ewsenspath}Cumulative available pathlength (\Dz) 
as a function of \W for our combined sample.
The left axis gives \cDz(\Wno) in units of\mms ($1\mmsno = 1000$\kmsno), 
while the right axis gives \Dz.
The solid (upper) line indicates the observed, uncorrected, \Dz(\Wno).
Dashed, dot-dashed, and dot-dot-dot-dashed lines indicate \cDz(\Wno) after 
correcting for spectral obscuration due to non-\lya and
intrinsic absorption systems (intrinsic+non-\lyano), Galactic and HVC 
absorption systems (Galactic+HVC), and our \prox
``proximity'' limit, respectively. The dotted line indicates the fully 
corrected ``effective'' \Dz(\Wno), after
the indicated spectral regions unavailable for intervening \lya 
detection have been removed.
The top axis indicates the \logNh corresponding to \W for 
$\bb=25$\kmsno. For example, $\lognh=12.3$ for $\Wno=10$\Mang 
(left-hand border of bottom axis).}
\end{figure}

In Figure~\ref{ewsenspath}, we display the available redshift pathlength
(\Dz) in terms of \W for our full GHRS+STIS sample. 
The solid line indicates the full observational pathlength, uncorrected for the regions
of spectra not available for \lya detection due to Galactic, HVC, intrinsic, and non-\lya
intervening absorption lines as well as our ``proximity effect'' limit for \lya absorbers.
While the ``proximity limit'' ($cz_{AGN} - cz_{abs}\gt1200$\kmsno) we have chosen is somewhat 
arbitrary, it does eliminate most \lya absorption systems ``intrinsic'' to the AGN 
\citep[][see Paper~I for a detailed description and justification]{Bajtlik88}. 
The dashed and dot-dashed lines indicate the ``effective'' or available pathlength after the removal of
 the spectral regions unavailable for \lya detection for the three reasons listed above.

The lowest (dotted) line is the cumulative available pathlength, 
after all the indicated corrections have been made. 
Our maximum available pathlength is $\Dz = 1.157$ for all features with $\Wno\ge150$\mang.
This pathlength is about three times less than the $\Dz\approx3.82$ at $\z\lt0.3$ for the Key 
Project data for strong absorbers ($\Wno\gt240$\mang), but the STIS velocity resolution and
\Nh sensitivity are much better.

Applying the effective pathlength correction of 
Figure~\ref{ewsenspath} to the \Nno(\Wno) distribution of Figure~\ref{ew_dist}, 
we obtain the true detected \W number density,
\begin{equation}{
\nWi=\Nno(\Wi)/\Dz(\Wi)\DW \approx \dtwondWdz~\vert_{\Wi},
}\end{equation} corrected for the pathlength,
\Dz(\Wi), available to detect features at each \Wi, without regard to \z. 
The approximation that \nW\ is equal to \dtwondWdz\ is limited by our sample
size (\Nno) and by our bin size ($\DW=20\mang$).

\begin{figure}[htb]\epsscale{0.67} 
\plotone{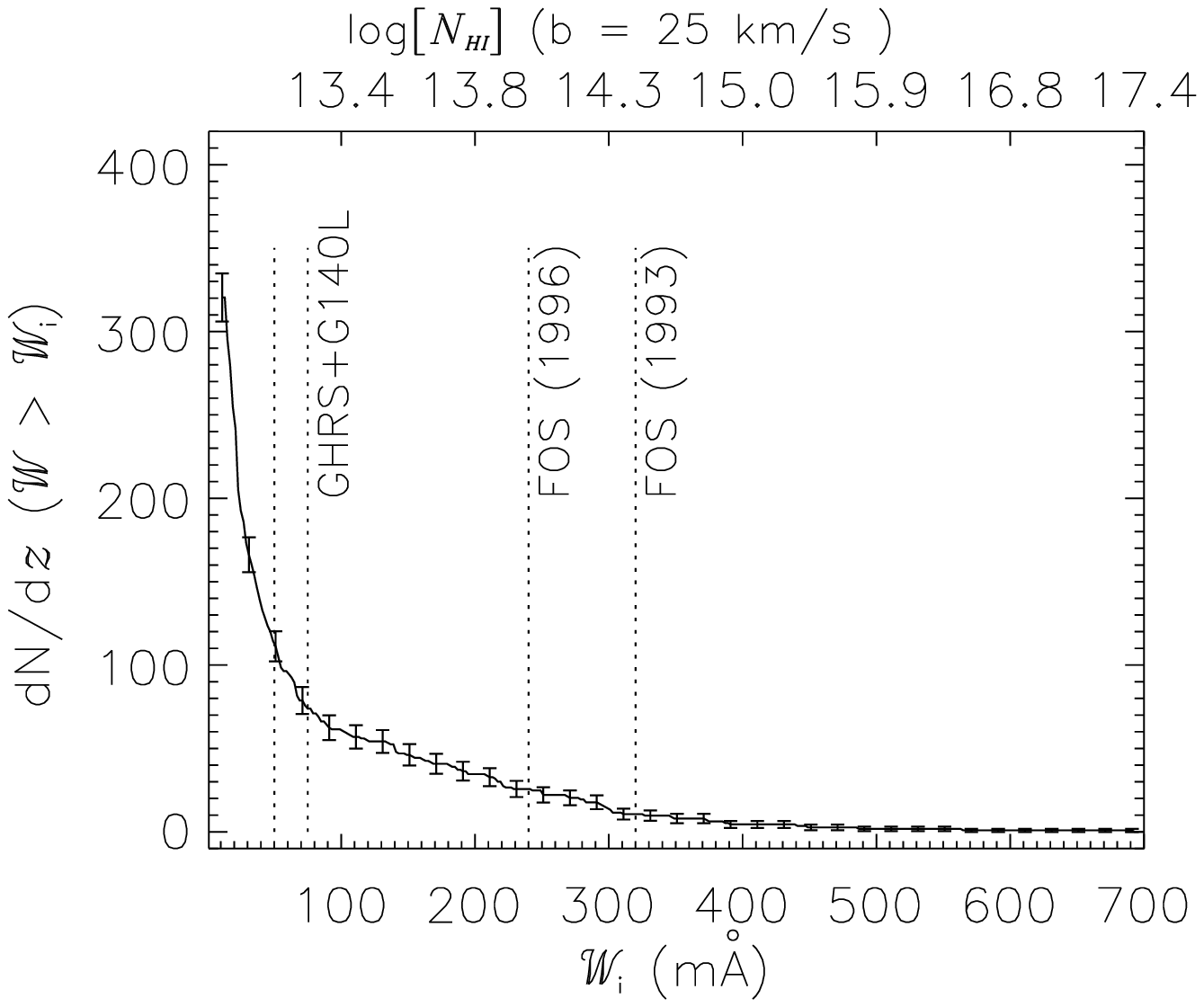} 
\caption{\label{int_dndz} The integrated \dndz above a given \Wi, $
\int_{\Wno_i}^\infty (\dtwondWdz) \dW$, for our GHRS+STIS \definite\ 
sample. For clarity, we plot every third error bar ($\sqrt{N}$ statistics).
The \W bins are 12\mang, connected by the solid line. Also shown by 
dotted vertical lines are the minimum
\W limits for other HST \lya surveys, detailed in Paper~II. The 
unlabeled vertical dashed line on the far left corresponds to the 
\protect{\citet{Tripp98}} G140L+G160M combined sample.
The first datapoint is at $\Wi = 13\Mang$ or $\lognh = 12.4$.}
\end{figure}

By integrating \dtwondWdz\ from any \Wi\ to \lognh=$\infty$ we can 
determine the number density of lines per unit redshift, \dndzno, stronger than \Wi, 
assuming no evolution with \z\ over our small range for \lya detection (\zrange).
Because of our very \lowz range, these values for \dndz are a good approximation to \dndzzerono.
Figure~\ref{int_dndz} shows \dndzno, defined as
\begin{equation}{
\dndz (\W >\Wno_i)=\int_{\Wno_i}^\infty \dtwondWdzover~\dW .
}\end{equation}
The vertical axis of Figure~\ref{int_dndz} gives \dndzzero in terms of both \W 
(lower axis) and \logNh (upper axis) assuming that all absorbers are 
single components with \bvalues of 25\kmsno.
The number density of lines rapidly 
increases at $\Wno\lt100$\Mang (\lognh=13.4), near the limit of the most sensitive surveys
previously conducted with HST. 
\section{Observed \hone Column Densities}\label{sec:Onh}
\begin{figure}[htb] \epsscale{0.67}	
\plotone{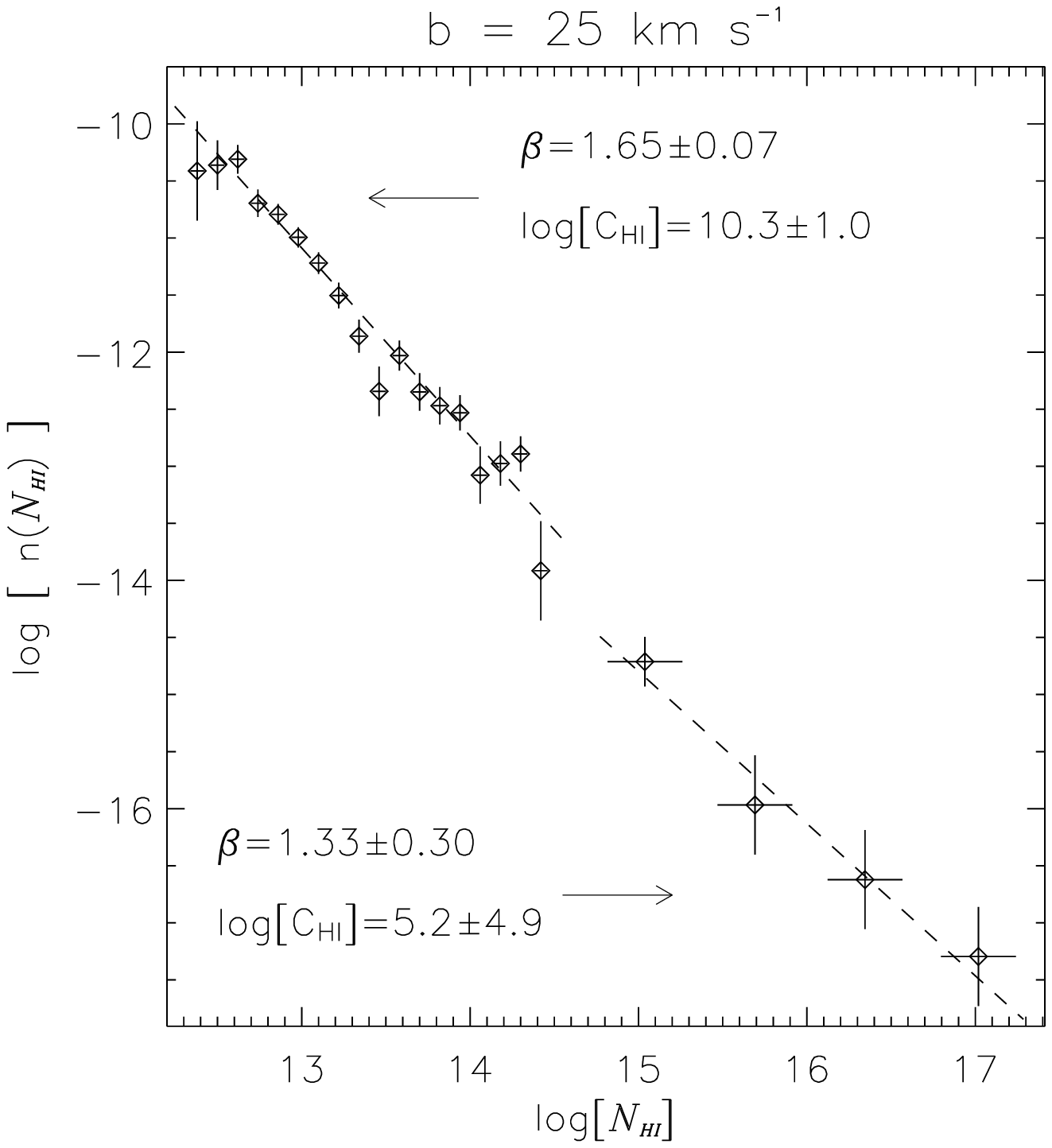}
\caption{\label{d2ndzdnh_full} The \lowzlya number distribution, 
for our combined sample, per unit redshift and column density, 
$\nNh=\int (\dtwondzdnhno) \dz$, for a constant \bvalue of \myb. 
The best-fit  power-laws for each \logNh regime, \ensuremath{12.3\leq\lognh\leq14.5}
and  \ensuremath{14.5\leq\lognh\leq17.5}, are indicated on the plot. 
In the lower \Nh regime, the \logNh bins are 0.12 dex, while in the upper regime the bins are 2/3 dex.
The \nNh\ error bars are based upon $\sqrt{N}$ statistics, while the \logNh error bars
are 1/3 of the bin size.}
\end{figure}

Because the \bvalues inferred from \lya line widths are unreliable for estimating \hone column densities, 
we assume a \bvalue of 25\kms for all absorbers. 
We recognize that this simplifying assumption can be  incorrect 
for individual absorbers, but we believe it to be the best choice 
based upon the limitations of these data.
The \Nh number density per unit redshift and column density is often 
modeled by a power-law distribution:
\begin{equation}{
 \dtwondnhdzover \sim{\mathcal{N}\left(\Nhno\right) \over 
\Dz\left(\Nhno\right)\DNh} \equiv \nNhno=\betaNh~~.}\end{equation}
In Figure~\ref{d2ndzdnh_full}, we display $\log\left[\nNhno\right]$ for 
our \definite\ \lya sample over the range $12.3\leq\lognh\leq17$.
For the column densities of $\lognh\lt13.5$,
the determination of $\beta$ and \NofNh is insensitive to \bvalueno,
since all these absorbers are on or near the linear portion of the \lya 
curve of growth.
Over the column density range \lowrange, a least-squares fit to $\log\left[\nNhno\right]=
\log\left[\NofNhno\right]-\beta$ \lognh \,
yields $\beta =\lowbeta\pm\Elowbeta$ and $\log\left[\NofNhno\right]=\lowconstant$.
There is no evidence for a turnover at $\lognh\lt12.5$ in our \lya sample,
even if we include the \expanded\, \lya sample. The value $\beta$=\lowbeta\ found here is
marginally (1.6\signo) flatter than the value $\beta=2.04\pm0.23$ found by \citet{Dave01}
and significantly flatter than the $\beta=2.15\pm0.04$ quoted for the simulated
column density distribution in that same paper.

Above \lognh\about 13.5, the \lya absorbers become partially 
saturated, and the choice of \bvalue becomes important in determining
\Nhno. Assuming \myb, we detect a break in the
$N^{-\beta}\subH$\ power-law above \lognh\about14.5 (see 
Figure~\ref{d2ndzdnh_full}) at the $\gt$ \twosig level. 
However, the choice for the  breakpoint in slope is arbitrary and could be as high as \lognh\about16, 
given the \bvalue uncertainty. 
A break in slope at \lognh\about16 is seen in the Key Project data \citep{Weymann98}, above which
they found a slope of $\beta=1.4$. 
Even with the better statistics at high equivalent width afforded by our combined 
GHRS+STIS samples, the break location and slope are still uncertain; our data are consistent with the better statistics of 
the Key Project data near the break in the \Nh distribution. As shown in Figure~\ref{d2ndzdnh_full}, 
for $14.5\lt\lognh\lt17.5$, we measure  $\beta =\highbeta\pm\Ehighbeta$ and $\log\left[\NofNhno\right]=\highconstant$. 
\section{Observed Redshift Distribution}\label{sec:Z}
In this section, we examine the redshift distribution of the \lowzlya forest, including
evidence for any structure at specific redshifts within our observed redshift range, and
the evolution of \dndz with \z\ obtained by comparing our results with those from other HST
and ground-based surveys. In the latter investigation, we pay particular attention to
the effects of spectral resolution and line blending.
\subsection{Local Variations in \dndzno.} \label{sec:dndzz}
\begin{figure}[htb] \epsscale{0.67} 
\plotone{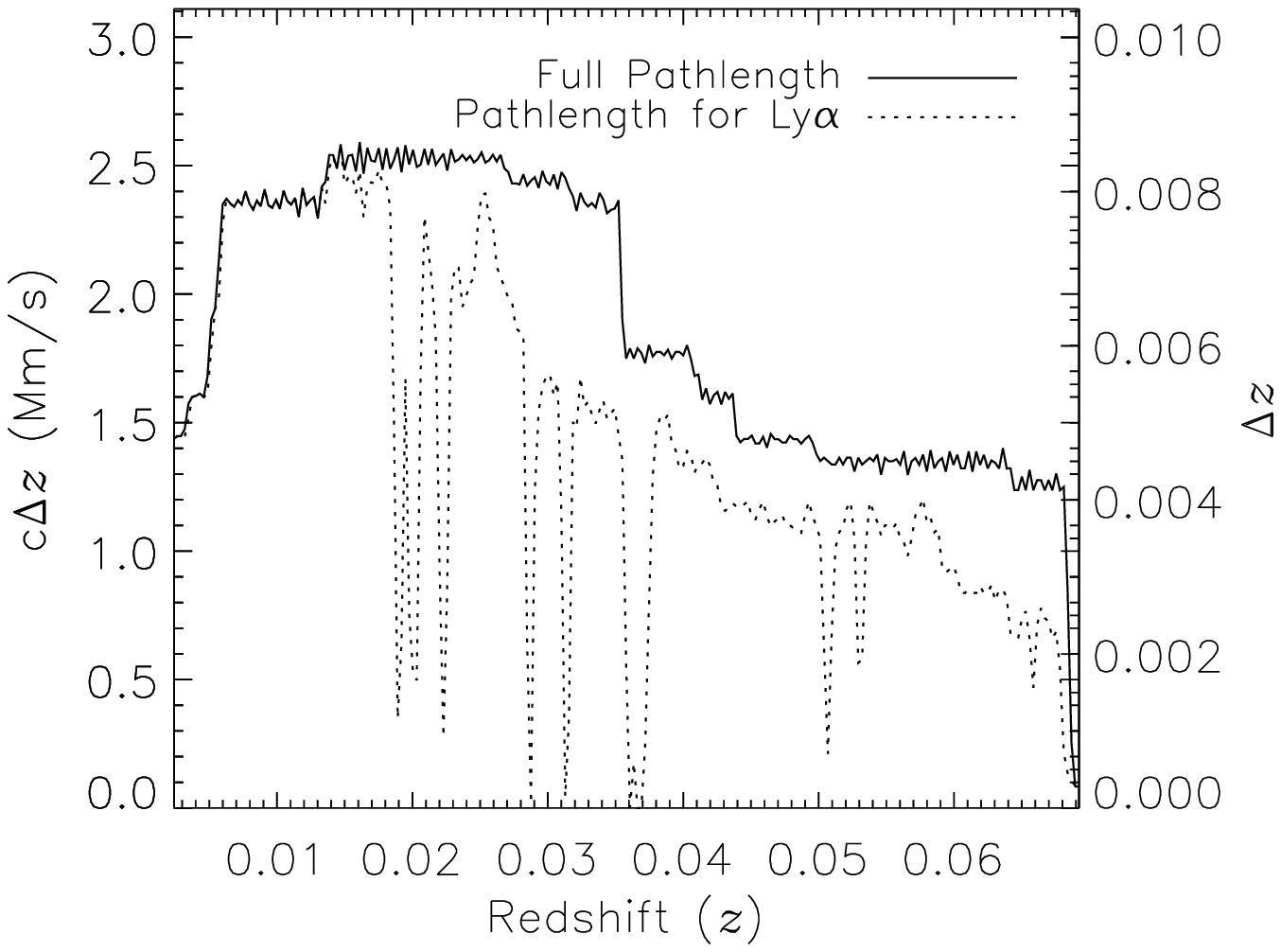}
\caption{\label{zsens_path}Cumulative pathlength (\Dz) ``available'' for detection of 
strong ($\Wno\gt150\mang$) absorbers for all 30 GHRS+STIS sightlines as a function of \z.
Left axis gives \cDz\ in units of\mms ($1\mmsno = 1000$\kmsno), while the right axis gives \Dz. 
The upper solid line presents the full \Dz, while the lower dotted curve is 
\Dz\ for detecting intervening \lya absorbers, after correcting for
obscuration by extragalactic non-\lya features, Galactic+HVC features, and the
(\proxno) proximity limit. The abrupt drop in pathlength at $\z\ge0.035$ indicates the
redshift limit of our GHRS spectra. The drop at $\z\le0.005$ indicates the onset of
Galactic damped \lya absorption. The redshift bins are 0.00028 in width.}
\end{figure}

The correction for our varying wavelength and sensitivity-corrected pathlength as a function of redshift, 
\Dz(\z), for the combined GHRS+STIS sample is shown in Figure~\ref{zsens_path}
for \lya absorbers with $\Wno\geq150\mang$.
The procedure for eliminating pathlength was described in detail in Paper~II.
Here we show the uncorrected total pathlength as a function of observed \lya 
absorber redshift (solid line) and the fully-corrected pathlength (dashed line)
after removing all spectral regions not suitable for detecting intervening \lya absorbers. 

\begin{figure}[htb] \epsscale{0.67}
\plotone{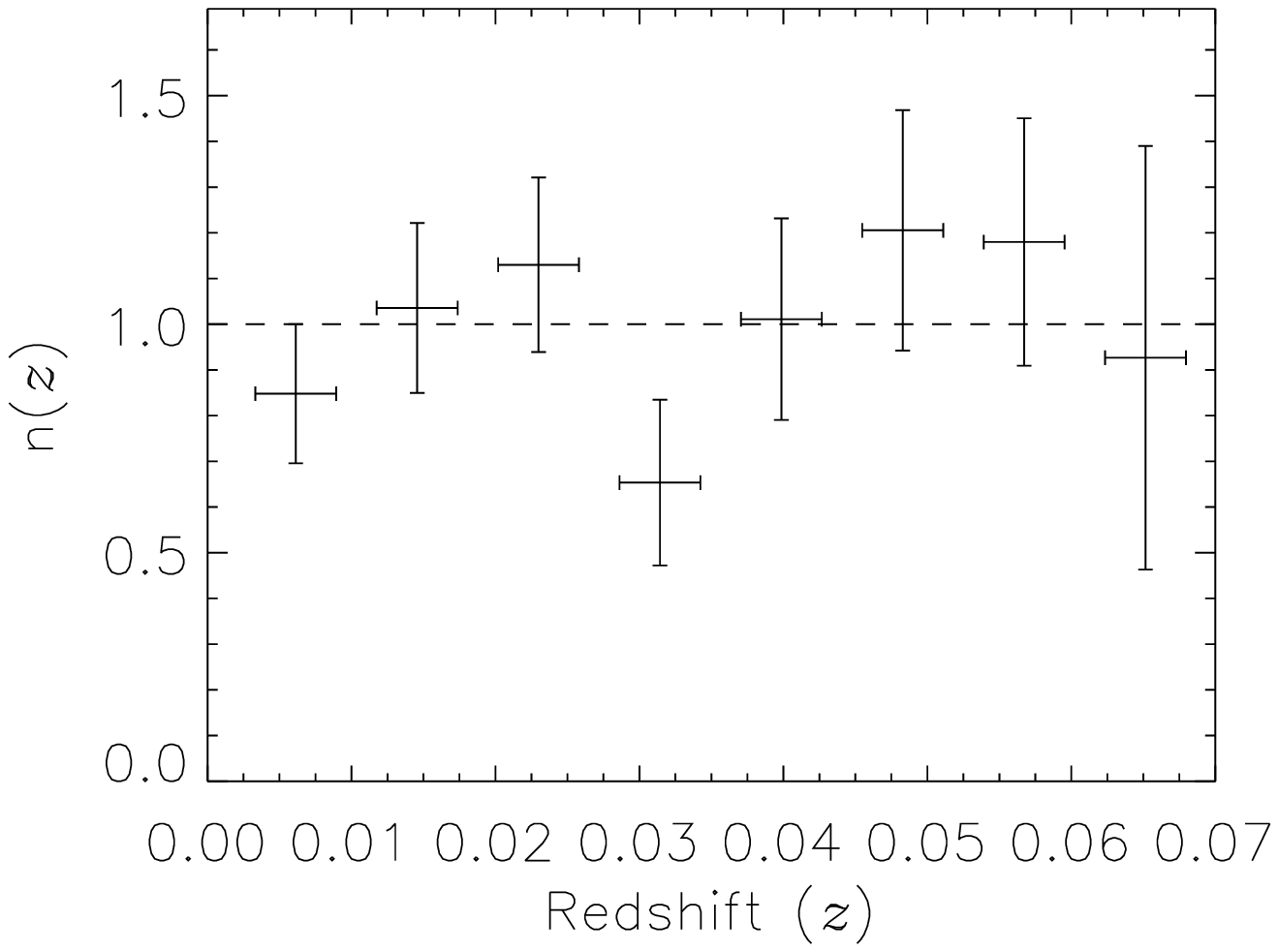}
\caption{\label{dndz_z}
The normalized \dndzz\ distribution, \nz, is plotted versus \z\ for the full (\real)  GHRS+STIS sample, 
excluding a portion of the \PKS\ sightline (see text). The distribution is normalized by integrating the
sensitivity function, using the previously derived \nNh\ distribution, to predict the expected \dndzz\ 
in the absence of evolution. No \z\ evolution of \dndz is observed over our small redshift range.
A value of \nz =1.0 corresponds to \dndz\about170 for $\lognh\ge13.0$, our normalized mean value.
Error bars for \nz\ are based upon Poisson statistics; redshift error bars are 1/3 of the bin size (\delz=0.0084).}
\end{figure}

Combining the fully-corrected pathlength of Figures~\ref{ewsenspath} and \ref{zsens_path}
yields a fully-bivariate (\z\ and \Nhno) sensitivity correction with which we can properly 
characterize the \lowzlya absorber distribution in redshift. 
A two-dimensional representation of this bivariate function is shown in
Paper~II for the GHRS sample alone. The GHRS+STIS sensitivity function 
is similar and is not presented here. Integrating the sensitivity function, using 
the previously derived \nNh\ distribution (for $\lognh \ge 12.3$), we can compare the observed \dndzz\ 
to that expected in the absence of evolution. The result of this analysis is shown in Figure~\ref{dndz_z}.
No  compelling evidence for inhomogeneities in the number density of local \lya absorbers is 
seen, consistent with our result from the GHRS data alone.
Error bars in this figure are based on Poisson statistics of the observed absorbers only with no contribution
from the uncertainties associated with predicting the expected \dndzz\ in the absence of evolution.
As in our GHRS analysis of Paper~II, we excluded a portion of our \PKS\ sightline
($\lambda\gt1280\ang$) since we specifically selected this sightline to observe 
a known complex of lines. An analysis of the distribution of deviations from the mean
finds a Gaussian distribution with a standard deviation of 4 lines per bin.
Thus, the deficiency in the $\z=0.030-0.035$ bin is 2.7\signo, and
 we see no evidence for any local structure variations or \z\ evolution over our small range 
in redshift ($\gamma$ is consistent with 0).
Also, since the deviations in Figure~\ref{dndz_z} are Gaussian
distributed, with a standard deviation comparable to the square root of the number of 
lines in each bin (0.25 in normalized units), 
this simple test suggests that our sample size may be large enough to have averaged over any cosmic
variance in line density in the local Universe \citep{Impey99}.

However, the above test is not very sensitive to the cosmic variance nor does it measure its amount.
Indeed, we expect that cosmic variance might be significant in our sample of sightlines because there are 
some STIS sightlines with only 1 (NGC985) or 2 (PG1351+640) absorbers, and five others with $\gt 10$.
In order to address this question in detail, we have constructed ensemble samples with 
varying numbers of sightlines, including the possibility that sightlines can be repeated within a given
sample. For each ensemble (i.e., each collection of samples with the same number of sightlines),
we computed the distribution of \dndzno. We find that for the ensemble with
15 sightlines (the STIS and GHRS sample sizes), the distribution of \dndz is twice as broad as
expected from $\sqrt{N}$ statistics (where N equals the number of absorbers).
By extending these results to our full sample size, we expect cosmic variance to contribute 
a systematic error which is slightly less than the statistical $\sqrt{N}$ error. 
Extrapolating these results, we estimate that the systematic error due to cosmic variance
will contribute $\lesssim 10\%$ to the total uncertainty in \dndz after the observation of 
$\gt 70$ sightlines similar to those in our GHRS+STIS survey. 
This number corresponds to a total pathlength of $\Dz \approx 3$.
Therefore, cosmic variance is an important, although minority, contributor to errors inherent in the results
presented herein.
\subsection{Redshift Evolution of \dndzno.} \label{sec:dndzz2}
\begin{figure}[htb] \epsscale{0.67}
\plotone{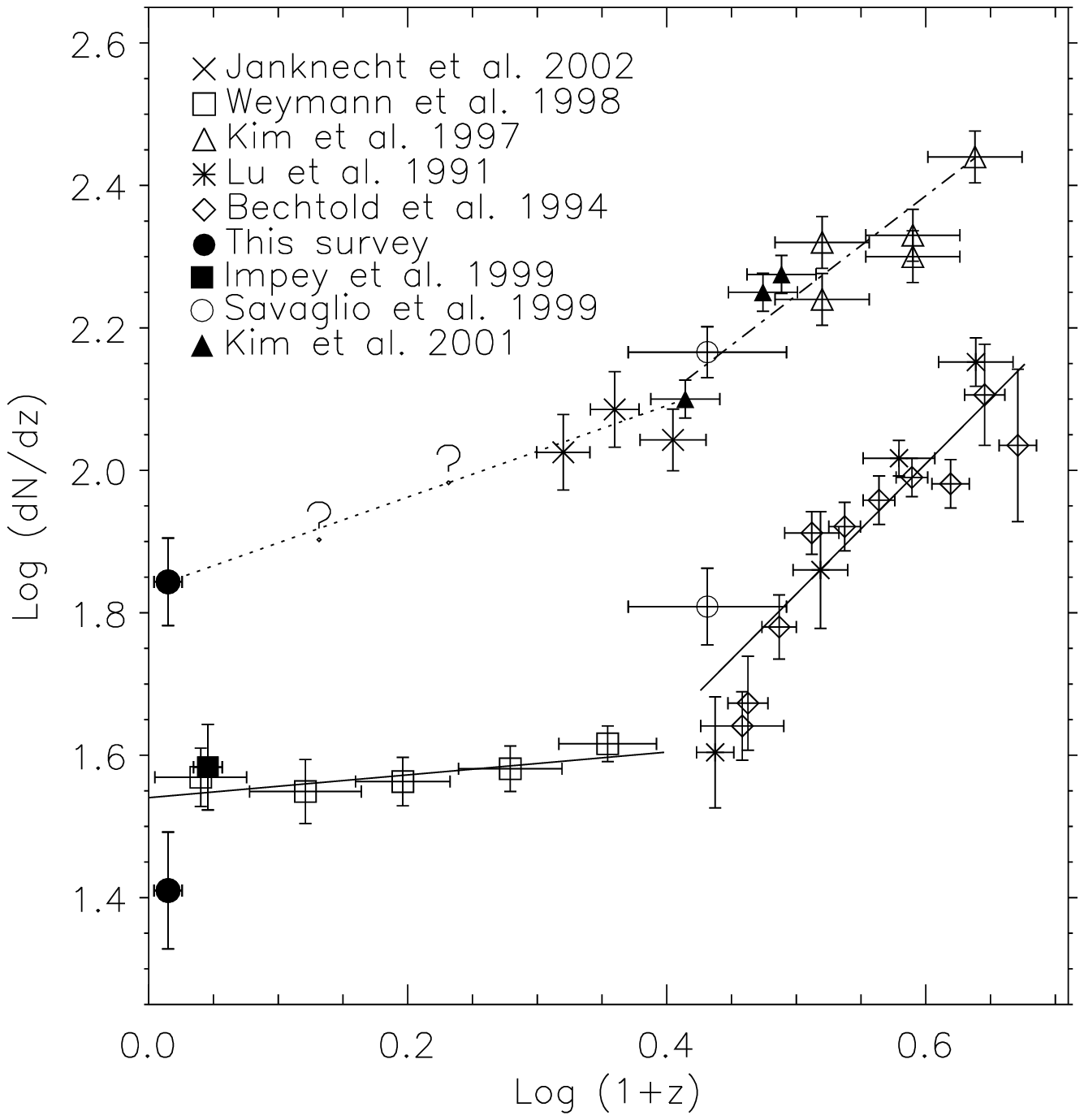}
\caption{\label{comp_dndz}
Comparison of \dndzlog\ versus \zonelog\ for two \Nh ranges.
Lower distribution corresponds to $\Wno\ge240$\Mang ($\lognh\ge14$ for \bb=25\kmsno).
Upper distribution corresponds to absorbers with \kimrange. 
The \zabout{0} points (solid circles) indicate the results of our survey
for each of the two \Nh ranges.
Solid lines are taken from \citet{Weymann98} and have slopes
of $\gamma=0.16$ (\zonelog\lt0.4) and $\gamma=1.85$ (\zonelog\gt0.4).
Connecting the lowest-\z\ (\kimrange) point from \citet{Kim01} to our datapoint
for the same range (upper dotted line) gives $\gamma=0.76\pm0.12$. 
The complete evolutionary picture for weak \lya absorbers is
not available, owing to the lack of data for $0.1\lt \zonelog \lt 0.3~(\z=0.26-1.0)$
indicated by the question marks.}
\end{figure}

Figure~\ref{comp_dndz}~displays \dndz over the redshift interval 
$0\lt\z\lt3$, for several studies over two \W ranges: \Wno$\ge$240\Mang
($\lognh\gt14$ for \bb=\myb) and $60\mang\le\Wno\le240\mang$ (\kimrange\ for \bb=25\kmsno).
The lower distribution in Figure~\ref{comp_dndz} is data normalized to 
$\Wno\ge240$\Mang by \citet{Weymann98}, while the upper
distribution corresponds to absorbers in the range \kimrange.
The data points indicated by open squares (\zonelog\lt0.4) were 
obtained as part of the HST/FOS Key Project \citep[][]{Weymann98}.
While normalized to $\Wno\ge240$\mang, these data do contain a few absorbers below that limit down to 60\mang.
The stars and diamonds correspond to ground-based data (\zonelog\gt0.4), taken
with an equivalent width limit of $\Wno\gt360$\mang, reported by 
\citet{Lu91} and \citet{Bechtold94}, but normalized to $\Wno\ge240$\Mang 
by \citet{Weymann98}. The \citet{Bechtold94} data mostly have a 
spectral resolution of $75-100$\kms while the Key Project data have a resolution of $230-270$\kmsno.

However, the actual FOS spectral resolution is nearly constant in\Ang for any one of the
three ``high-resolution" FOS gratings (G130H, G190H and G270H), so that the resolution
in\kms varies from \about270\kms at the blue end to \about200\kms at the red end of each grating. 
The \zabout{0} point from the STIS/G140L study of \citet{Impey99} has a spectral resolution only marginally 
better than the quoted Key Project resolution.
The lower column density points at \highz are high resolution data (\about7\kmsno) from the {\it Keck Telescope} 
\citep[open triangles;][]{Kim97} and the {\it Very Large Telescope} (VLT) Ultraviolet Echelle Spectrograph (UVES) \citep[filled 
triangles;][]{Kim01}.
The dash-dot line is a best-fit to the {\it Keck} data and has $\gamma=1.19\pm0.45$.
The two \lowz points indicated by filled circles are taken from the 
current survey, one point for each of the two sensitivity ranges ($\logdndzzero=1.40\pm(0.08, 0.07)$ for $\lognh\gt14$,
and $1.85\pm (0.06, 0.05)$ for \kimrange).
The two errors quoted are the statistical error and our estimate of the systematic error due
to cosmic variance (see \S~\ref{sec:dndzz}), respectively. Figure~\ref{comp_dndz} and all subsequent figures and tables
show only the statistical errors for all data points. 
Both of our \dndz points in Figure~\ref{comp_dndz} are lower (but within \onesigno) than the line densities determined from our GHRS data alone.
The solid lines indicate the best fits to the $\Wno\gt240$\Mang data,
above and below \zonelog=0.4, and have slopes of
$\gamma=0.16\pm0.16$ \citep[\zonelog\lt0.4, ][]{Weymann98} and 
$\gamma=1.85\pm0.27$ \citep[\zonelog\gt0.4, ][]{Bechtold94}. 

In our enlarged GHRS+STIS sample, we find a minor discrepancy in \dndzzero
between our high-\W point and the values from both the Key Project and from \citet{Impey99}. 
Because we suspect that this difference is due to the differing spectral 
resolutions of these studies, we smoothed our data to the spectral resolutions of both the 
\citet{Bechtold94} compilation (75\kmsno) and the Key Project data (\about240\kmsno). 
This smoothing produced a net increase in the number of $\Wno\geq240$\Mang lines in our sample, owing to 
blending of closely spaced pairs and groups of lines in our data.
The 75\kms smoothing resulted in only a marginal increase to \dndzlog, 
but the 240\kms smoothing increased our \zzero value to $\dndzlog=(1.46-1.50)$. 
We have quoted the range for this value that spans our various attempts to match both the precise
LSF and SNR of the Key Project data. 
While this range is still below the Key Project data and the \citet{Impey99} points at \zzerono, 
the error bars overlap at the \onesig level.
This exercise shows that, by using data of poorer resolution than the \highz
data, the Key Project may have somewhat overestimated the number of high-\W lines because 
lower-\W lines are blended at 240\kms but not at 75\kmsno. By this reasoning, the 
correct line density at
\zzero in Figure~\ref{comp_dndz} for comparison with the \highz data of 
\citet{Bechtold94} is the value, \dndzlog=1.40, obtained from our data using a 75\kms smoothing.
By assuming that the internal consistency of the Key Project data has
derived the correct slope for the line densities at \lowzno, the lower 
absolute value for the
\zzero line density that we derive here suggests
that the break from faster to slower evolution occurred at a later cosmic 
time; i.e., at \zabout{1.0} rather than \about1.5. Alternately, the evolution may slow more 
gradually with time, without a sharp break at a single redshift. 
Our best value of \logdndzzero=1.40 is in close agreement with the \zzero value predicted by \citet{Dave99} for a 
$\Lambda$CDM simulation with an assumed \Jnuv\ from \citet{Haardt96}.

Thus, our new result and analysis are further support for the accuracy of the \citet{Dave99} simulations.
These $\Lambda$CDM simulations also exhibit a more gradual slowing of the \dndz
evolution with redshift below \z=1.6. Spectroscopy at a common velocity
resolution over the full range in observable redshifts is required to make the
most accurate \dndz evolutionary plot. 

At lower column densities, our data have \about3 times poorer resolution than the 
\citet{Kim97} and \citet{Kim01} data, so STIS echelle spectra are needed for the most accurate comparison.
However, we can apply the same argument to the lower column density regime to conclude that
our point is too high relative to the \highz points in Figure~\ref{comp_dndz}. However, we do not 
expect that the downwards correction required for our low column density point
is as large as the upward correction ($\delta(\dndzlog)=0.06$)
that we made to the high column density absorbers.
This is because the lower column density lines do not 
cluster as strongly as the higher column density lines (see \S~\ref{sec:TPCF}). 
Thus, a slight \dndz break is still required for the lower column density absorbers. 
So, either the break in the slope  of \dndz is much more modest for the low column
density clouds, or the break occurs at a substantially smaller 
look-back-time; i.e., \zabout{0.5} compared to \zabout{1.0-1.5}. New observations made recently with 
HST/STIS+E240M (Jannuzi, PI: PG~1634+706, PG~1630+377, PKS~0232-04, PG~0117+213)
will be important in determining the amount and position 
of the break in slope in Figure~\ref{comp_dndz} by determining the line density of lower column
density absorbers at \z= 0.5 - 1.5. Also, archival
HST/STIS E140M spectra of a few very bright AGN may have sufficient 
pathlength to verify the low column density \zzero line density we find in 
Figure~\ref{comp_dndz}, but at a spectral resolution comparable to the {\it Keck} and VLT data.
\subsection{Evolution of the Column Density Distribution}\label{sec:evonh}
\begin{figure}[htb] \epsscale{0.7}
\plotone{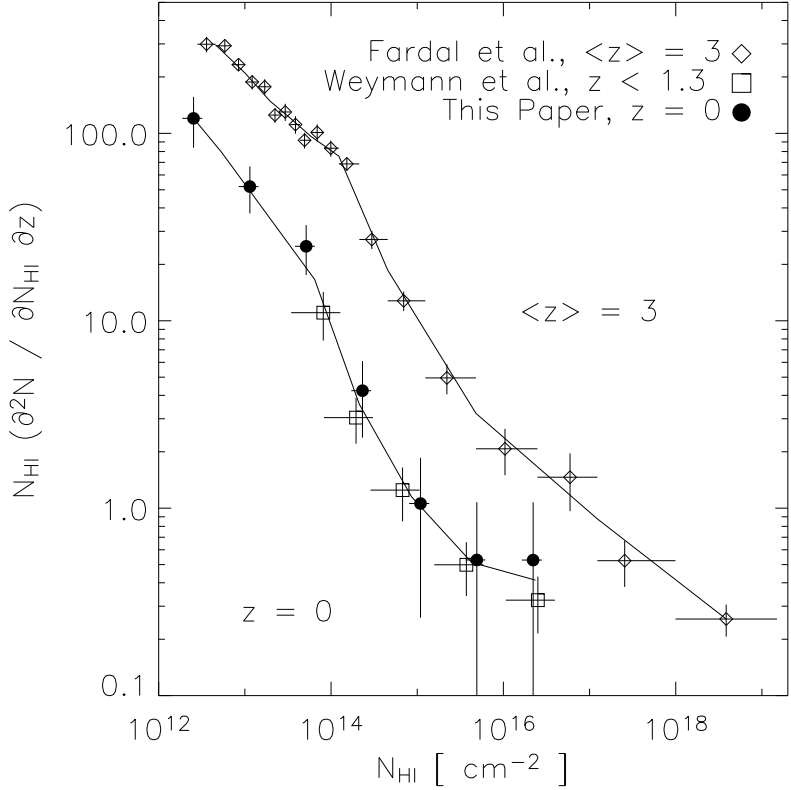}
\caption{\label{d2ndzdnh} The column density distribution at low (closed circles and boxes)
and high (diamonds) redshift plotted in a manner so as to accentuate deviations from a
single power-law distribution. For this presentation only, we have used a constant \bvalue
of 30\kms to match our data to that of the Key Project.}
\end{figure}

A more detailed method for investigating the redshift evolution of the \lya absorbers is to compare the
column density distributions at high and low redshifts. 
Figure~\ref{d2ndzdnh} compares the column
density distribution, \dtwondzdnhno, at \z=3 from \citet{Fardal98}
with similar data at lower redshift from HST. The \lowzno, lower column density data 
($\lognh\leq14$) come from the current survey (closed circles in Figure~\ref{d2ndzdnh}),
while the open boxes are from the HST Key Project \citep[][normalized to \zzerono]{Weymann98}.

It is tempting to relate a specific column density at \highz with a specific resulting
column density at \lowz (i.e., a ``cloud'' of a certain overdensity diminishes in column density
by a specific amount). However, the subtle and uncertain locations of breaks in the slope of the
column density distributions make these associations uncertain. 
At \z=3, breaks appear to be present at $\lognh \approx 14.5$ and $15.5-16.0$, while the \lowz breaks
appear at $\lognh \approx 13.5-14.0$ and 15. 
This column density change of a factor of 3-10 from \z=3 to 0 is somewhat less than the theoretical 
expectation of a factor of 20 \citep{Dave99,Bryan99,Schaye99,Schaye01}.
The \lowzlya column density slope at high column density ($\beta=\highbeta\pm\Ehighbeta$)
 is consistent with values found at \highz \citep{Hu95,Lu96,Kirkman97,Kim97},
 suggesting, but certainly not proving, that these are the same population of
 clouds seen at different redshifts.
\subsection{The Opacity of the Low-\z\ \lya Forest}
\begin{figure}[htb] \epsscale{0.67} 
\plotone{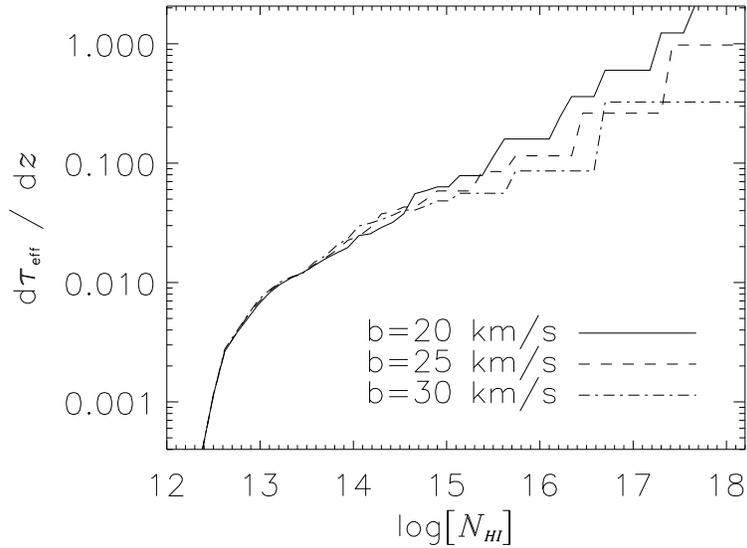}
\caption{\label{dtaueff_dz} Cumulative opacity, \dtaudz, at low redshift 
at the Lyman edge (912\ang) as a function of
\Nhno, calculated for three different \bvaluesno: 20, 25, and 30\kmsno. 
The crossover of the curves arises from small number statistics and
the \lya curve of growth (see text).}
\end{figure}

Models of the radiative transfer of ionizing radiation through the IGM \citep{Fardal98,Shull99b}
require the knowledge of the photoelectric opacity of the \lya forest.
Following the methods outlined in Paper~II, we present, in Figure~\ref{dtaueff_dz}, the 
cumulative opacity, \dtaudz, for our combined GHRS+STIS sample
at low redshift. The opacity is computed at the Lyman edge (912\ang) as a function of
\Nhno, for three different \bvaluesno: 20, 25, and 30\kmsno.
The jaggedness in the \dtaudz\ curves at $\lognh \geq 14$
arises from the \lya curve of growth and small number statistics.
A lower assumed value of \bb~will systematically increase the inferred 
column density for lines above $\lognh \approx 14$.
At $\lognh \gt 15$, \dtaudz~becomes uncertain due to poor
number statistics in our sample and to the large dependence on \bvalueno.
For a constant \bb=\myb\ for all \lowzlya absorbers, $\dtaudz \about 0.2$ 
for $\lognh \leq 16$.
However, if \bb=20\kms is a more representative Doppler parameter for the higher column
density lines \citep[e.g., the 1586\kms absorber in the 3C~273 sightline;][]{Sembach01},
$\dtaudz \about 0.4$ for $\lognh \leq 16$. 
If some \lya clouds have \bb\about15\kmsno, 
the cumulative \lya cloud opacity in the local  Universe could
approach unity for $\lognh\about16$ and over a redshift path of $\Dz\about1$.

As Figure~\ref{dtaueff_dz} indicates,
\lya absorbers with $15\lt\lognh\lt18$ probably dominate the continuum 
opacity of the \lowzlya forest and could impact the level of the ionizing background. 
However, the number density of these high column 
density systems is so low,
that the median opacity of the \lowz forest is $\leq 0.1$, and the probability of higher 
opacity is stochastic. 
Characterizing the distribution of these $\lognh\leq16$ absorbers
accurately at \lowz will remain a challenge, even for HST/COS,
but would be important in understanding the extragalactic ionizing 
background in the current epoch.
\section{The Baryon Density in the Local Intergalactic Medium}\label{sec:omega_b}
In Paper~II we described a method for estimating the baryon content of the local \lya forest
based upon the observed column density distribution (updated in \S~\ref{sec:evonh}) and
a number of simplifying assumptions in the context of optically-thin photoionized clouds. These
simplifying assumptions included: 
(1) spherical geometry, 
(2) an isothermal density profile, 
(3) characteristic absorber sizes of 100~\hsfi~kpc, based upon quasar pair experiments \citep{Dinshaw97,Rosenberg03}, and 
(4) a value and slope for the extragalactic ionizing flux based upon local Seyfert and quasar space
densities and ionizing spectra. When this model was integrated (Paper~II) over our observed column density
distribution from $13.5\leq\lognh\leq16.0$, we found \about$20$\% of all local baryons
in the low column density \lya forest. Specific dependences of this value on measurable parameters were:
\begin{equation}
\omegalya=(0.008\pm0.001) [\Jm b_{100} (4.8/\alpha_s + 3)]^{1/2} \hsfi ~~,
\end{equation}
where J$_{-23}$ is the extragalactic ionizing radiation intensity (\Jnuv) at 912\Ang in units of 10$^{-23}$\cgs; 
$b_{100}$ is the absorber characteristic radius in units of 100 
\hsfi kpc; and $\alpha_s = 1.8$ is the spectral index of the ionizing radiation. 
A larger value of the \Jnuv\ \citep[J$_{-23}$=1.3; ][]{Shull99b}
and a larger integration range over \Nh increases the Paper~II value of 20\% significantly (see below).

More recently, \citet{Schaye01} developed a different method for estimating the baryon content which makes
somewhat different assumptions. Like our method of Paper~II, this model assumes photoionization of
optically-thin absorbers, but also assumes gravitationally-bound clouds whose observed
column densities are equal to their characteristic column densities over a Jeans length. This
also requires these absorbers to be nearly spherical. These assumptions yield the following
expression for the baryon content of the \lya forest:
\begin{equation}
\omegalya=(2.2 \times 10^{-9})~ \hohi \Gamma_{-12}^{1/3} T_4^{0.59} \int \Nhno^{1/3} f(\Nhno, z) dN\subH,
\end{equation} where $\Gamma_{-12}$ is the \hone photoionization rate in units of 10$^{-12}$\persecondno, 
T$_4$ is the IGM temperature in units of 10$^4$~K, and $f$(\Nhno,\z) is the column density
distribution \dtwondnhdz for the special case of \zzerono, assumed to be a power-law \betaNh. 
Here, we have chosen $\hhi = 0.7$, $\Gamma_{-12}=0.03$ (equivalent to $\Jno =1.3\times 10^{-23}~\cgs$), 
$\beta = 1.7$, and $T_4=0.5$ at \zzerono.
The comparison by \citet{Dave01} between simulations and STIS echelle observations finds $T_4 \simeq 0.5$ at \zzerono.
The above expression assumes that the \lya
forest absorbers have the universal ratio of baryons to dark matter; i.e., they have no bias.

Here we employ both of these methods to estimate the baryon content of the local \lya absorbers
based upon the same input data, namely our column density distribution as displayed in 
Figures~\ref{d2ndzdnh_full} and \ref{d2ndzdnh}. 
The power-law slope that we obtain above a column density of 10$^{14.5}$\percmtwo is
 similar to that obtained by the Key Project team \citep{Weymann98}.
Specifically, we assume $\beta=\lowbeta$ and \highbeta\ below and above \Nh=10$^{14.5}$\percmtwono. Based upon our enlarged
sample, we are now confident that the \lowbeta\ slope extends at least to $\lognh=12.5$, 
so we integrate these best-fit distributions from $\lognh=12.0 - 17.5$. 
At or near our adopted lower limit (which corresponds approximately to an 
overdensity of about 3 at the current epoch), some of the assumptions of these methods may 
break down. 
Above 10$^{17}$\percmtwono, absorbers will become optically thick and can ``hide'' additional gas mass. 
Therefore, neither method can hope to derive an accurate measurement for the baryon content 
of the \lya forest, even if the column density distribution and other absorbers properties 
(size, ionizing flux, temperature) are known to very high precision. 
Paper~II comments on the systematic uncertainties that arise in the
application of the first method. 
Many of the same uncertainties plague the second method
as well; e.g., absorber shapes, presence of varying temperature/ionization state of the gas
(specifically the amount of the ``warm-hot'' medium). To these substantial systematic
uncertainties we add the uncertainty in converting equivalent width to column density for
partially saturated lines. For this calculation we adopt $\bb=25$\kms for all clouds, as
described previously. Using higher resolution STIS echelle spectra, \citet{Dave01}
find a median $\bb=22$\kms for a sample of \about 90 absorbers in two
sightlines, so our assumed value is consistent with the best current measurements.

Table~\ref{omega_lya} shows the resulting  \lya absorber 
baryon densities expressed as percentages of the total baryon
density from WMAP and other measurements \citep[\omegab $=0.047\pm0.006$; ][]{Spergel03},
assuming $\Jm=1.3$ and $h=0.71$.
Regardless of the methodology employed, the baryon content of the local IGM is dominated
by the lower column density absorbers. 
The \citet{Schaye01} method has an inherent dependence of $\omegab$ on column density 
that scales as \omegab$ \propto \Nmin^{-(\beta - 4/3)} \sim \Nmin^{-0.37}$, 
so that the baryon fraction of the lowest column densities begins to diverge weakly by that model. 
On the other hand, at the higher column densities, the Paper~II assumption of a constant 100~kpc size 
is probably an overestimate of the size and thus the baryon content.
For example, \citet{Tripp02} and \citet{Rosenberg03} find one-dimensional absorber sizes of 1-30~kpc at $\lognh= 15.5-16.5$.
Thus, the most conservative, and we believe most accurate, approach is to use the constant-size
assumption at $\lognh\leq14.5$ and the \citet{Schaye01} method at higher columns. 
This combined method finds \lowomega\ at low column densities and \highomega\ at higher column densities, for
a total baryon fraction in the local, photoionized IGM of \totalomega.
This estimate depends weakly on the assumed ionizing background \citep[$\Jno^{1/2}$, see ][]{Shull99b,PaperII}.
However, we have not attempted to determine the size of the systematic error due to cosmic variance,
since the baryon density depends both on the number density of absorbers and their distribution in column density.
This total is somewhat less than (but within the errors of) the value presented by \citet{Stocke03}
because a slightly steeper low column density slope ($\beta=1.8$ compared to the \lowbeta\ assumed here) was 
used in that work.
Also, \citet{Dave01} find $\beta=2.0\pm0.2$, consistent with simulations
and 1.5\sig steeper than the slope found here, but with a similar normalization. 
If the \citet{Dave01} $\beta$ value is used, a larger baryon fraction results, even 
more dominated by the lower column density absorbers. 

Photoionized absorbers identified in the \citet{Dave99} simulations account for \about30\% of 
the baryons at $z\approx0$, consistent with all of these observational estimates. 
But, while \citet{Dave01} claim that the observable \lowz
\lya forest column density distribution and resulting baryon fraction
are consistent with detailed numerical simulations 
\citep[e.g.,][]{Dave99,Dave01b}, it is not obvious how to relate precisely the simulations with
the observations. For example, at what temperature
does \lya absorption become difficult/impossible to detect in HST spectra? 
We have shown in \S~\ref{sec:broad} that there is no strong evidence for a substantial number of very broad absorbers
($\Dv \geq 100$\kmsno) which could arise in warm-hot gas. Nevertheless, the simulations predict that this gas is 
a substantial baryon reservoir (i.e., at $T = 10^{5-6}~{\rm K}$, collisionally
ionized gas has neutral fractions $\fHI\leq10^{-5}$).
\citet{Dave99} divide absorbers into photoionized and shock-ionized clouds, 
but what fraction of the latter population is detected in \lya absorber surveys
like the current one ?
This detail will need to be addressed before the inherent strengths of the simulations, in conjunction 
with the observations, can be used to measure accurately the baryon content of the IGM. 
At this point, it is the accuracy and range of applicability of the models, not the uncertainties in the
column density distribution, that limit the accuracy of the baryon fractions computed.
\begin{deluxetable}{lcccc}
\tablecaption{Baryon Content of the Local \lya Forest \label{omega_lya}\tablenotemark{a}}
\tabletypesize{\footnotesize}
\tablecolumns{5}
\tablewidth{0pt}
\tablehead{
\colhead{Log Column Density}&
\multicolumn{2}{c}{Paper~II Method}&
\multicolumn{2}{c}{Schaye Method}\\
\colhead{Range (\nopercmtwono)}&
\colhead{$\Omega_{\lya}$\tablenotemark{b}}&
\colhead{$100 \times \Omega_{\lya}$/$\Omega_{b}$\tablenotemark{c}}&
\colhead{$\Omega_{\lya}$\tablenotemark{b}}&
\colhead{$100 \times \Omega_{\lya}$/$\Omega_{b}$\tablenotemark{c}}
}
\startdata
12.50 $-$ 13.50 &   0.0062 $\pm$   0.0006 &  13.1 $\pm$   1.3 &   0.0088 $\pm$   0.0009 &  18.7 $\pm$   2.0 \\
13.50 $-$ 14.50 &   0.0041 $\pm$   0.0006 &   8.7 $\pm$   1.3 &   0.0039 $\pm$   0.0006 &   8.2 $\pm$   1.2 \\
14.50 $-$ 15.50 &   0.0021 $\pm$   0.0008 &   4.5 $\pm$   1.7 &   0.0013 $\pm$   0.0005 &   2.9 $\pm$   1.0 \\
15.50 $-$ 16.50 &   0.0007 $\pm$   0.0007 &   1.5 $\pm$   1.5 &   0.0003 $\pm$   0.0003 &   0.7 $\pm$   0.7 \\
16.50 $-$ 17.50 &   0.0059 $\pm$   0.0046 &  12.5 $\pm$   9.7 &   0.0017 $\pm$   0.0012 &   3.5 $\pm$   2.6 \\
\tableline
\tableline
12.50 $-$ 14.50 &   0.0101 $\pm$   0.0009 & [21.5 $\pm$   1.8] &   0.0125 $\pm$   0.0011 &  26.6 $\pm$   2.3 \\
14.50 $-$ 17.50 &   0.0087 $\pm$   0.0047 &  18.5 $\pm$  10.0 &   0.0033 $\pm$   0.0014 &   [7.1 $\pm$   2.9] \\
\tableline
12.50 $-$ 17.50 &   0.0188 $\pm$   0.0048 &  40.0 $\pm$  10.2 &   0.0158 $\pm$   0.0017 &  33.6 $\pm$   3.7 \enddata
\tablenotetext{a}{Best estimate of total $\Omega_{\lya}$ is given in brackets ($21.5\% + 7.1\% = 28.6\%$ total).}
\tablenotetext{b}{$\Omega_{\lya}$ is the baryon density contained within this column density bin.}
\tablenotetext{c}{Percentage in bin of $\Omega_{b}~(0.047 \pm 0.006$; Spergel \etl 2003) represented by $\Omega_{\lya}$.}
\end{deluxetable}

\section{\lya Absorber Linear Two-Point Correlation Function}\label{sec:TPCF}
The two-point correlation function (TPCF, $\xi$) for \lya absorbers
can be estimated from the pair
counts of \lya absorption lines along each line of sight in our data 
according to :
\begin{equation}{\label{xi} \xi(\Dv)={\Nobs(\Dv) \over \Nran(\Dv)} - 1~~.}\end{equation}
Here, \Nobs\ is the number of observed pairs and \Nran\ is the number of 
pairs that would be
expected randomly in the absence of clustering, in a given velocity 
difference bin, \Dv.
We determine \Nran~ from Monte Carlo simulations based upon
our determined number density, \dtwondnhdz, as well as the wavelength 
extent and
sensitivity limit of our observations. Like the pathlength normalization 
vector, we include only those portions of the spectra not obscured by Galactic 
lines, non-\lya lines, and spectral regions blueward of \prox of the target.

At each position along the spectrum, the probability of finding an 
absorber is calculated by:
\begin{equation}{ \Plam=\int_{\z} \int^{\infty}_{N_{min}(\lambda)} 
\dtwondnhdzover~\dNh \dz \approx \Dz(\lambda) 
\int^{\infty}_{N_{min}(\lambda)}
\betaNh~\dNh~,}\end{equation}
where $N_{min}(\lambda)$ is based upon the sensitivity limit of the 
spectrum.
The integral in \z\ can be replaced by the \z\ width of each pixel,
$\Dz(\lambda)$, since there appears to be no \z\ evolution between 
\zrange\ (i.e., $\gamma=0$ over the very small range of redshifts
in these spectra).
The values of $\beta$ and log[$\NofNhno$] were taken from our full 
GHRS+STIS sample as shown in Figure~\ref{d2ndzdnh_full}.
The probability, \Plam, is then compared to a uniformly distributed 
random number between zero and one.
If the probability exceeds the random number, an absorber is inserted 
into the Monte Carlo simulation at this position ($\lambda$).
To correct for blending effects, once an absorber is inserted into the 
Monte Carlo simulation,
\Plam~is set to zero for the adjacent 50\kmsno, since no pairs were 
observed at our resolution with
separations less than 50\kmsno. Undoubtedly, such closer pairs exist, 
but at our resolution we are insensitive to them due to blending, 
particularly with the complex LSF of STIS, as discussed in \S~\ref{sec:spec}.
Because \Plam~depends exclusively on our observed distribution, any 
blended lines counted as a single absorption system will
affect $\Nobs$ and $\Nran$ identically, leaving $\xi(\Dv)$ unchanged.
One would expect to begin resolving pairs separated by 1.5--2 Gaussian 
widths, which is in agreement with our
observed 50\kms cutoff. Our GHRS pre-COSTAR and STIS \bvalues have a higher 
median value of \about40\kmsno.
Therefore, in our Monte Carlo simulations, we may have somewhat overestimated 
the number of random pairs $\le 70$\kmsno. 
For each sightline, we performed 1,000 simulations($N_s$) and combined them to form \Nran. 
The error in \Nran(\Dv), denoted $\sigma_{ran}$, is taken to be $\sqrt{\Nran(\Dv)}$. 

\begin{figure}[htb]
\epsscale{0.8} 
\plotone{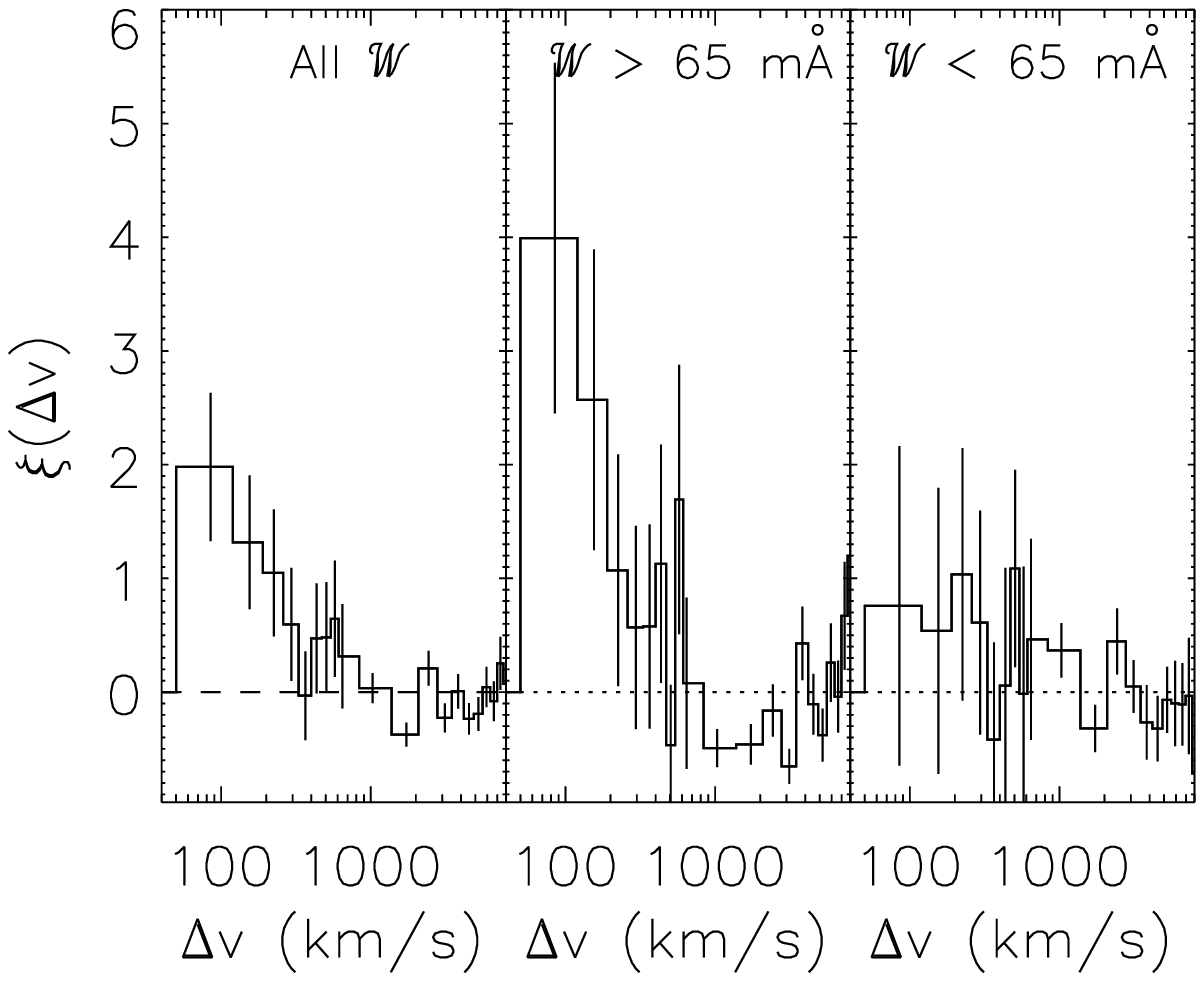}
\caption{\label{twopoint_no} Two-point correlation function (TPCF, 
$\xi$) of the \lya absorbers as a function of velocity pair separation, \Dv, in 
70\kms bins (1~\hsfi~Mpc for a pure Hubble flow) starting at $\Dv=50$\kmsno. 
The corrected TPCF, $\xi(\Dv)$, has been normalized by \Nran(\Dv)
accounting for the varying available spectral regions so that the 
horizontal line is the expectation for a random distribution. The left panel
shows the TPCF for all \lya absorbers, the center panel shows the TPCF for
the stronger half of our sample ($\Wno\gt65\mang$), and the right panel shows the residual TPCF after subtracting
the stronger half of the sample from the full TPCF (weighted by the number of pairs, see text). 
In the full sample, 20 line pairs contribute
to the first bin, while 10 pairs each contribute in the other samples. The error bars are 
statistical $\sqrt{N}$ errors only.}
\end{figure}

%
Table~\ref{pairs} lists all absorber pairs found in the STIS spectra
with velocity separations of $50\le\Dv\le150$\kmsno. A similar table for
line pairs found in the GHRS spectra can be found in Paper~II (Table~6). 
Figure~\ref{twopoint_no} display the results of our TPCF, $\xi(\Dv)$, analysis.
Table~\ref{pairs} lists by column: 
(1) The central wavelength of the line pair; 
(2-3) the wavelength and rest-frame velocity separation of the pair; 
(4-5) the equivalent widths of the two absorbers; 
(6-7) the observed \bvalues of the two absorbers; and (8) the target sightline.
We have visually inspected all line pairs in 
Table~\ref{pairs} and find that all but two entries are 
pairs of distinct lines. The other two pairs consist of two strong lines that 
are already both quite broad (50\kmsno). 
Both of these line pairs, the 1238.4\Ang pair in \objectname[]{PG~0804+761} and the 1225.5\Ang
pair in the \objectname[]{VII~ZW~118} sightline, are strong lines with asymmetric wings.
Thus, we believe that our line pair identification has been conservative,
and we have not included line pairs created by an arbitrary division of a single
broad line into two parts.
The GHRS data also had 2 of 13 marginally resolved line pairs, so
at most, the TPCF peak at $\Dv\lt190$\kms may be overestimated by 
15\% (4/26). As mentioned in Paper~II, we have also removed the $1280\ang\lt\lambda\lt1295\Ang$
portion of the GHRS \PKS~sightline because we purposely reobserved that pathlength to 
study a strong cluster of lines 
\citep{pks}. For the entire dataset, there are 18 spectra with no line pairs,
6 with 1, 5 with 2, 2 with 3, and 
\objectname[]{PG~1211+143} with 5 independent line pairs (i.e., no absorber duplications).
All line pairs in the \objectname[]{PG~1211+143} sightline have been verified
by inspection of a STIS E140M spectrum \citep{pg1211}.
We have used the same strategy as employed in \S~\ref{sec:dndzz} to estimate the
effect that cosmic variance has on the TPCF results by creating all
possible subsamples with 15 sightlines from our full sample and determining
the line pair distribution for this ensemble. By this process we estimate
that cosmic variance could add an uncertainty to the TPCF 
equal to the statistical errors shown in Figure~\ref{twopoint_no}.

Figure~\ref{twopoint_no} displays the TPCF obtained for our full \real\ GHRS+STIS dataset (left panel),
the strongest half of our sample ($\Wno\gt65\mang$, center panel), and the residual TPCF after subtracting
the strongest half of the sample from the full TPCF (labelled $\Wno\lt65\mang$, right panel).
The velocity separation bins in Figure~\ref{twopoint_no} are \Dv=70\kmsno. 
There is a 5.3\sig excess in the first two bins at $\Dv\leq190$\kmsno, and
a 7.2\sig excess at $\Dv\leq260$\kms in our full sample
(3.7\sig and 5.1\sig when cosmic variance is taken into account).
No other \lowz study has the velocity resolution and sample size required to see 
this peak clearly  \citep{Impey99,Ulmer96,Weymann98} because the line pairs in Table~\ref{pairs} would not be fully 
resolved by HST/GHRS+G140L or FOS spectra.
Therefore, we believe that the TPCF of the \lya forest at \lowz has the 
same general characteristics as at \highzno; i.e., no excess power at large
\Dv, but with an excess at $\Dv\lesssim200$\kmsno.
The excess power we find in our sample at low \Dv\ is somewhat larger than the
slight excess seen in the \highz data \citep{Rauch92}.
Other features in the full TPCF at larger separations are less than 3\sig and are not seen in other data \citep{Impey99}. 

The center panel of Figure~\ref{twopoint_no} indicates that the majority of
the TPCF power arises from the stronger half of our sample ($\Wno\gt65\Mang ;~\lognh\gt13.1$).
Lower number statistics results in only a 4.5\sig excess at $\Dv\leq190$\kms (5.6\sig at $\Dv \leq260$\kmsno),
but a much higher $\xi(\Dv)$.
Thus, the excesses seen in the full sample (Figure~\ref{twopoint_no}, left panel) are due mainly to the higher column density absorbers.
This is shown explicitly in the right panel of Figure~\ref{twopoint_no}, 
which is the residual TCPF, $\xi(\Dv)$, when subtracting the
center panel ($\Wno\gt65\mang$) from the full sample, appropriately weighted by the number of pairs in each bin.
Although labelled as $\Wno\lt65\mang$ in  Figure~\ref{twopoint_no}, this sample is the TPCF of our weak absorbers to
themselves and to the strong, $\Wno\gt65\mang$, absorbers. The weak absorbers do not show any evidence of clustering
with either the high or low column density absorbers at any scale. 
On the other hand, the TPCF for the stronger half sample exhibits a low statistical
excess (3\signo) at $260\lt\Dv\lt680$\kms as well as a deficit at $700\lt\Dv\lt4000$\kmsno, 
similar to the \lowz\ galaxy-galaxy TPCF result
presented in Paper~III. In that paper we interpreted the excess as
due to clustering of galaxies within supercluster filaments at 
$\Dv\lt1000\kms$ and the deficit as due to the presence of voids in the galaxy
distribution. We believe that we are beginning to see a hint of these same
signals in the high column density TPCF in Figure~\ref{twopoint_no}.

In Paper~III, we split our absorber sample into two subsets at approximately the same \W
as here. We found that the stronger absorbers were much more closely related to galaxies than the weaker absorbers,
which were much more randomly distributed in space. These two results support the two-population model for \lya absorbers
\citep{Sargent88} with the transition between these two populations occurring in the $\lognh=13-14$ regime.
The breaks in the low- and \highz column density distributions near $\lognh=14$ could be due to these more
numerous, more uniformly distributed \lyano-only absorbers. However suggestive, our current data cannot rule out a single population
of absorbers whose clustering properties and proximity to galaxies change monotonically from low to high column density within
our sample.
\begin{deluxetable}{cccccccl}
\tablecaption{\lya Line Pairs with Velocity Separations of 50--150\kms \label{pairs}}
\tabletypesize{\footnotesize}
\tablecolumns{8}
\tablewidth{0pt}
\tablehead{
\colhead{$\lam_c$\tablenotemark{a}} &
\colhead{$\Dlam$\tablenotemark{b}} &
\colhead{$\Delta V$\tablenotemark{c}} &
\colhead{$\Wno_1$} &
\colhead{$\Wno_2$} &
\colhead{$\bb_1$} &
\colhead{$\bb_2$} &
\colhead{Sightline} \\
\colhead{(\noang)} &
\colhead{(\nomang)} &
\colhead{(\nokmsno)} &
\colhead{(\nomang)} &
\colhead{(\nomang)} &
\colhead{(\nokmsno)}&
\colhead{(\nokmsno)}&
\colhead{}
}
\startdata
1278.19& 0.39&          90&          30&         277&          50&          48&HE1029-140\\
1224.85& 0.22&          53&          39&          51&          21&          35&MR2251-178\\
1222.06& 0.37&          90&          77&          40&          51&          40&PG0804+761\\
1238.40& 0.43&         104&         324&          27&          56&          48&PG0804+761\\
1277.97& 0.50&         118&         308&         691&          60&          99&PG1211+143\\
1235.91& 0.38&          90&         188&         153&          62&          42&PG1211+143\\
1268.54& 0.23&          54&         214&          91&          59&          37&PG1211+143\\
1294.33& 0.56&         130&         563&         249&          77&          61&PG1211+143\\
1294.89& 0.56&         128&         249&          65&          61&          64&PG1211+143\\
1235.96& 0.46&         112&         299&         280&          64&          47&PKS2005-489\\
1268.34& 0.63&         149&         106&         139&          85&          57&TON-S180\\
1225.49& 0.32&          77&          67&         267&          55&          51&VIIZW118\\
1234.50& 0.40&          97&          34&          45&          32&          60&VIIZW118\enddata
\tablenotetext{a}{Central wavelength.}
\tablenotetext{b}{Wavelength difference.}
\tablenotetext{c}{Rest-frame velocity difference.}
\end{deluxetable}

\clearpage
\section{Conclusions}\label{sec:conc}
We have observed 15 new sightlines with HST/STIS+G140M to investigate further the properties
of the \lowzno, low column density \lya forest. These new sightlines, together with
15 HST/GHRS+G160M sightlines, increase our sample size to \comboNabs\ \lya absorbers over
a pathlength of $\Dz = 1.157$. Using permutations of the available sightlines, we have tested our sample
for the effects of cosmic variance in \S~\ref{sec:dndzz}
and found that it still remains an important source of error in these results.
While our previous GHRS sample was insufficient to adequately sample the cosmic variance,
our new sample is close to being large enough. Cosmic variance is now
a minority contributor to the uncertainties on the results reported herein.
Observation of $\gtrsim 70$ sightlines with pathlength comparable to these STIS/GHRS spectra 
(total pathlength of $\Dz \approx 3$) would be
required to reduce the effects of cosmic variance to insignificance
($\leq 10\%$ of the total uncertainty).

The major conclusions of our analysis of this new, larger sample are:
\begin{enumerate}
\item While the \lya absorption lines observed in the STIS/G140M spectra are
similar to those found in our previous GHRS sample,
we have identified a few possible broad ($\bobs\gt100$\kmsno), shallow
absorption features. It is not known whether these are real absorbers or
undulations in the target continuum. 
Although their presence is intriguing for detections of the predicted
``warm-hot'' IGM \citep{Cen99,Dave01b}, we conclude that these features are more likely
blended, narrow absorbers or are not real.
There are too few of these absorbers in this sample to bias the statistics presented.
\lya absorption line studies are sensitive to gas
at $T\leq150,000~{\rm K}$ with neutral fraction $\fHI\gt10^{-5}$, and so the bulk of the warm-hot medium is best detected using other means.
\item At \zzerono, the column density distribution
for $\lognh\le14.5$, \dtwondzdnh $\propto \Nhno^{-\beta}$ with $\beta=\lowbeta\pm\Elowbeta$.
There is some evidence for a break at $\lognh=14.5$, above which $\beta=\highbeta\pm\Ehighbeta$. 
The location of  this break is uncertain owing to small 
number statistics amongst the higher column density lines
and to the dependence of column density on our selection of $\bb = \myb$.
It is consistent with the break at $\lognh=16$ 
and $\beta=1.3$ found by the Key Project team \citep{Weymann98}.
Identifying this subtle break with a similar break found in the \highz
column density distribution \citep{Fardal98,Hu95}, suggests that a typical 
high \Nh \lya absorber has diminished in \hone column density by a 
factor of $3-10$ from $\z=3$ to the present. 
\item We calculate that the \lowzlya forest produces a Lyman-continuum opacity 
at 1~ryd of \dtaudz$\approx$0.01 for $\lognh\le13$ and \about0.1 for $\lognh\le15$.
Most of the intergalactic opacity probably arises from rare, higher-\Nh 
absorbers, and is, therefore, highly stochastic.
\item We have compared the redshift evolution of \dndz between $\z = 3$ and 0
(Figure~\ref{comp_dndz}) at low and high \Nhno.
We find that the significant evolution can be explained by hydrodynamical simulations of the 
\lowzlya forest for the predicted evolution of the ionizing background \citep{Haardt96}. 
Our \zzerono, high column density point is significantly lower than that measured by the Key Project team using FOS
spectra. 
However, when line blending is taken into account, our revised \dndzzero value is
in close agreement with the $\Lambda$CDM simulation of \citet{Dave99}.
\item The \z\ evolution of \dndz for the \lownh 
(\kimrange) absorbers is qualitatively similar to the results at higher \Nhno.
The \lownh absorbers appear either to have more modest evolution than the \highnh absorbers, or
the break from faster to slower evolution occurs at a later time for the \lownh absorbers.
These conclusions are tentative, pending new observations
of \lownh absorbers at intermediate redshifts now being made with HST.
\item We find a 5.3\sig excess in the TPCF of \lowz \lya absorbers for velocity separations $\Dv\le190$\kmsno,
and a 7.2\sig excess at $\Dv \leq260$\kmsno, consistent with,
 but slightly larger than, results from higher redshift studies \citep{Rauch92}.
The effects of cosmic variance could reduce the significance of these results by $\sim\sqrt{2}$.
The excess signal at $\Dv\le260$\kms is due exclusively to the higher column 
density absorbers ($\Wno\ge65\Mang ;~\lognh\gt13.1$).
This result, taken together with a similar result at \highz \citep{Womble96} is
new evidence supporting the two-cloud population model for \lya absorbers introduced by
\citet{Sargent88}. 
In Paper~III, we showed that the higher \Nh absorbers are more closely
associated with galaxies than the \lownh absorbers. 
Based upon our present TPCF result, and the galaxy proximity results from Paper~III,
the more uniformly-distributed \lya forest absorbers become the dominant population
below $\lognh \sim 14$ at \zzerono.
\item In Paper~II, we applied a photoionization correction
to find that the \lowzlya forest may contain $\ge 20$\% of the total
number of baryons, with closure parameter
$\omegalya=(0.008\pm0.001)\hsfi$, for a standard absorber size of 100~kpc and an ionizing
radiation field of intensity $\Jno\approx10^{-23}$\cgs\ and spectral index $\alpha_s =1.8$.
In this paper, by assuming $\Jno = 1.3\times10^{-23}$\cgs, we have used both our original method and a different accounting method \citep{Schaye01} 
to our enlarged dataset, finding a somewhat larger value than above (\totalomega). 
\end{enumerate}
The steep slope of the \lowz column density distribution 
\citep[$\beta=\lowbeta$ compared to $\beta=1.4$ at \highzno;][]{Hu95} 
means that the baryon content of the \lowzlya forest is even
more dominated by the lower column density absorbers.
Despite recent suggestions that a warm-hot medium
may contain a substantial baryon fraction in the current epoch \citep{Tripp03,Dave01b,Cen99}, we argue that
some of this warm-hot medium is already accounted for in our derived baryon fraction.
\lya absorption is still detectable for absorbers with $T\leq150,000~{\rm K}$.
Above this temperature, shallow broad \lya absorption becomes too difficult to detect in
typical HST/STIS or GHRS spectra, owing to intrinsic AGN continuum fluctuations 
with wavelength. Sensitive searches for potential broad, shallow absorbers can be made with COS by
targeting BL~Lac objects, whose continua are simpler to model and are not expected to have such fluctuations.
\acknowledgements
This work was supported by HST guest observer
grant GO-06593.01-95A, the HST COS project (NAS5-98043), the Colorado
astrophysical theory program (NASA grant NAGW-766 and NSF grant AST02-0642).
We thank Ken Sembach for providing us with the PG~1116+215 first-order STIS spectrum,
Mary Putman for providing the LDS and HIPASS Galactic \hone velocity maps, and
Jessica Rosenberg for providing \bvalues of \lya absorbers from STIS echelle data prior
to publication. We thank Chris Impey for emphasizing to us the importance of cosmic 
variance on our analysis, and we thank our referee, Joop Schaye, for his many helpful
suggestions.
\appendix
\section{STIS+G140M Spectra and Absorption Line Lists}\label{sec:ApA}
In this Appendix we present composite HST/STIS+E140M spectra, and line 
lists (Table~\ref{linelist_all}) describing the detected
absorption features. Descriptions of features unique to 
individual objects are presented at the end of this Appendix. 

\begin{figure}[hb] \epsscale{0.64}
\plotone{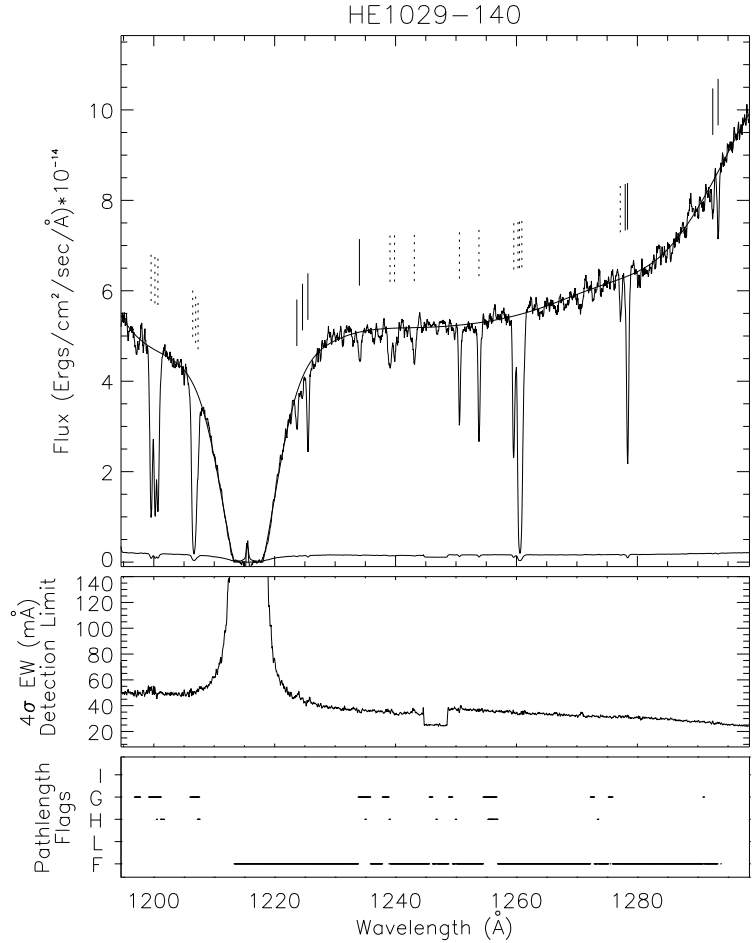}
	\caption{\label{HE1029-140}STIS spectra, sensitivity limits, and pathlength flags for the HE1029-140 sightline.}
\end{figure}

\begin{figure}[hb] \epsscale{0.64}
\plotone{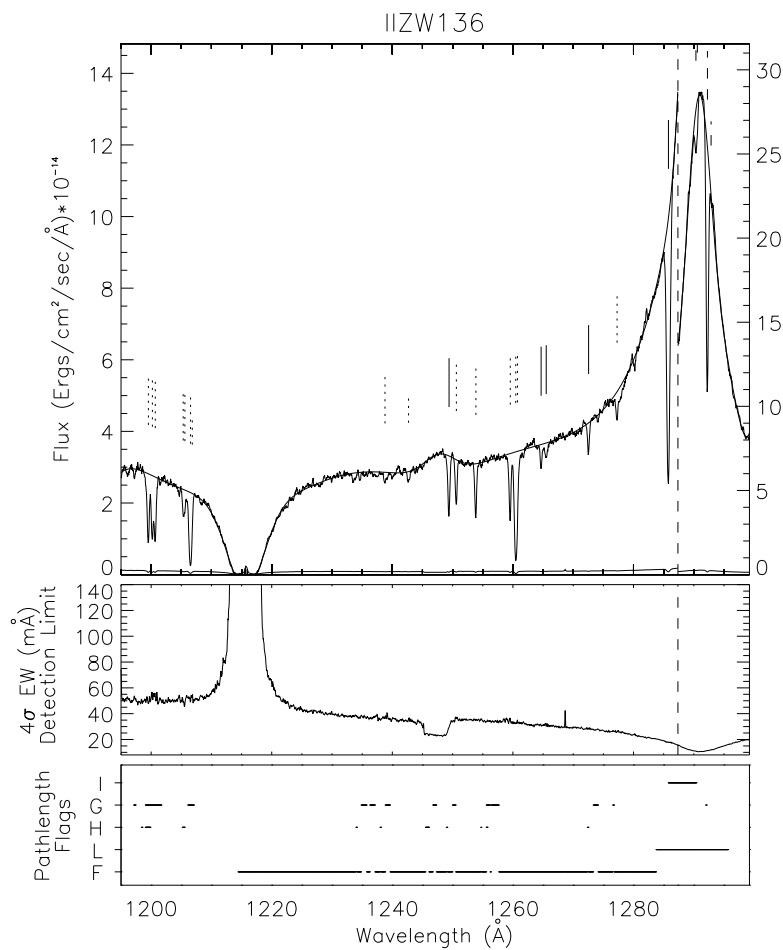}
	\caption{\label{IIZW136} Spectrum, sensitivity limits, and pathlength flags for the  II~ZW~136 sightline.}
\end{figure}

\begin{figure}[hb] \epsscale{0.62}
\plotone{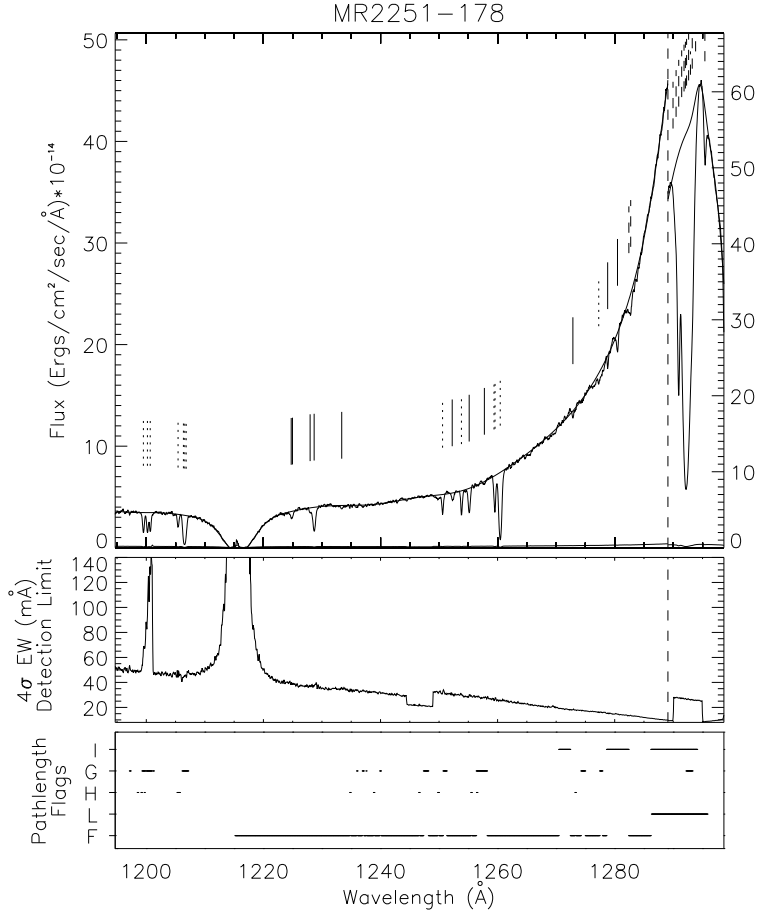}
	\caption{\label{MR2251-178}Spectrum, sensitivity limits, and pathlength flags for the MR~2251-178 sightline.}
\end{figure}

\begin{figure}[hb] \epsscale{0.64}
\plotone{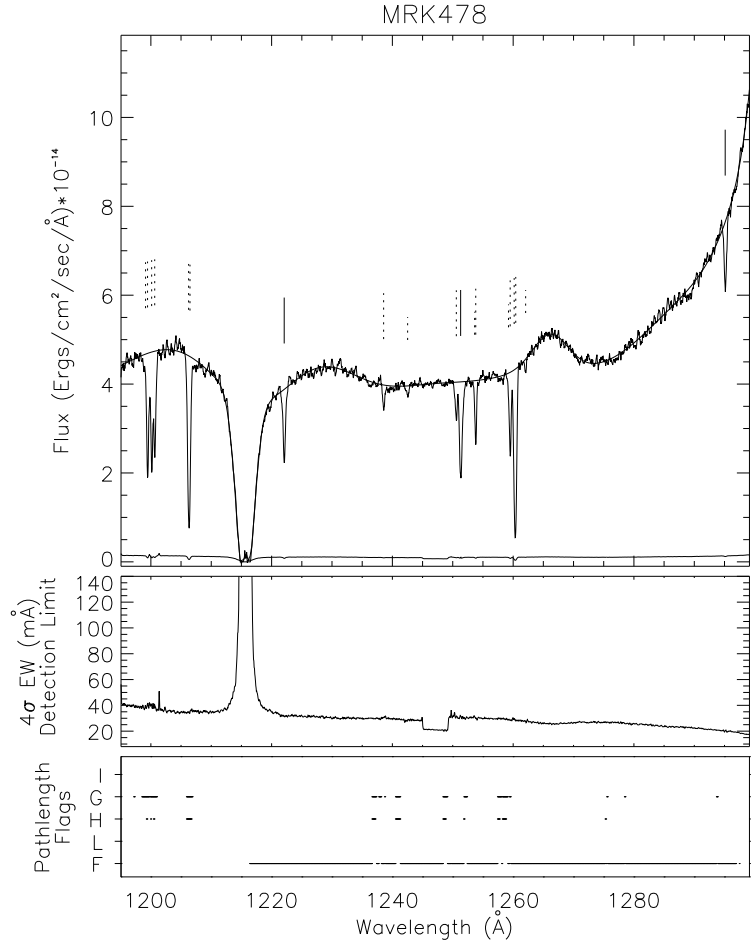}
	\caption{\label{MRK478} Spectrum, sensitivity limits, and pathlength flags for the  MRK~478 sightline.}
\end{figure}

\begin{figure}[hb] \epsscale{0.64}
\plotone{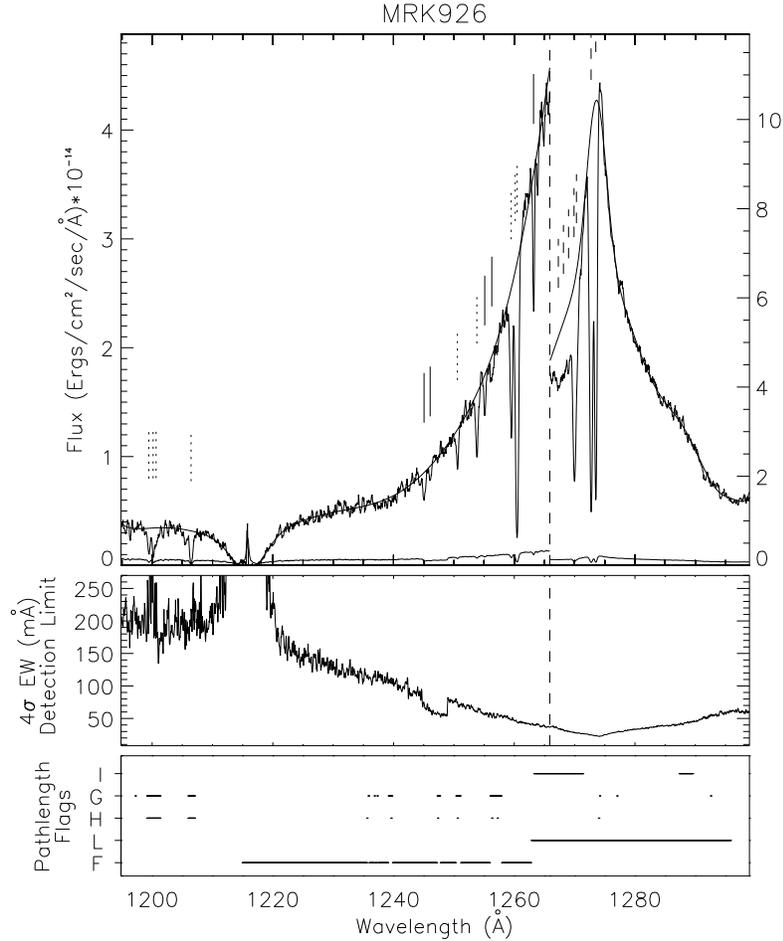}
	\caption{\label{MRK926} Spectrum, sensitivity limits, and pathlength flags for the  MRK~926 sightline.}
\end{figure}

\begin{figure}[hb] \epsscale{0.64}
\plotone{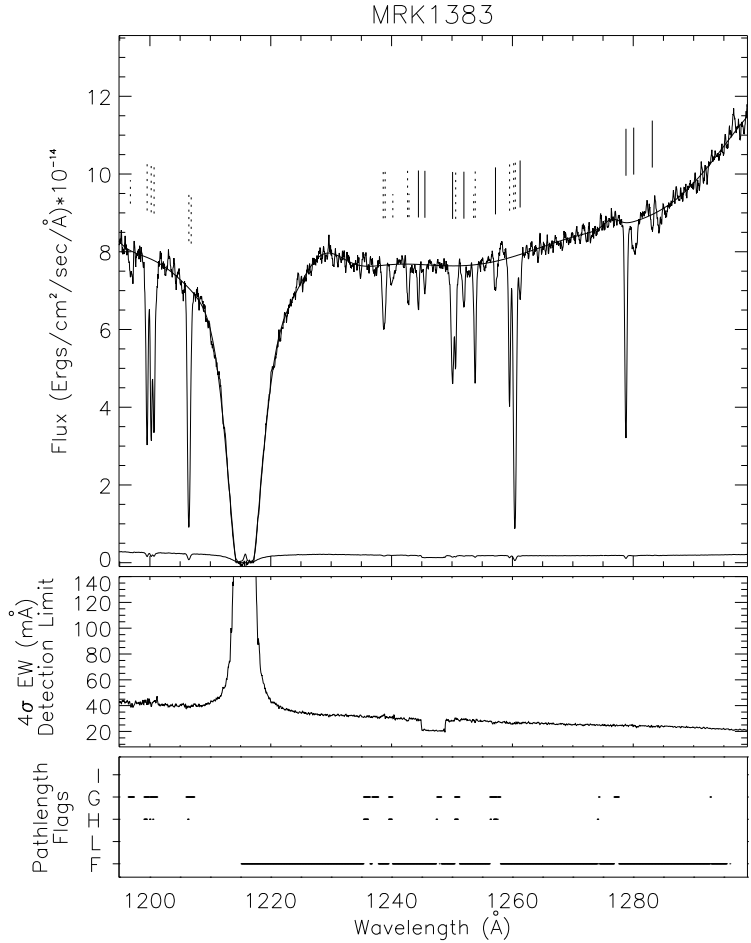}
	\caption{\label{MRK1383} Spectrum, sensitivity limits, and pathlength flags for the  MRK~1383 sightline.}
\end{figure}

\begin{figure}[hb] \epsscale{0.64}
\plotone{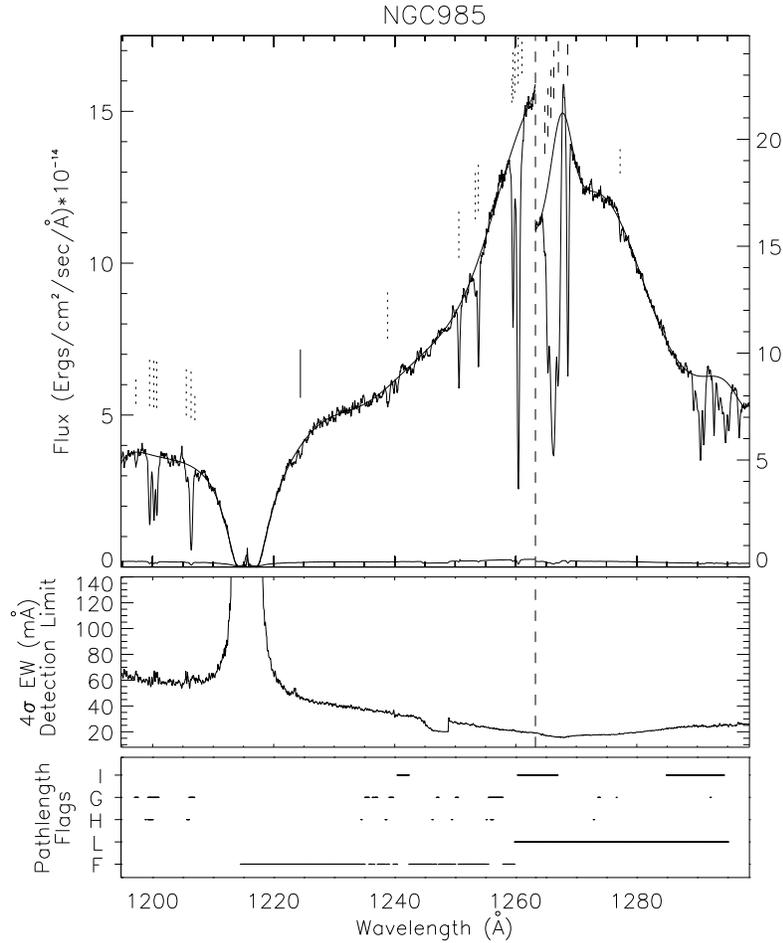}
	\caption{\label{NGC985} Spectrum, sensitivity limits, and pathlength flags for the  NGC~985 sightline.}
\end{figure}

\begin{figure}[hb] \epsscale{0.64}
\plotone{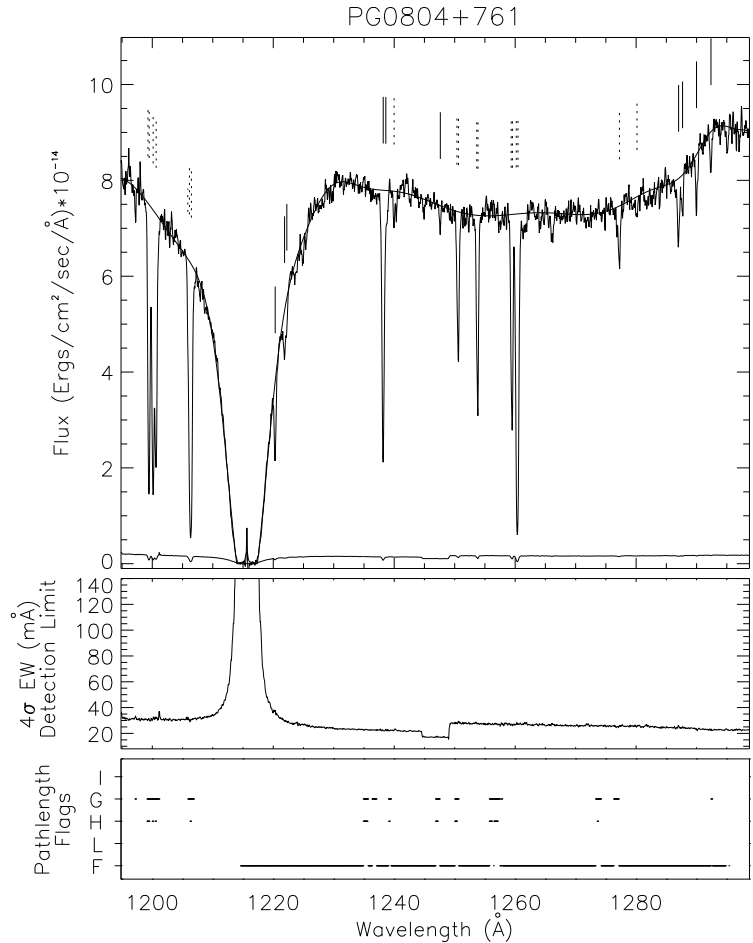}
	\caption{\label{PG0804+761} Spectrum, sensitivity limits, and pathlength flags for the  PG~0804+8761 sightline.}
\end{figure}

\begin{figure}[hb] \epsscale{0.64}
\plotone{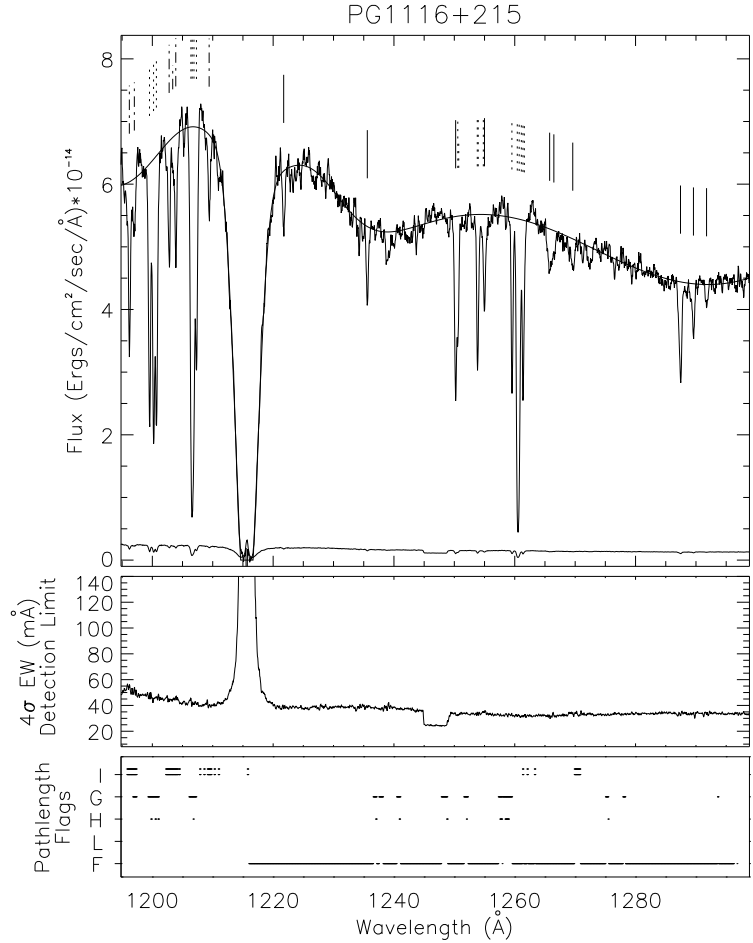}
	\caption{\label{PG1116+215} Spectrum, sensitivity 
limits, and pathlength flags for the  PG~1116+215 sightline.}
\end{figure}

\begin{figure}[hb] \epsscale{0.64}
\plotone{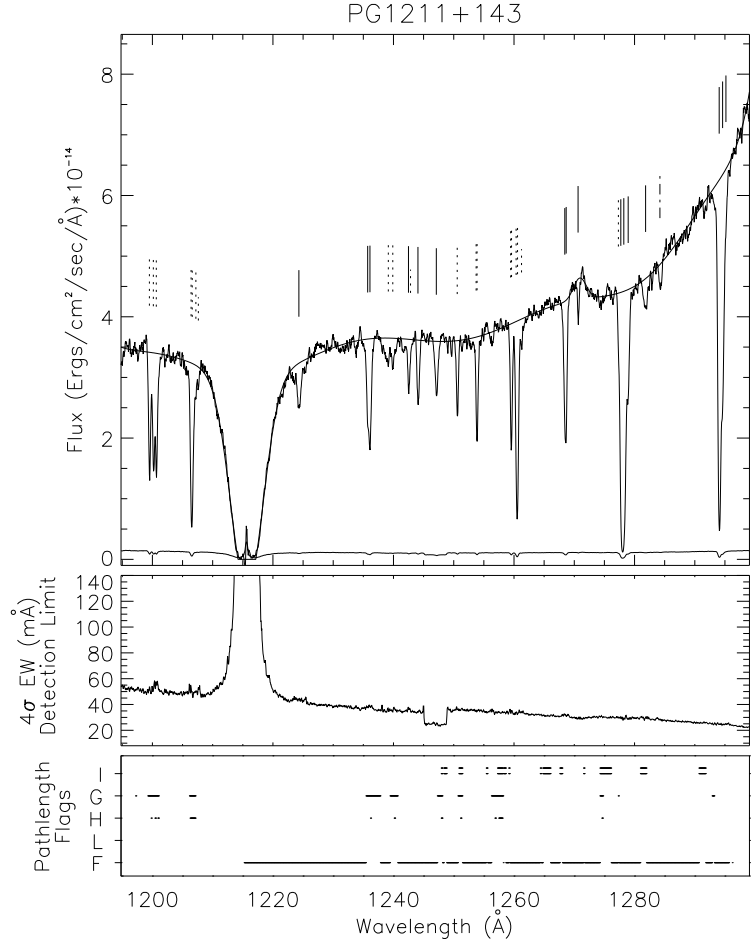}
	\caption{\label{PG1211+143} Spectrum, sensitivity limits, and pathlength flags for the  PG~1211+143 sightline.}
\end{figure}

\begin{figure}[hb] \epsscale{0.64}
\plotone{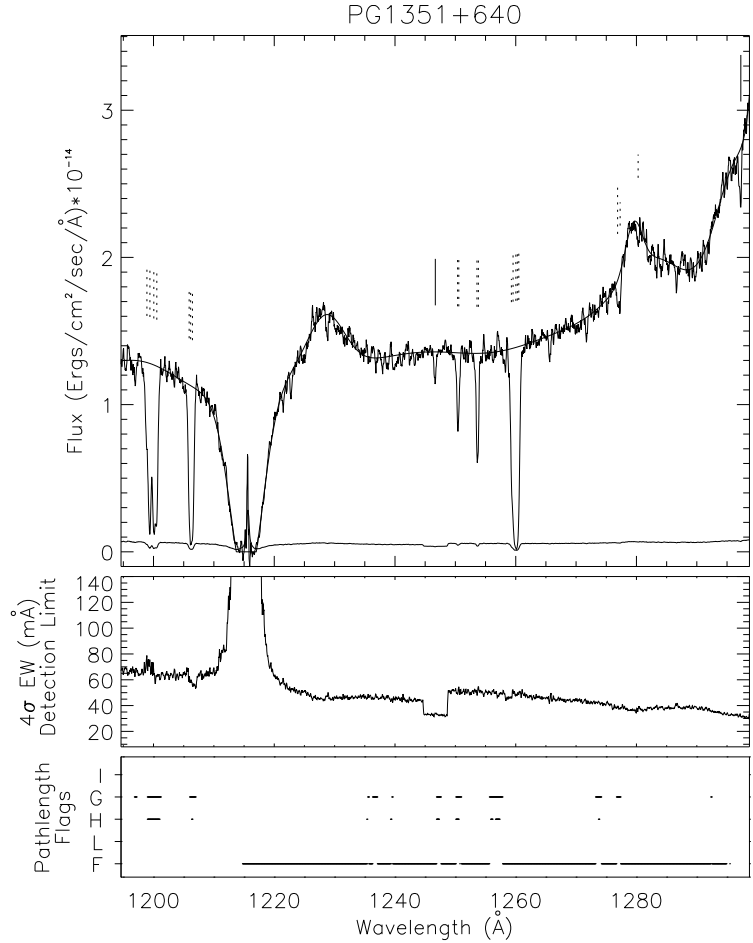}
	\caption{\label{PG1351+640} Spectrum, sensitivity limits, and pathlength flags for the  PG~1351+640 sightline.}
\end{figure}

\begin{figure}[hb] \epsscale{0.64}
\plotone{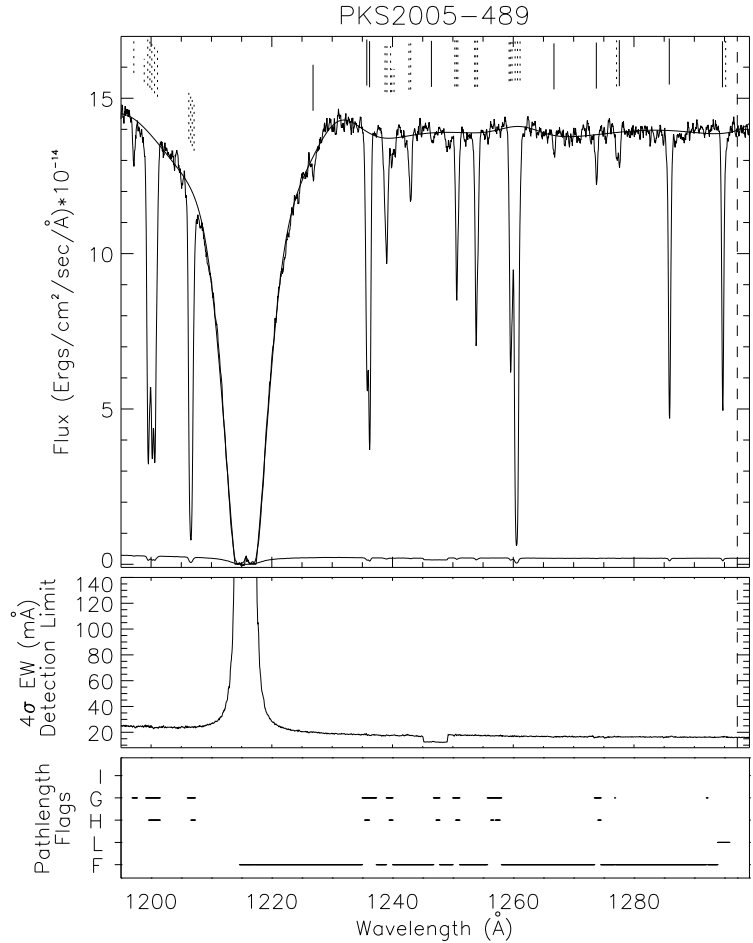}
	\caption{\label{PKS2005-489} Spectrum, sensitivity limits, and pathlength flags for the  PKS~2005-489 sightline.}
\end{figure}

\begin{figure}[hb] \epsscale{0.64}
\plotone{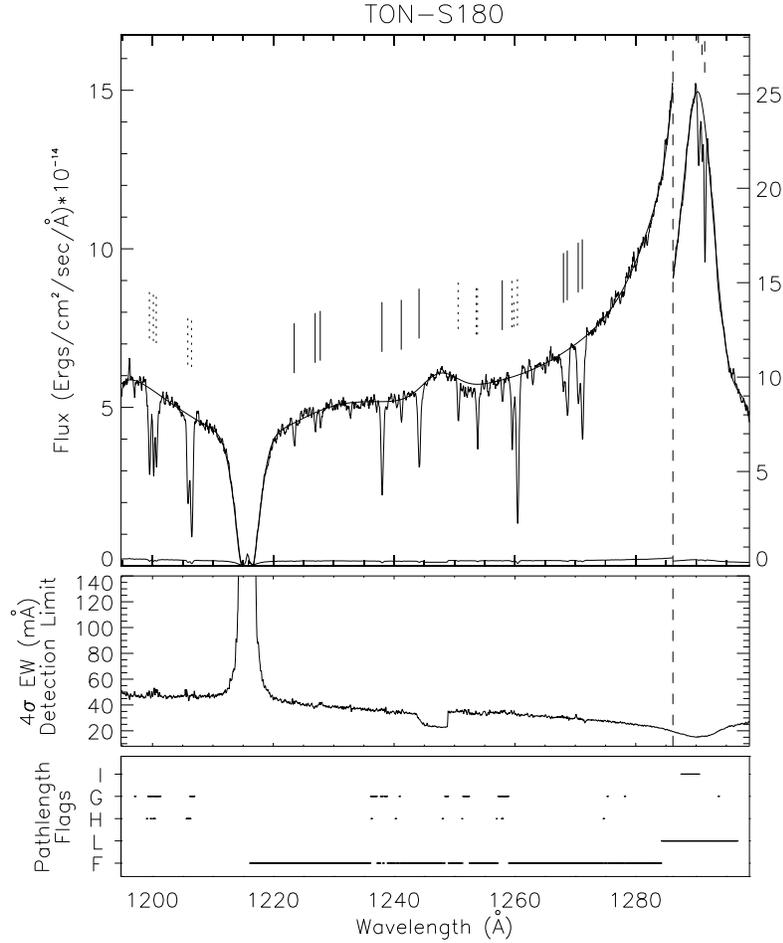}
	\caption{\label{TONS180} Spectrum, sensitivity limits, and pathlength flags for the TON~S180 sightline.}
\end{figure}

\begin{figure}[hb] \epsscale{0.64}
\plotone{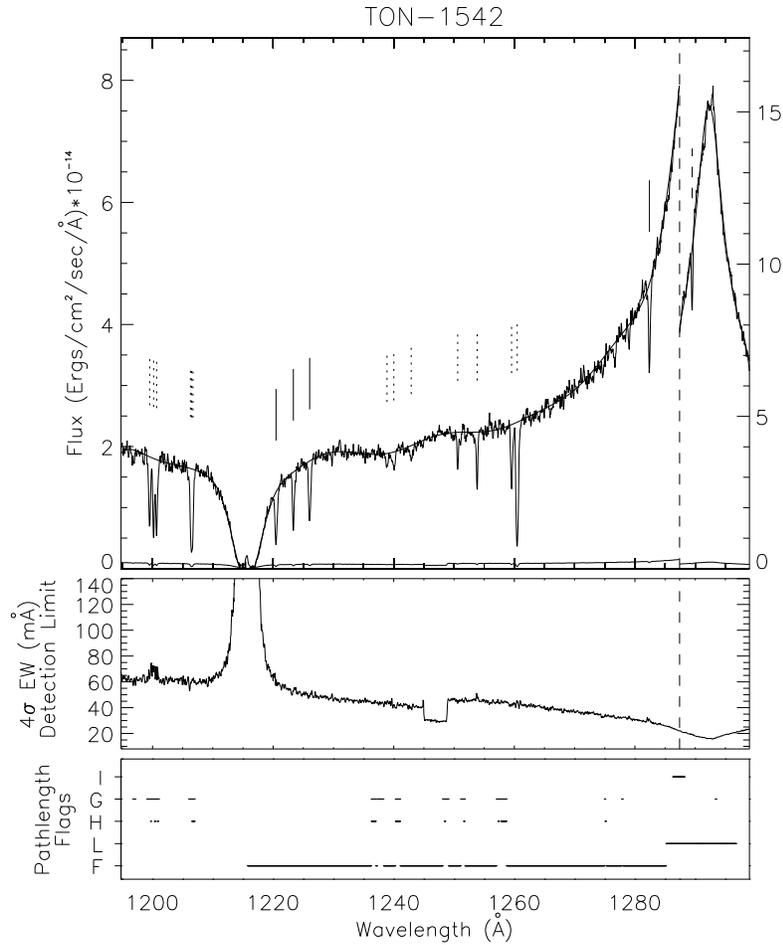}
	\caption{\label{TON1542}Spectrum, sensitivity limits, and pathlength flags for the  TON~1542 sightline.}
\end{figure}

\begin{figure}[hb] \epsscale{0.64}
\plotone{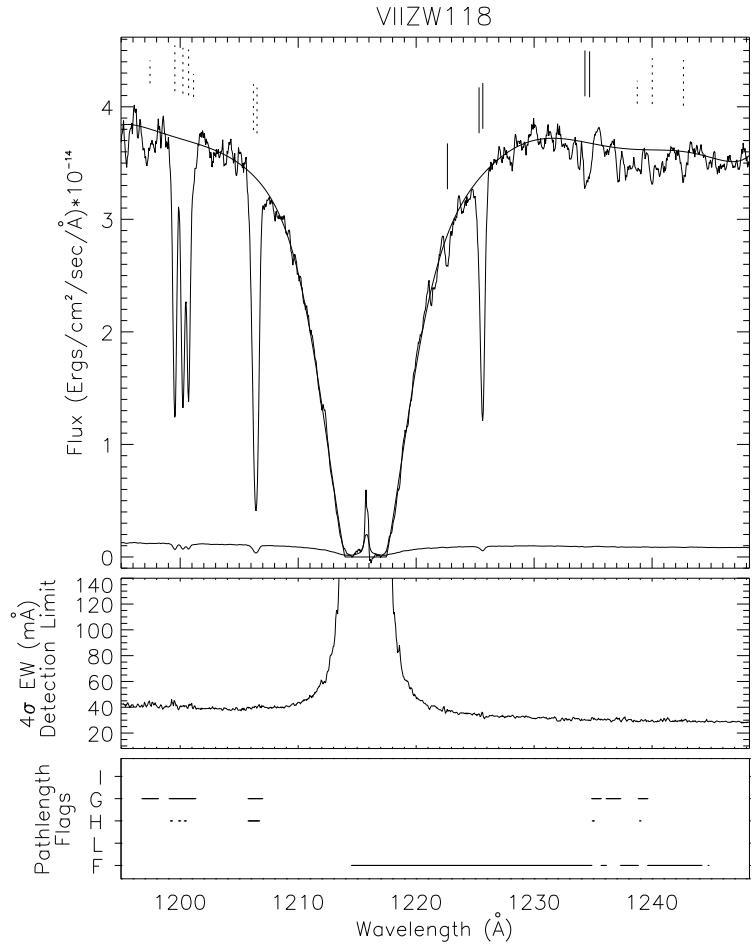}
	\caption{\label{VIIZW118} Spectrum, sensitivity limits, and pathlength flags for the VII~ZW~118 sightline.}
\end{figure}

The STIS+G140M spectra, error vectors, and pathlength accountings are 
presented in Figures~\ref{HE1029-140}--\ref{VIIZW118} for all targets.
\notetoeditor{We are open to suggestions as to the best way to combine figures 12-26 into
a more economical format (including moving the appendix "paragraphs" into the figure captions).
If figures 12-26 are converted into a single figure, then please change the above sentence to
"The STIS+G140M spectra, error vectors, and pathlength accountings are 
presented in Figure~12 for all targets in alphabetical order."}
The upper panel of each figure presents the composite spectrum and the
\onesig error values, which are plotted below the flux vector.
As discussed in Paper~I, our list of intergalactic
\lya absorbers is restricted to those more than 1200\kms blueward of 
the target \lya emission.
This limit is indicated by the dashed vertical line in the upper and 
middle panels of the figures. All non-Galactic features
redward of this line are classified as intrinsic absorbers. 
While we realize that the 1200\kms limit is somewhat arbitrary, our choice is 
based upon HST work on intrinsic absorbers by
\citet{Crenshaw99} and is justified in detail in Paper~I.
All otherwise unidentified absorption features blueward
are taken to be intergalactic \lya absorption features, and are 
indicated by a solid vertical line above the feature.
\lya absorbers with significance level (\SL)~$\gt4\sig$ 
are plotted with a solid vertical line. 
Galactic and high-velocity cloud (HVC) absorption 
lines are similarly represented with a dotted
line (\real\ longer line, \tent\ shorter line), 
intrinsic absorption lines with a dashed line, and intervening 
higher-\z\ absorption (\lyb and/or \ion{O}{6}) lines with a dot-dashed line.
HVCs are identified by their metal absorption lines and have $|cz| \le 500\kmsno$.
As described in Paper~I and \S~\ref{sec:spec}, the global continuum fits are a combination 
of a polynomial of order given in 
Table~\ref{stis_objects} for each object plus Gaussian components for \lya and weaker emission
lines from the target and Galactic damped \lya absorption. The weak emission
lines modeled by the continuum fits are described for each target individually 
in this Appendix.
Once the continuum fits are made and the absorption features
removed, the remaining continuum fluctuations are Gaussian distributed
with a standard deviation equal to the noise vector plotted at the bottom of each spectrum. 
Given the number of independent REs in our full
dataset, we expect that $\about 10\%$ of the \tent\ \lya absorption features
are spurious, but that $\lesssim 0.1\%$ of the \real\ absorption lines are spurious.
However, the specified wavelengths of the Galactic features means that most of the 
\tent\ Galactic absorptions are real. Therefore, we identify all of them and list
them in Table~\ref{linelist_all}.

In several cases, where \lya emission occurs in 
the observed band, the portion of the spectrum redward of the (\proxno) proximity 
limit (vertical dashed line) is excluded from the region best fitted by the 
continuum. Thus, any intrinsic or Galactic absorptions
redward of the ``proximity limit'' may not have their rest-frame 
equivalent widths (\Wno) most accurately represented. 
The middle panel shows the 4\sig \W sensitivity 
detection limit per RE as a function of wavelength for each target. 
As with the upper panel, we include in our statistics only the portion of the spectrum that
lies $\gt1200$\kms blueward of the target \lya emission. Finally, the 
bottom panel of each figure summarizes the
available pathlength for inclusion in our sample and line statistics. 
The row marked {\bf `I'} indicates the portion of
the spectrum removed due to specific features intrinsic to the target or 
non-\lya lines associated with known
intervening systems at higher redshift not associated with the target 
AGN. The pathlength removed due to higher
redshift intervening absorbers (PG1116+215 only) is indicated by a double line to 
differentiate it from that removed due to intrinsic
absorption features. The other rows are marked as follows: {\bf `G'} 
indicates pathlength attributed to non-HVC Galactic features such as
\SIItriplet, {\bf `H'} indicates pathlength attributed to HVCs, {\bf 
`L'} indicates the portion of the spectrum that is redward of \proxno\ for \lyano, 
and {\bf `F'} indicates those regions of the
spectrum that are available for the detection of intergalactic \lya 
absorbers. 
The {\bf `F'} row also corrects for our
inability to detect and model features near the edge of our 
wavelength coverage; we remove 10 pixels on the edges
of our wavelength coverage. Only for \objectname[]{PG~1116+215} does the 
intrinsic/intervening {\bf `I'} accounting remove any significant pathlength not already 
removed by the {\bf `L'} row. Although not explicitly marked, spectral regions with 
$\lambda\lt1218\ang$ are unavailable for the detection of intergalactic \lya absorbers,
owing to Galactic \lya absorption.

For a few objects, the \prox ``proximity limit'' lies within the 
observed spectral band and is indicated by a dashed vertical line.
For targets with strong intrinsic \lya emission
(\objectname[]{II~ZW~136}, \objectname[]{MR~2251-178}, \objectname[]{MRK~926},
\objectname[]{NGC~985}, \objectname[]{TON~1542}, and \objectname[]{TON~S180}), fluxes are 
scaled differently above and below this convenient breakpoint. For these spectra, the left 
axis corresponds to flux blueward of the proximity limit, while the right axis 
corresponds to flux redward of this limit. In all spectra, the \onesig error vector
is plotted at the bottom.

The \real\ line lists for each individual spectrum are presented sequentially in Table~\ref{linelist_all};
Galactic absorptions at \tent\ are also listed.
The first column of Table~\ref{linelist_all} indicates the LSR-adjusted
wavelength and wavelength uncertainty for each feature. The wavelength
scale of each spectrum was corrected to the LSR using the Galactic 
\hone 21~cm emission, \NItriplet, and \SIItriplet, where available. 
The second column lists the non-relativistic velocity (\cz\ in\kmsno)
relative to the LSR. For \lya absorbers judged to be intrinsic to the 
AGN, velocities are listed relative to the AGN narrow-line region (see \S~\ref{sec:spec} and Paper~I). 
For intervening non-\lya absorbers, we report the velocity (\cz) of the absorber. 
We also provide velocity or redshift uncertainties, 
based upon the total wavelength uncertainties (see \S~\ref{sec:spec} and Paper~I). 
The third column provides the single-component Doppler widths (\bb\ in\kmsno) and uncertainties
for each feature as estimated from the Gaussian line widths. 
While the Doppler widths have been corrected for the spectral
resolution of STIS, the single-component Gaussian fits were 
restricted to the range $12\lt\bobs\lt100$\kmsno. However, we neither use nor recommend 
the use of these derived \bvaluesno, due both to the complex LSF
of the STIS instrument and to an unknown amount of ``jitter'' during
these long exposures (see \S~\ref{sec:spec}).
The fourth column lists the rest-frame equivalent width (\W in\mang) and its 
uncertainty for each absorption feature. 
This uncertainty includes both the statistical error of the $\chi^2$ fit and our 
conservative estimate of the systematic error in the continuum placement (4.2\%).
The fifth column indicates the significance level (\SL\, in \signo) of each feature.
The \SL\ is calculated by integrating the signal-to-noise ratio (per RE) of the best-fit 
Gaussian for each feature.
The final two columns present the identification (Id) and alternative 
identification (Alt~Id), if applicable for each feature.
Absorption lines that are determined not to be intergalactic \lya have 
their identifications prefaced by {\bf g:}
(Galactic), {\bf h:} (HVC), {\bf i:} (intrinsic), or {\bf z:} 
(intervening). Alternate line identifications that lie between
0.2--0.4\Ang from the expected location are indicated as speculative by 
the inclusion of a {\bf ?} following the identification.
%
\clearpage
\paragraph{HE~1029-140}\label{sec:HE}
The field of this radio-quiet QSO ($\z=0.086$) has been studied in
detail by \citet{Wisotzki94} and \citet{Bahcall97}. Surveying a
$10\arcmin.2 \times 16\arcmin.2$ field, \citet{Wisotzki94}
report a field galaxy, which they designate G10, 4\arcmin.2
off the \objectname[]{HE~1029-140} line-of-sight at $\z=0.0511\pm
0.0002$ ($\cz=15,319\pm60$\kmsno). 
This places the galaxy 268\hsfi\ kpc from the sightline with a
projected \lya expected at 1277.8\ang. We report two \lya features in this
region: a weak absorber ($\Wno=30\mang$) at \cz=15,369\kms and a strong
absorber ($\Wno=278\mang$) at $\cz=15,464$\kmsno.
Thus, further investigations of the galaxy field at this redshift are warranted. 
Two HVCs are detected in \SiIII{1206.5} and \SFsixty, at velocities of
approximately 100\kms and 200\kmsno.
\clearpage
\paragraph{II~ZW~136}The spectrum of this Seyfert~1 galaxy (also known as 
\objectname[]{UGC~11763}, \objectname[]{MRK~1513}, and
\objectname[]{PG~2130+099}) shows the usual Galactic absorption features 
plus two HVCs at $-225$ and $-295$\kmsno. The absorption feature at
1249.4\Ang is certainly \lya and not HVC
\SII{1250}, owing to the  lack of any absorption associated with
\SII{1253.8} or 1259.5. The Galactic \NV{1242.8} line is
peculiarly strong and asymmetric, indicating possible, but unclaimed
blending with extragalactic \lyano. The small bump in the continuum
near 1248\Ang is intrinsic \CIII{1175} emission.
Both the shape of the 1285.8\Ang \lya line and its intergalactic/intrinsic
nature are uncertain, owing to its proximity to the peak of the AGN \lya emission.
\clearpage
\paragraph{MR~2251-178}
The spectrum of this $\z=0.0644$ QSO shows strong intrinsic 
absorption from $-50$ to $-800$\kmsno, and HVC gas at $-270$\kms in \SiIII{1206.5} and \SFsixty.
The intrinsic absorber also contains \Sithree\ at two wavelengths: 1282.4 and 1282.7\ang.
{\it Far Ultraviolet Spectroscopic Explorer} (FUSE) data indicate \ion{O}{6} 
HVC gas at $-340$ to $-180$\kms and $-145$ to $-65$\kms \citep{Sembach03}.
The 9021\kms absorber is possibly associated with gas
in the vicinity of ESO~603-G025 ($cz=9030$\kmsno, 322\hsfi\ kpc off the line of sight).
The strong 3205\kms absorber is possibly associated with gas
in the vicinity of ESO~603-G027 ($cz=3267$\kmsno, 354\hsfi\ kpc off the line of sight). 
In fitting the strong intrinsic 
\lyano, our restriction of \bvalue$\le100$\kms (appropriate for the
intervening absorbers; see Paper~I) results in numerous non-unique 
components.
There is no physical reason to divide the intrinsic absorption like this, but a case
can be made for two components: one at 832\kms and a broad blend of the
remaining components.
\clearpage
\paragraph{MRK~478}
The spectrum of this Seyfert~1 galaxy shows IVC absorption at $-90$\kms in \SiIII{1206.5}. 
Asymmetries in the line profiles indicate additional possible IVC absorption in \SIItriplet\
 and \SFsixty. FUSE data indicate \ion{O}{6} HVC gas at 340--435\kms \citep{Sembach03}.
The continuum bumps at \about 1227\Ang and 1267\Ang are due to intrinsic \NI{1134.9}
and \CIII{1175.7}, respectively.
\clearpage
\paragraph{MRK~926}
The spectrum of this QSO shows possible IVC absorption in \SFsixty\ at \about $-80$\kmsno. 
FUSE data indicate \ion{O}{6} HVC gas at $-395$ to $-65$\kms \citep{Sembach03}.
The spectrum shows extensive intrinsic absorption
up to $\sim -1400$\kms from the recession velocity.
In this one case, we have identified the highest relative velocity component in
this absorption trough as intrinsic \lya despite it exceeding the -1200\kms limit we 
use to identify probable intrinsic absorbers. We feel that
this exception is justified in this case because the component which
exceeds -1200\kms is part of a contiguous absorption trough extending down
to $\sim -600$\kms (see Figure~\ref{MRK926} and Table~\ref{linelist_all}).
Although the absorption at 1263.19\Ang is at approximately the 
correct wavelength to be intrinsic \Sithree, it is much too strong.
Therefore, we identify this absorber as intergalactic \lyano.
\clearpage
\paragraph{MRK~1383}
The spectrum of this Seyfert~1 galaxy shows an IVC
absorption at \about $-45$\kms in \NVdoublet\ and \SII{1253.8}, barely resolved from the Galactic
absorption. Asymmetries in the line profiles also suggest
possible IVC absorption in \SiIII{1206.5} and \SFsixty. 
FUSE data indicate \ion{O}{6} HVC gas at 100--160\kms \citep{Sembach03}.
There is a very broad trough (FWHM \about 130\kmsno) of
absorption near the location of Galactic \CI{1280.1} that we identify
as extragalactic \lya due to the absence of Galactic
\CI{1277.2} absorption (see \S~\ref{sec:broad} and Table~\ref{broad_lya}). 
Similarly, we identify the 1250.1\Ang feature as \lya and not HVC \SII{1250.6} due to
 the absence of \SII{1253.8} and \SII{1259.5} HVC absorption.
The continuum bumps near \about 1230\Ang and 1277\ang, respectively, are intrinsic
\NI{1134.9} and \CIII{1175.7} emission.
\paragraph{NGC~985}
NGC~985 (MRK~1048) is a composite system of two
interacting galaxies with nuclei separated by \about 3\arcsec, 
only one of which is an AGN \citep{Marcum93}. 
Strong intrinsic absorption systems with velocities relative to
NGC~985 between $-200$ and $-800$\kms are detected in our spectrum in
\lya and \NVdoublet\ \citep{Appleton02}. \citet{Bowen02} report an association between
a \lya absorber and the galaxy \objectname[]{NGC~988} (\cz=1504\kmsno,
36.2\arcmin\ and thus 266\hsfi~kpc off the line of sight).
We do not confirm this \lya absorber, since our nearest \real\ \lya absorber is at 2155\kmsno.
HVC gas at $-230$ to $-110$\kms is seen in
\SiIII{1206.5}, \SII{1253.8}, \SII{1259.5}, and \SFsixty. 
The 1238.8\Ang ($\Wno=36\pm27\mang$) absorption line is 
identified as Galactic \NV{1238.8}, despite undetected \NV{1242.8}. 
\paragraph{PG~0804+761}
The spectrum of this \z=0.100 QSO contains a strong IVC, seen in most of
the  Galactic features at $-60$ to $-20$\kmsno, consistent with 
\citet{Lockman}.
The accuracy of the central wavelengths and equivalent widths of the
blended Galactic \NV{1238.8} absorption and the corresponding HVC absorption are
compromised by the presence of a nearby strong \lya feature at
1238.2\ang.  The continuum bumps at \about 1232\ang, 1250\ang, and
1293\Ang are intrinsic emission of the \ion{C}{1} blend, \NI{1134.9}, and
\CIII{1175.7}, respectively.
Galactic \CI{1277.2} is particularly strong in this sightline, and the
identification is supported by 
\CI{1280.1}. The broad absorption features near 1222\Ang 
are possibly associated with gas in the neighborhood of
 UGC 04238 ($\cz=1544$\kmsno, 146\hsfi\ kpc off the line of sight). 
This is one of the closest \lya absorber-galaxy associations in our entire survey.
\paragraph{PG~1116+215}
The spectrum of this $\z=0.176$ Seyfert~1 
shows a strong HVC at 100--215\kms in \SiIII{1206.5} and \SFsixty.
The continuum bumps at 1208\Ang and 1225\Ang are intrinsic 
\lyb and \OVIdoublet\ emission, respectively.
The 1238--1241\Ang spectral region is complicated, with numerous weak absorption features
that appear to be a combination of Galactic and HVC \NV{1238.8}
and \MGdoublet.
Our data support the suspicion of \citet{Tripp98}
that their absorption feature at 5846\kms 
is not \lyano, but is instead a combination of weak Galactic and HVC absorption. 
FUSE data indicate \ion{O}{6} HVC gas at 115--310\kms \citep{Sembach03}.
Intervening \lyb absorption is detected at
1196.2\Ang (\z=0.1661), 1197.0\Ang (\z=0.1670), and 1203.9\Ang 
(\z=0.1736), and intervening \OVI{1031.9} absorption is seen at \z=0.1656
and \z=0.1661, and \OVI{1037.6} at \z=0.165 (identifications consistent with Tripp \etl 1998).
\paragraph{PG~1211+143}
\citet{Lockman} reports $-62$ to +36\kms IVC gas along the sightline to this \z=0.0809
Seyfert. We confirm the blueshifted IVC and note
two possible HVC absorptions at +180 and +280\kms in \SiIII{1206.5}.
The IVCs cause additional uncertainty in the LSR velocity scale.
We estimate this additional uncertainty to be $\pm 10$\kms
and have added it in quadrature for all absorptions.
The strong absorption systems have been studied by \citet{Bowen02} and \citet{Impey99},
who discuss potential galaxy associations.
The continuum bumps near 1236\Ang and 1270\Ang are intrinsic 
\NI{1134.9} and \CIII{1175.7} emission, respectively.
A STIS+E140M spectrum of \lya and a FUSE spectrum of higher Lyman lines 
\citep{pg1211} confirm the pair of lines near 1268.5\Ang as \lyano, and not 
\SiIII{1206.5}, related to the \lya pair at 1278\ang. 
 We identify the absorber at 1284.2\Ang as
\SiIII{1206.5} associated with the \lya absorber at 1294\ang.
\paragraph{PG~1351+640}
The spectrum of this \z=0.0882 Seyfert~1 contains evidence for IVC
and HVC gas at \about $-50$ and $-150$\kmsno. The \about $-150$\kms HVC has previously been
reported by \citet{Lockman}. The \about $-50$\kms IVC causes
additional uncertainty of the LSR velocity scale.
We estimate this uncertainty to be $\pm 10$\kms
and have added it in quadrature to the statistical uncertainty
for all absorptions.
FUSE data indicate \ion{O}{6} HVC gas at $-160$ to $-100$\kms
and 100 to 160\kms \citep{Sembach03}.
The continuum bumps near 1228\Ang and 1280\Ang are due to
intrinsic \NI{1134.9} and \CIII{1175.7}, respectively.
This sightline includes the northwest portion of the Bootes void at \cz=14,500--17,000\kms
\citep{Kirshner87}; we detect no \lya absorbers within this void.
\paragraph{PKS~2005-489}
The spectrum of this \z=0.071 BL~Lac object contains strong
IVC absorption at 60--90\kms in \ion{N}{1}, \ion{N}{5}, \ion{Si}{2}, and \ion{S}{2}.
These IVCs introduce error in the LSR wavelength scale, resulting in higher than average velocity errors.
FUSE data indicate \ion{O}{6} HVC gas at 120--225\kms \citep{Sembach03}.
There is an indication of Galactic and IVC \CI{1277.2}
absorption, but these are not supported by \CI{1280.2}.
The continuum bumps near 1228\Ang and 1260\Ang are intrinsic \NI{1134.9} and \CIII{1175.7} emission.
These weak emissions are the first broad lines detected in this BL~Lac object 
(\citet{Falomo87} detected narrow H$\alpha$ and [\ion{N}{2}] optically).
The Galactic \ion{N}{5} absorption in this sightline, which is $\sim 30^{\circ}$ above the
Galactic center, is the second strongest (after 3C~273) \ion{N}{5} observed along any
of our sightlines. Higher resolution studies of this sightline are on-going in HST Cycle~12 to probe
the outflow associated with the nuclear activity in our own Galaxy.
\clearpage
\paragraph{TON~S180}
The spectrum of this \z=0.06198 Seyfert~1.2 shows HVC gas in \SiIII{1206.5} and 
\SFsixty\ at \about $-150$\kmsno.
FUSE data indicate \ion{O}{6} HVC gas at $-265$ to $-100$\kms and 
220 to 280\kms \citep{Sembach03}. 
The continuum bump near 1248\Ang is due to intrinsic \CIII{1175.7}.
\clearpage
\paragraph{TON~1542}
The spectrum of this \z=0.06301 Seyfert~1 galaxy, also known as 
MRK~771 and PG~1229+204, does not show any significant HVC absorbers.
The 1895\kms \lya absorber appears to be associated with the galaxy 
CGCG~129-006, which lies 20.0\arcmin\ (160\hsfi\ kpc) off the sightline at 1921\kmsno.
The 1226.0\Ang (2536\kmsno) \lya absorber appears to be associated with
the galaxy NGC4529, which lies 11.4\arcmin\ (120\hsfi\ kpc)
off the line of sight at 2536\kmsno.
It is unclear whether the absorption feature at 1240\Ang
is Galactic \MGdoublet\ or \lyano. To be conservative, we have 
identified this absorber as Galactic \ion{Mg}{2}.
The small spectral continuum bump near 1248\Ang is due to intrinsic \CIII{1175.7}.
\clearpage
\paragraph{VII~ZW~118}\label{sec:VIIZW118}
The spectrum of this \z=0.07966 Seyfert~1 galaxy shows some
evidence of IVC gas in \SiIII{1206.5} at \about $-70$\kmsno,
which is consistent with \citet{Lockman}.
This target was observed near the beginning of our program, when the STIS
calibration was not as well defined as for later observations. The continuum
fit longward of 1230\Ang has larger uncertainty than for other objects, 
owing to the absence of the longer wavelength setting.
The continuum bump near 1230\Ang is intrinsic \NI{1134.9}.
The 2426 and 2469\kms absorptions are within a galaxy void.
No galaxies closer to these absorbers than 3\hsfi\ Mpc have been found, 
despite considerable efforts to do so \citep{McLin02}.
\begin{deluxetable}{lcccrcc}
\tableheadfrac{0.02}
\tablecolumns{7}
\tablecaption{HST/STIS Absorption Features\label{linelist_all}}
\tablewidth{0pt}
\tabletypesize{\footnotesize}
\tablehead{
\colhead{Wavelength} &
\colhead{Velocity} &
\colhead{\bb} &
\colhead{\ensuremath{\mathcal{W}}} &
\colhead{SL} &
\colhead{Id} &
\colhead{Alt Id}\\
\colhead{(\AA)} &
\colhead{({\rm km~s}\ensuremath{^{-1}})} &
\colhead{({\rm km~s}\ensuremath{^{-1}})} &
\colhead{(m\AA)} &
\colhead{($\sigma$)} &
\colhead{ } &\colhead{ }
}
\startdata
\cutinhead{HE~1029-140}
1199.556 $\pm$  0.036&  2 $\pm$  9&  64 $\pm$  4& 437 $\pm$  51& 36.1&g:NI1199.5&\nodata \\
1200.190 $\pm$  0.037&  -7 $\pm$  9&  47 $\pm$  6& 332 $\pm$  67& 27.6&g:NI1200.2&\nodata \\
1200.686 $\pm$  0.038&  -6 $\pm$  9&  56 $\pm$  6& 374 $\pm$  67& 31.2&g:NI1200.7&\nodata \\
1206.442 $\pm$  0.041&  -14 $\pm$  10&  71 $\pm$  5& 470 $\pm$  67& 39.4&g:SiIII1206.5&\nodata \\
1206.927 $\pm$  0.039&  106 $\pm$  10&  78 $\pm$  9& 478 $\pm$ 103& 39.6&h:SiIII1206.5&\nodata \\
1207.332 $\pm$  0.041&  207 $\pm$  10&  $\lt 20$ &  62 $\pm$  46&  5.1&h:SiIII1206.5&\nodata \\
1223.664 $\pm$  0.044&  1971 $\pm$  11&  53 $\pm$  10& 110 $\pm$  39& 10.1&Ly$\alpha$&\nodata \\
1224.601 $\pm$  0.041&  2202 $\pm$  10&  35 $\pm$  17&  45 $\pm$  31&  4.5&Ly$\alpha$&\nodata \\
1225.496 $\pm$  0.037&  2423 $\pm$  9&  42 $\pm$  5& 183 $\pm$  32& 17.9&Ly$\alpha$&\nodata \\
1234.013 $\pm$  0.060&  4523 $\pm$  15&  52 $\pm$  19&  59 $\pm$  37&  6.4&Ly$\alpha$&\nodata \\
1239.073 $\pm$  0.050&  66 $\pm$  12&  75 $\pm$  15& 109 $\pm$  41& 12.6&g:NV1238.8&\nodata \\
1239.828 $\pm$  0.048&  -23 $\pm$  12&  44 $\pm$  13&  61 $\pm$  29&  7.0&g:MgII1239.9&\nodata \\
1243.099 $\pm$  0.047&  72 $\pm$  11&  64 $\pm$  12&  88 $\pm$  30& 10.6&g:NV1242.8&\nodata \\
1250.562 $\pm$  0.038&  -5 $\pm$  9&  27 $\pm$  4& 148 $\pm$  26& 15.1&g:SII1250.6&\nodata \\
1253.795 $\pm$  0.038&  -4 $\pm$  9&  38 $\pm$  3& 202 $\pm$  27& 22.7&g:SII1253.8&\nodata \\
1259.554 $\pm$  0.042&  13 $\pm$  10&  43 $\pm$  6& 236 $\pm$  57& 24.4&g:SII1259.5&\nodata \\
1260.250 $\pm$  0.043&  179 $\pm$  10&  33 $\pm$  27& 139 $\pm$ 167& 14.6&h:SII1259.5&\nodata \\
1260.507 $\pm$  0.134&  2 $\pm$  9&  72 $\pm$  38& 438 $\pm$ 526& 46.2&g:SiII+FeII1260.5&\nodata \\
1260.892 $\pm$  0.042&  93 $\pm$  10&  70 $\pm$  10& 352 $\pm$ 108& 37.5&h:SiII+FeII1260.5&\nodata \\
1277.173 $\pm$  0.050&  -17 $\pm$  12&  27 $\pm$  15&  59 $\pm$  39&  7.4&g:CI1277.2&Ly$\alpha$ \\
1277.991 $\pm$  0.042& 15369 $\pm$  10&  44 $\pm$  26&  30 $\pm$  32&  4.2&Ly$\alpha$&h:CI1277.2 \\
1278.379 $\pm$  0.038& 15464 $\pm$  9&  42 $\pm$  3& 278 $\pm$  26& 38.0&Ly$\alpha$&\nodata \\
1292.475 $\pm$  0.049& 18941 $\pm$  12&  36 $\pm$  13&  38 $\pm$  20&  5.5&Ly$\alpha$&\nodata \\
1293.352 $\pm$  0.041& 19157 $\pm$  10&  30 $\pm$  7&  62 $\pm$  18&  9.4&Ly$\alpha$&\nodata \\
\cutinhead{II~ZW~136}
1199.545 $\pm$  0.030&  -1 $\pm$  7&  44 $\pm$  3& 302 $\pm$  37& 25.4&g:NI1199.5&\nodata \\
1200.202 $\pm$  0.030&  -5 $\pm$  8&  41 $\pm$  4& 268 $\pm$  43& 22.4&g:NI1200.2&\nodata \\
1200.687 $\pm$  0.030&  -6 $\pm$  8&  41 $\pm$  5& 274 $\pm$  49& 22.7&g:NI1200.7&\nodata \\
1205.313 $\pm$  0.035&  -295 $\pm$  9&  $\lt 20$ &  66 $\pm$  37&  5.2&h:SiIII1206.5&\nodata \\
1205.593 $\pm$  0.035&  -225 $\pm$  9&  $\lt 20$ &  63 $\pm$  39&  5.2&h:SiIII1206.5&\nodata \\
1206.533 $\pm$  0.037&  8 $\pm$  9&  65 $\pm$  10& 502 $\pm$ 157& 39.9&g:SiIII1206.5&  \\
1206.823 $\pm$  0.035&  80 $\pm$  9&  24 $\pm$  27&  46 $\pm$  55&  3.5&h:SiIII1206.5&\nodata \\
1238.783 $\pm$  0.050&  -9 $\pm$  12&  44 $\pm$  17&  47 $\pm$  29&  5.3&g:NV1238.8&h:MgII1239.9 \\
1242.683 $\pm$  0.036&  -29 $\pm$  9&  12 $\pm$  26&  31 $\pm$  26&  3.6&g:NV1242.8&\nodata \\
1249.406 $\pm$  0.030&  8320 $\pm$  7&  33 $\pm$  4& 192 $\pm$  22& 27.2&Ly$\alpha$&\nodata \\
1250.622 $\pm$  0.031&  9 $\pm$  7&  31 $\pm$  4& 150 $\pm$  23& 17.1&g:SII1250.6&\nodata \\
1253.856 $\pm$  0.031&  11 $\pm$  7&  29 $\pm$  4& 180 $\pm$  24& 20.0&g:SII1253.8&\nodata \\
1259.558 $\pm$  0.031&  9 $\pm$  7&  29 $\pm$  4& 206 $\pm$  24& 25.1&g:SII1259.5&\nodata \\
1260.467 $\pm$  0.034&  2 $\pm$  8&  62 $\pm$  5& 471 $\pm$  74& 59.4&g:SiII+FeII1260.5&\nodata \\
1260.759 $\pm$  0.036&  14 $\pm$  9&  25 $\pm$  8& 102 $\pm$  52& 13.0&g:CI1260.7&h:SiII+FeII1260.5 \\
1264.672 $\pm$  0.036& 12084 $\pm$  9&  29 $\pm$  7&  71 $\pm$  21&  9.6&Ly$\alpha$&\nodata \\
1265.522 $\pm$  0.036& 12294 $\pm$  9&  44 $\pm$  14&  52 $\pm$  27&  7.0&Ly$\alpha$&\nodata \\
1272.547 $\pm$  0.033& 14026 $\pm$  8&  27 $\pm$  6&  72 $\pm$  19& 10.4&Ly$\alpha$&\nodata \\
1277.269 $\pm$  0.036&  6 $\pm$  9&  31 $\pm$  10&  57 $\pm$  24&  8.2&g:CI1277.2&Ly$\alpha$ \\
1285.796 $\pm$  0.031& 17293 $\pm$  7&  79 $\pm$  3& 486 $\pm$  23&105.3&Ly$\alpha$&\nodata \\
1290.345 $\pm$  0.036&  -437 $\pm$  8&  25 $\pm$  9&  22 $\pm$  9&  8.3&i:Ly$\alpha$&\nodata \\
1290.589 $\pm$  0.037&  -381 $\pm$  9&  30 $\pm$  11&  22 $\pm$  12&  8.2&i:Ly$\alpha$&\nodata \\
1292.261 $\pm$  0.031&  7 $\pm$  7&  49 $\pm$  3& 268 $\pm$  11& 95.3&i:Ly$\alpha$&\nodata \\
\cutinhead{MR~2251-178}
1199.501 $\pm$  0.039&  -12 $\pm$  10&  33 $\pm$  5& 222 $\pm$  44& 14.1&g:NI1199.5&\nodata \\
1200.189 $\pm$  0.040&  -8 $\pm$  10&  32 $\pm$  7& 218 $\pm$  63&  9.6&g:NI1200.2&\nodata \\
1200.672 $\pm$  0.041&  -9 $\pm$  10&  25 $\pm$  9& 189 $\pm$  70&  6.8&g:NI1200.7&\nodata \\
1205.420 $\pm$  0.039&  -268 $\pm$  10&  26 $\pm$  6& 134 $\pm$  33& 11.5&h:SiIII1206.5&\nodata \\
1206.334 $\pm$  0.042&  -41 $\pm$  11&  38 $\pm$  5& 256 $\pm$  50& 21.6&h:SiIII1206.5&\nodata \\
1206.574 $\pm$  0.042&  18 $\pm$  11&  34 $\pm$  9& 215 $\pm$  74& 18.5&g:SiIII1206.5&\nodata \\
1206.809 $\pm$  0.042&  77 $\pm$  11&  35 $\pm$  7& 170 $\pm$  46& 14.9&h:SiIII1206.5&\nodata \\
1224.743 $\pm$  0.044&  2237 $\pm$  11&  $\lt 20$ &  39 $\pm$  34&  4.1&Ly$\alpha$&\nodata \\
1224.960 $\pm$  0.058&  2291 $\pm$  14&  26 $\pm$  20&  52 $\pm$  46&  5.6&Ly$\alpha$&\nodata \\
1227.978 $\pm$  0.052&  3035 $\pm$  13&  46 $\pm$  15&  60 $\pm$  32&  6.5&Ly$\alpha$&\nodata \\
1228.667 $\pm$  0.039&  3205 $\pm$  10&  69 $\pm$  4& 349 $\pm$  37& 38.7&Ly$\alpha$&\nodata \\
1233.381 $\pm$  0.043&  4368 $\pm$  11&  49 $\pm$  19&  40 $\pm$  28&  4.8&Ly$\alpha$&\nodata \\
1250.605 $\pm$  0.039&     5 $\pm$  9&  $\lt 20$ & 119 $\pm$  24& 15.2&g:SII1250.6&\nodata \\
1252.249 $\pm$  0.044&  9021 $\pm$  11&  32 $\pm$  13&  51 $\pm$  28&  6.6&Ly$\alpha$&\nodata \\
1253.823 $\pm$  0.039&  3 $\pm$  9&  $\lt 20$ & 136 $\pm$  21& 17.3&g:SII1253.8&\nodata \\
1255.145 $\pm$  0.040&  9735 $\pm$  10&  43 $\pm$  3& 181 $\pm$  23& 24.0&Ly$\alpha$&\nodata \\
1257.741 $\pm$  0.044& 10375 $\pm$  11&  65 $\pm$  25&  38 $\pm$  30&  5.5&Ly$\alpha$&\nodata \\
1259.391 $\pm$  0.044&  -26 $\pm$  10&  25 $\pm$  7&  76 $\pm$  24& 11.3&h:SII1259.5&\nodata \\
1259.561 $\pm$  0.044&  10 $\pm$  10&  $\lt 20$ & 124 $\pm$  22& 18.7&g:SII1259.5&\nodata \\
1260.453 $\pm$  0.040&  -11 $\pm$  9&  76 $\pm$  3& 580 $\pm$  33& 89.4&g:SiII+FeII1260.5&\nodata \\
1272.860 $\pm$  0.044& 14103 $\pm$  11&  $\lt 20$ &  20 $\pm$  10&  4.4&Ly$\alpha$&\nodata \\
1277.276 $\pm$  0.044&  18 $\pm$  10&  $\lt 20$ &  17 $\pm$  12&  4.1&g:CI1277.2&\nodata \\
1278.801 $\pm$  0.042& 15569 $\pm$  10&  $\lt 20$ &  17 $\pm$  8&  4.2&Ly$\alpha$&\nodata \\
1280.476 $\pm$  0.043& 15982 $\pm$  10&  26 $\pm$  7&  30 $\pm$  9&  8.2&Ly$\alpha$&i:SiIII1206.5 \\
1282.417 $\pm$  0.044& 16460 $\pm$  11&  30 $\pm$  17&  15 $\pm$  12&  4.5&i:SiIII1206.5&Ly$\alpha$\\
1282.732 $\pm$  0.044& 16538 $\pm$  11&  $\lt 20$ &  20 $\pm$  11&  6.1&i:SiIII1206.5&Ly$\alpha$\\
1290.401 $\pm$  0.058&  -824 $\pm$  14&  97 $\pm$  3& 139 $\pm$  18& 23.3&i:Ly$\alpha$&\nodata \\
1290.934 $\pm$  0.045&  -701 $\pm$  10&  59 $\pm$  6& 287 $\pm$  61& 42.1&i:Ly$\alpha$&\nodata \\
1291.442 $\pm$  0.045&  -583 $\pm$  10&  48 $\pm$  9& 136 $\pm$  76& 20.2&i:Ly$\alpha$&\nodata \\
1291.842 $\pm$  0.045&  -490 $\pm$  10&  71 $\pm$  22& 394 $\pm$ 215& 58.6&i:Ly$\alpha$&\nodata \\
1292.085 $\pm$  0.292&  -434 $\pm$  68&  30 $\pm$  44&  41 $\pm$  50&  6.2&i:Ly$\alpha$&\nodata \\
1292.283 $\pm$  0.817&  -388 $\pm$  189&  53 $\pm$  57& 187 $\pm$ 224& 27.9&i:Ly$\alpha$&\nodata \\
1292.622 $\pm$  0.045&  -310 $\pm$  10&  78 $\pm$  77& 386 $\pm$ 463& 57.8&i:Ly$\alpha$&\nodata \\
1292.942 $\pm$  0.045&  -236 $\pm$  10&  45 $\pm$  52&  30 $\pm$  37&  4.6&i:Ly$\alpha$&\nodata \\
1293.240 $\pm$  0.430&  -167 $\pm$  100&  86 $\pm$  72& 298 $\pm$ 358& 45.3&i:Ly$\alpha$&\nodata \\
1293.822 $\pm$  0.045&  -32 $\pm$  10&  75 $\pm$  41&  52 $\pm$  62&  8.0&i:Ly$\alpha$&\nodata \\
1295.407 $\pm$  0.041&  335 $\pm$  10&  48 $\pm$  5&  57 $\pm$  11& 26.0&i:Ly$\alpha$&\nodata \\
\cutinhead{MRK~478}
1199.121 $\pm$  0.042&  -95 $\pm$  11&  97 $\pm$  3&  49 $\pm$  19&  5.2&h:NI1199.5&\nodata \\
1199.457 $\pm$  0.038&  -23 $\pm$  10&  36 $\pm$  3& 223 $\pm$  31& 24.0&g:NI1199.5&\nodata \\
1200.133 $\pm$  0.038&  -22 $\pm$  10&  39 $\pm$  3& 229 $\pm$  30& 24.9&g:NI1200.2&\nodata \\
1200.637 $\pm$  0.038&  -18 $\pm$  10&  39 $\pm$  4& 202 $\pm$  29& 22.3&g:NI1200.7&\nodata \\
1206.149 $\pm$  0.042&  -87 $\pm$  11&  31 $\pm$  6& 170 $\pm$  47& 20.0&h:SiIII1206.5&\nodata \\
1206.400 $\pm$  0.039&  -25 $\pm$  10&  42 $\pm$  4& 290 $\pm$  49& 33.6&g:SiIII1206.5&\nodata \\
1222.085 $\pm$  0.040&  1582 $\pm$  10&  45 $\pm$  4& 194 $\pm$  31& 22.5&Ly$\alpha$&\nodata \\
1238.547 $\pm$  0.045&  -61 $\pm$  11&  36 $\pm$  10&  55 $\pm$  23&  7.5&h:NV1238.8&\nodata \\
1242.528 $\pm$  0.076&  -66 $\pm$  18&  52 $\pm$  25&  30 $\pm$  26&  3.9&h:NV1242.8&\nodata \\
1250.592 $\pm$  0.041&  2 $\pm$  10&  $\lt 20$ &  66 $\pm$  21&  8.6&g:SII1250.6&\nodata \\
1251.305 $\pm$  0.040&  8788 $\pm$  10&  58 $\pm$  3& 290 $\pm$  30& 39.3&Ly$\alpha$&\nodata \\
1253.624 $\pm$  0.044&  -66 $\pm$  10&  $\lt 20$ &  24 $\pm$  21&  3.1&h:SII1253.8&\nodata \\
1253.809 $\pm$  0.040&  -1 $\pm$  10&  $\lt 20$ & 103 $\pm$  22& 12.8&g:SII1253.8&\nodata \\
1259.229 $\pm$  0.044&  -69 $\pm$  10&  $\lt 20$ &  25 $\pm$  22&  3.3&h:SII1259.5&\nodata \\
1259.509 $\pm$  0.040&  -2 $\pm$  10&  27 $\pm$  4& 158 $\pm$  23& 20.7&g:SII1259.5&\nodata \\
1260.169 $\pm$  0.044&  -79 $\pm$  10&  38 $\pm$  3& 216 $\pm$  33& 28.3&h:SiII+FeII1260.5&\nodata \\
1260.435 $\pm$  0.040&  -15 $\pm$  10&  41 $\pm$  3& 293 $\pm$  35& 38.2&g:SiII+FeII1260.5&\nodata \\
1295.119 $\pm$  0.042& 19593 $\pm$  10&  43 $\pm$  6&  84 $\pm$  18& 16.2&Ly$\alpha$&\nodata \\
\cutinhead{MRK~926}
1199.457 $\pm$  0.051&  -11 $\pm$  12&  66 $\pm$  15& 382 $\pm$ 161&  8.8&g:NI1199.5&\nodata \\
1200.154 $\pm$  0.046&  -11 $\pm$  11&  36 $\pm$  13& 261 $\pm$ 137&  5.6&g:NI1200.2&\nodata \\
1200.655 $\pm$  0.040&  -11 $\pm$  10&  97 $\pm$  3& 342 $\pm$  77&  7.0&g:NI1200.7&\nodata \\
1206.456 $\pm$  0.046&  -11 $\pm$  11&  80 $\pm$  12& 642 $\pm$ 190& 13.9&g:SiIII1206.5&\nodata \\
1245.050 $\pm$  0.051&  7245 $\pm$  12&  66 $\pm$  14& 179 $\pm$  75&  9.7&Ly$\alpha$&\nodata \\
1246.068 $\pm$  0.060&  7496 $\pm$  15&  45 $\pm$  20&  72 $\pm$  53&  4.4&Ly$\alpha$&\nodata \\
1250.586 $\pm$  0.044&  1 $\pm$  10&  31 $\pm$  11& 109 $\pm$  51&  5.8&g:SII1250.6&\nodata \\
1253.797 $\pm$  0.043&  -3 $\pm$  10&  50 $\pm$  9& 175 $\pm$  55& 10.3&g:SII1253.8&\nodata \\
1255.108 $\pm$  0.048&  9726 $\pm$  12&  33 $\pm$  14&  77 $\pm$  44&  4.8&Ly$\alpha$&\nodata \\
1256.279 $\pm$  0.095& 10015 $\pm$  23&  73 $\pm$  34&  66 $\pm$  61&  4.4&Ly$\alpha$&\nodata \\
1259.498 $\pm$  0.038&  -5 $\pm$  9&  59 $\pm$  5& 295 $\pm$  46& 24.1&g:SII1259.5&\nodata \\
1260.156 $\pm$  0.041&  -82 $\pm$  10&  32 $\pm$  13&  83 $\pm$  66&  6.9&h:SiII+FeII1260.5&Ly$\alpha$ \\
1260.481 $\pm$  0.041&  -5 $\pm$  9&  66 $\pm$  6& 529 $\pm$  96& 44.3&g:SiII+FeII1260.5&\nodata \\
1263.193 $\pm$  0.038& 11720 $\pm$  9&  22 $\pm$  5& 117 $\pm$  25& 11.4&Ly$\alpha$&i:SiIII1206.5\\
1267.312 $\pm$  0.064& -1384 $\pm$  15&  97 $\pm$  3& 122 $\pm$  16& 13.8&i:Ly$\alpha$&\nodata \\
1268.163 $\pm$  0.078& -1184 $\pm$  18&  97 $\pm$  3& 128 $\pm$  16& 14.6&i:Ly$\alpha$&\nodata \\
1268.993 $\pm$  0.058&  -988 $\pm$  14&  87 $\pm$  21& 131 $\pm$  62& 15.2&i:Ly$\alpha$&\nodata \\
1269.950 $\pm$  0.051&  -763 $\pm$  12&  82 $\pm$  11& 449 $\pm$ 136& 54.4&i:Ly$\alpha$&\nodata \\
1270.483 $\pm$  0.116&  -638 $\pm$  27&  62 $\pm$  26&  90 $\pm$  99& 11.4&i:Ly$\alpha$&\nodata \\
1272.722 $\pm$  0.037&  -110 $\pm$  9&  83 $\pm$  3& 596 $\pm$  32&103.7&i:Ly$\alpha$&\nodata \\
1273.507 $\pm$  0.037&  75 $\pm$  9&  60 $\pm$  3& 453 $\pm$  29& 82.0&i:Ly$\alpha$&\nodata \\
\cutinhead{MRK~1383}
1196.770 $\pm$  0.035&  -108 $\pm$  9&  52 $\pm$  34&  37 $\pm$  40&  3.5&h:MnII1197.2&\nodata \\
1199.541 $\pm$  0.030&  -2 $\pm$  8&  40 $\pm$  4& 253 $\pm$  35& 26.0&g:NI1199.5&\nodata \\
1200.211 $\pm$  0.030&  -2 $\pm$  8&  35 $\pm$  3& 231 $\pm$  31& 24.3&g:NI1200.2&\nodata \\
1200.684 $\pm$  0.030&  -6 $\pm$  8&  40 $\pm$  4& 233 $\pm$  34& 24.9&g:NI1200.7&\nodata \\
1206.446 $\pm$  0.031&  -14 $\pm$  8&  69 $\pm$  4& 509 $\pm$  61& 50.1&g:SiIII1206.5&\nodata \\
1206.827 $\pm$  0.036&  81 $\pm$  9&  70 $\pm$  46&  42 $\pm$  50&  4.2&h:SiIII1206.5&\nodata \\
1238.615 $\pm$  0.036&  -45 $\pm$  9&  47 $\pm$  12&  74 $\pm$  32&  9.3&h:NV1238.8&\nodata \\
1238.935 $\pm$  0.036&  33 $\pm$  9&  46 $\pm$  13&  66 $\pm$  30&  8.4&g:NV1238.8&\nodata \\
1240.210 $\pm$  0.128&  75 $\pm$  31&  53 $\pm$  51&  24 $\pm$  29&  3.1&g:MgII1239.9&\nodata \\
1242.645 $\pm$  0.036&  -37 $\pm$  9&  30 $\pm$  15&  40 $\pm$  25&  5.1&h:NV1242.8&\nodata \\
1242.885 $\pm$  0.036&  21 $\pm$  9&  $\lt 20$ &  25 $\pm$  17&  3.1&g:NV1242.8&\nodata \\
1244.438 $\pm$  0.035&  7094 $\pm$  8&  26 $\pm$  8&  54 $\pm$  19&  7.3&Ly$\alpha$&\nodata \\
1245.510 $\pm$  0.037&  7359 $\pm$  9&  31 $\pm$  10&  35 $\pm$  15&  6.8&Ly$\alpha$&\nodata \\
1250.096 $\pm$  0.033&  8490 $\pm$  8&  65 $\pm$  5& 218 $\pm$  30& 30.7&Ly$\alpha$&\nodata \\
1250.614 $\pm$  0.031&  7 $\pm$  8&  $\lt 20$ & 102 $\pm$  19& 14.1&g:SII1250.6&\nodata \\
1251.968 $\pm$  0.039&  8951 $\pm$  10&  49 $\pm$  10&  66 $\pm$  22&  9.5&Ly$\alpha$&\nodata \\
1253.585 $\pm$  0.036&  -51 $\pm$  9&  30 $\pm$  20&  28 $\pm$  25&  3.8&h:SII1253.8&\nodata \\
1253.845 $\pm$  0.032&  8 $\pm$  8&  28 $\pm$  4& 141 $\pm$  23& 19.5&g:SII1253.8&\nodata \\
1257.200 $\pm$  0.050& 10242 $\pm$  12&  53 $\pm$  15&  54 $\pm$  27&  7.7&Ly$\alpha$&\nodata \\
1259.528 $\pm$  0.037&  2 $\pm$  9&  29 $\pm$  4& 181 $\pm$  22& 24.3&g:SII1259.5&\nodata \\
1260.208 $\pm$  0.037&  -93 $\pm$  9&  34 $\pm$  5& 121 $\pm$  42& 16.2&h:SiII+FeII1260.6&\nodata \\
1260.478 $\pm$  0.034&  4 $\pm$  8&  57 $\pm$  4& 421 $\pm$  63& 56.5&g:SiII+FeII1260.5&\nodata \\
1261.278 $\pm$  0.037& 11247 $\pm$  9&  42 $\pm$  10&  61 $\pm$  24&  8.5&Ly$\alpha$&\nodata \\
1278.793 $\pm$  0.031& 15566 $\pm$  8&  46 $\pm$  3& 282 $\pm$  20& 45.9&Ly$\alpha$&\nodata \\
1280.077 $\pm$  0.072& 15883 $\pm$  18&  27 $\pm$  34&  26 $\pm$  31&  4.2&Ly$\alpha$&g:CI1280.1 \\
1283.147 $\pm$  0.061& 16640 $\pm$  15&  44 $\pm$  20&  25 $\pm$  20&  4.1&Ly$\alpha$&\nodata \\
\cutinhead{NGC~985}
1199.482 $\pm$  0.040&  -17 $\pm$  10&  46 $\pm$  4& 277 $\pm$  44& 19.1&g:NI1199.5&\nodata \\
1200.189 $\pm$  0.040&  -8 $\pm$  10&  33 $\pm$  5& 218 $\pm$  41& 15.0&g:NI1200.2&\nodata \\
1200.668 $\pm$  0.040&  -10 $\pm$  10&  30 $\pm$  5& 198 $\pm$  40& 13.6&g:NI1200.7&\nodata \\
1205.551 $\pm$  0.043&  -236 $\pm$  11&  24 $\pm$  21&  56 $\pm$  52&  4.0&h:SiIII1206.5&\nodata \\
1206.344 $\pm$  0.040&  -39 $\pm$  10&  61 $\pm$  4& 452 $\pm$  52& 33.2&g:SiIII1206.5&\nodata \\
1206.982 $\pm$  0.043&  120 $\pm$  11&  97 $\pm$  3&  45 $\pm$  23&  3.3&h:SiIII1206.5&\nodata \\
1224.411 $\pm$  0.078&  2156 $\pm$  19&  61 $\pm$  27&  51 $\pm$  42&  4.5&Ly$\alpha$&\nodata \\
1238.805 $\pm$  0.053&  1 $\pm$  10&  28 $\pm$  17&  36 $\pm$  27&  4.0&g:NV1238.8&\nodata \\
1250.596 $\pm$  0.041&  3 $\pm$  10&  $\lt 20$ & 100 $\pm$  17& 14.6&g:SII1250.6&\nodata \\
1253.324 $\pm$  0.054&  -114 $\pm$  13&  45 $\pm$  16&  38 $\pm$  22&  6.0&h:SII1253.8&Ly$\alpha$ \\
1253.817 $\pm$  0.041&  1 $\pm$  10&  24 $\pm$  4& 120 $\pm$  17& 19.0&g:SII1253.8&\nodata \\
1259.522 $\pm$  0.041&  1 $\pm$  10&  28 $\pm$  4& 154 $\pm$  16& 28.4&g:SII1259.5&\nodata \\
1259.888 $\pm$  0.045&  -145 $\pm$  11&  $\lt 20$ &  23 $\pm$  13&  4.2&h:SiII+FeII1260.5&h:SII1259.5 \\
1260.398 $\pm$  0.040&  -15 $\pm$  10&  55 $\pm$  3& 426 $\pm$  22& 78.9&g:SiII+FeII1260.5&\nodata \\
1261.020 $\pm$  0.045&  76 $\pm$  11&  38 $\pm$  21&  26 $\pm$  21&  4.8&g:CI1260.7&h:SiII+FeII1260.5 \\
1264.785 $\pm$  0.045&  -788 $\pm$  11&  80 $\pm$  13&  79 $\pm$  28& 16.8&i:Ly$\alpha$&\nodata \\
1265.315 $\pm$  0.045&  -663 $\pm$  11&  63 $\pm$  4& 248 $\pm$  28& 54.9&i:Ly$\alpha$&\nodata \\
1265.813 $\pm$  0.045&  -545 $\pm$  11&  54 $\pm$  6& 183 $\pm$  62& 42.3&i:Ly$\alpha$&\nodata \\
1266.271 $\pm$  0.048&  -437 $\pm$  11&  90 $\pm$  6& 492 $\pm$  82&118.3&i:Ly$\alpha$&\nodata \\
1267.045 $\pm$  0.045&  -254 $\pm$  11&  94 $\pm$  3& 431 $\pm$  30&108.9&i:Ly$\alpha$&\nodata \\
1268.576 $\pm$  0.041&  108 $\pm$  10&  38 $\pm$  3& 232 $\pm$  13& 59.1&i:Ly$\alpha$&\nodata \\
1277.242 $\pm$  0.046&  10 $\pm$  11&  $\lt 20$ &  17 $\pm$  10&  3.7&g:CI1277.2&\nodata \\
1289.358 $\pm$  0.043&  -670 $\pm$  10&  23 $\pm$  6&  63 $\pm$  17&  9.8&i:NV1238.8&\nodata \\
1290.059 $\pm$  0.046&  -507 $\pm$  11&  84 $\pm$  14& 122 $\pm$  41& 19.2&i:NV1238.8&\nodata \\
1290.518 $\pm$  0.042&  -401 $\pm$  10&  35 $\pm$  3& 162 $\pm$  22& 25.6&i:NV1238.8&\nodata \\
1291.067 $\pm$  0.042&  -273 $\pm$  10&  38 $\pm$  3& 150 $\pm$  19& 23.9&i:NV1238.8&\nodata \\
1292.739 $\pm$  0.042&  114 $\pm$  10&  $\lt 20$ & 100 $\pm$  15& 15.5&i:NV1238.8&\nodata \\
1293.502 $\pm$  0.047&  -674 $\pm$  11&  43 $\pm$  9&  69 $\pm$  24& 10.9&i:NV1242.8&\nodata \\
1294.109 $\pm$  0.046&  -534 $\pm$  11&  56 $\pm$  15&  82 $\pm$  39& 12.8&i:NV1242.8&\nodata \\
1294.638 $\pm$  0.044&  -412 $\pm$  10&  50 $\pm$  5& 156 $\pm$  30& 24.3&i:NV1242.8&\nodata \\
1295.192 $\pm$  0.043&  -283 $\pm$  10&  35 $\pm$  5&  96 $\pm$  21& 14.7&i:NV1242.8&\nodata \\
1296.894 $\pm$  0.042&  110 $\pm$  10&  $\lt 20$ &  73 $\pm$  16& 11.0&i:NV1242.8&\nodata \\
\cutinhead{PG~0804+761}
1199.286 $\pm$  0.045&  -53 $\pm$  11&  $\lt 20$ &  42 $\pm$  49&  5.6&h:NI1199.5&\nodata \\
1199.488 $\pm$  0.044&  -15 $\pm$  11&  45 $\pm$  6& 327 $\pm$  79& 43.4&g:NI1199.5&\nodata \\
1200.137 $\pm$  0.041&  -21 $\pm$  10&  54 $\pm$  4& 375 $\pm$  43& 50.2&g:NI1200.2&\nodata \\
1200.638 $\pm$  0.041&  -18 $\pm$  10&  48 $\pm$  3& 311 $\pm$  38& 41.7&g:NI1200.7&\nodata \\
1205.870 $\pm$  0.044&  -156 $\pm$  11&  $\lt 20$ &  29 $\pm$  22&  3.8&h:SiIII1206.5&\nodata \\
1206.141 $\pm$  0.046&  -89 $\pm$  11&  39 $\pm$  7& 217 $\pm$  61& 28.1&h:SiIII1206.5&\nodata \\
1206.463 $\pm$  0.042&  -9 $\pm$  10&  52 $\pm$  3& 385 $\pm$  43& 50.2&g:SiIII1206.5&\nodata \\
1220.322 $\pm$  0.046&  1147 $\pm$  11&  38 $\pm$  4& 165 $\pm$  29& 18.9&Ly$\alpha$&\nodata \\
1221.872 $\pm$  0.048&  1530 $\pm$  12&  45 $\pm$  10&  78 $\pm$  28& 10.5&Ly$\alpha$&\nodata \\
1222.242 $\pm$  0.046&  1621 $\pm$  11&  32 $\pm$  15&  41 $\pm$  27&  5.6&Ly$\alpha$&\nodata \\
1238.182 $\pm$  0.043&  5552 $\pm$  11&  50 $\pm$  4& 324 $\pm$  44& 58.5&Ly$\alpha$&\nodata \\
1238.613 $\pm$  0.047&  5658 $\pm$  11&  41 $\pm$  51&  28 $\pm$  33&  5.0&Ly$\alpha$&h:NV1238.8 \\
1239.984 $\pm$  0.046&  20 $\pm$  11&  26 $\pm$  10&  34 $\pm$  14&  6.2&g:MgII1239.9&\nodata \\
1247.624 $\pm$  0.045&  7880 $\pm$  11&  $\lt 20$ &  18 $\pm$  9&  4.2&Ly$\alpha$&\nodata \\
1250.377 $\pm$  0.047&  -53 $\pm$  11&  $\lt 20$ &  36 $\pm$  22&  5.0&h:SII1250.6&\nodata \\
1250.625 $\pm$  0.043&  10 $\pm$  10&  29 $\pm$  4& 134 $\pm$  24& 18.8&g:SII1250.6&\nodata \\
1253.644 $\pm$  0.047&  -37 $\pm$  11&  $\lt 20$&  68 $\pm$  20&  9.9&h:SII1253.8&\nodata \\
1253.858 $\pm$  0.043&  11 $\pm$  10&  20 $\pm$  5& 154 $\pm$  20& 22.4&g:SII1253.8&\nodata \\
1259.353 $\pm$  0.047&  -35 $\pm$  11&  21 $\pm$  5& 105 $\pm$  24& 15.2&h:SII1259.5&\nodata \\
1259.566 $\pm$  0.043&  16 $\pm$  10&  22 $\pm$  4& 162 $\pm$  22& 23.3&g:SII1259.5&\nodata \\
1260.188 $\pm$  0.047&  -74 $\pm$  11&  41 $\pm$  4& 185 $\pm$  27& 26.5&h:SiII+FeII1260.5&\nodata \\
1260.468 $\pm$  0.047&  2 $\pm$  11&  53 $\pm$  3& 380 $\pm$  33& 54.5&g:SiII+FeII1260.5&\nodata \\
1277.244 $\pm$  0.047&  0 $\pm$  11&  46 $\pm$  8&  77 $\pm$  22& 11.7&g:CI1277.2&\nodata \\
1280.139 $\pm$  0.048&  1 $\pm$  11&  42 $\pm$  21&  26 $\pm$  21&  3.9&g:CI1280.1&\nodata \\
1287.025 $\pm$  0.048& 17597 $\pm$  12&  47 $\pm$  8&  72 $\pm$  21& 12.0&Ly$\alpha$&\nodata \\
1287.681 $\pm$  0.052& 17758 $\pm$  13&  36 $\pm$  11&  38 $\pm$  18&  6.3&Ly$\alpha$&\nodata \\
1289.983 $\pm$  0.046& 18326 $\pm$  11&  $\lt 20$ &  37 $\pm$  14&  6.1&Ly$\alpha$&\nodata \\
1292.381 $\pm$  0.048& 18918 $\pm$  12&  32 $\pm$  11&  34 $\pm$  17&  5.8&Ly$\alpha$&\nodata \\
\cutinhead{PG~1116+215}
1196.194 $\pm$  0.028& 49824 $\pm$  6&  34 $\pm$  6& 149 $\pm$  35& 15.1&z:Ly$\beta$&\nodata \\
1197.026 $\pm$  0.059& 50068 $\pm$  16&  41 $\pm$  21&  46 $\pm$  39&  4.4&z:Ly$\beta$&\nodata \\
1199.531 $\pm$  0.027&  -5 $\pm$  7&  45 $\pm$  4& 293 $\pm$  39& 25.1&g:NI1199.5&\nodata \\
1200.201 $\pm$  0.027&  -5 $\pm$  7&  38 $\pm$  5& 277 $\pm$  57& 24.4&g:NI1200.2&\nodata \\
1200.676 $\pm$  0.028&  -8 $\pm$  7&  47 $\pm$  4& 296 $\pm$  46& 26.6&g:NI1200.7&\nodata \\
1202.781 $\pm$  0.030& 49636 $\pm$  6&  40 $\pm$  7& 109 $\pm$  28& 12.1&z:OVI1031.9&\nodata \\
1203.373 $\pm$  0.033& 49817 $\pm$  7&  45 $\pm$  28&  39 $\pm$  39&  3.7&z:OV1031.9&\nodata \\
1203.882 $\pm$  0.030& 52072 $\pm$  6&  38 $\pm$  7& 106 $\pm$  29& 11.8&z:Ly$\beta$&\nodata \\
1206.394 $\pm$  0.033&  -26 $\pm$  8&  22 $\pm$  28&  53 $\pm$  63&  5.4&h:SiIII1206.5&z:OVI1031.9 \\
1206.601 $\pm$  0.149&  25 $\pm$  26&  87 $\pm$  25& 578 $\pm$ 327& 58.5&g:SiIII1206.5&\nodata \\
1206.894 $\pm$  0.033&  98 $\pm$  8&  25 $\pm$  41&  63 $\pm$  76&  6.4&h:SiIII1206.5&\nodata \\
1207.308 $\pm$  0.035&  201 $\pm$  9&  41 $\pm$  6& 227 $\pm$  75& 22.8&h:SiIII1206.5&\nodata \\
1209.403 $\pm$  0.034& 49633 $\pm$  8&  46 $\pm$  10&  76 $\pm$  27&  9.1&z:OVI1037.6&\nodata \\
1221.749 $\pm$  0.038&  1499 $\pm$  9&  52 $\pm$  12&  82 $\pm$  33&  8.6&Ly$\alpha$&\nodata \\
1235.594 $\pm$  0.032&  4913 $\pm$  8&  30 $\pm$  9&  90 $\pm$  32&  9.4&Ly$\alpha$&\nodata \\
1250.212 $\pm$  0.033&  8518 $\pm$  8&  45 $\pm$  4& 227 $\pm$  32& 27.2&Ly$\alpha$&\nodata \\
1250.575 $\pm$  0.033&  -6 $\pm$  8&  $\lt 20$ &  80 $\pm$  30&  9.1&g:SII1250.6&\nodata \\
1250.724 $\pm$  0.033&  30 $\pm$  8&  $\lt 20$ &  27 $\pm$  21&  3.0&h:SII1250.6&\nodata \\
1253.799 $\pm$  0.033&  0 $\pm$  7&  22 $\pm$  5& 115 $\pm$  27& 14.1&g:SII1253.8&\nodata \\
1253.949 $\pm$  0.033&  36 $\pm$  8&  42 $\pm$  11&  68 $\pm$  33&  8.2&h:SII1253.8&\nodata \\
1254.799 $\pm$  0.033&  239 $\pm$  8&  58 $\pm$  24&  48 $\pm$  40&  6.0&h:SII1253.8&\nodata \\
1254.989 $\pm$  0.033&  9696 $\pm$  8&  25 $\pm$  11&  68 $\pm$  34&  8.8&Ly$\alpha$&\nodata \\
1259.525 $\pm$  0.028&  6 $\pm$  7&  25 $\pm$  4& 180 $\pm$  30& 21.3&g:SII1259.5&\nodata \\
1260.415 $\pm$  0.034&  -20 $\pm$  8&  47 $\pm$  3& 378 $\pm$  36& 44.4&g:SiII+FeII1260.5&\nodata \\
1260.730 $\pm$  0.049&  55 $\pm$  12&  36 $\pm$  11& 224 $\pm$ 257& 26.2&h:SiII+FeII1260.5&g:CII1260.7 \\
1261.085 $\pm$  0.034&  139 $\pm$  8&  64 $\pm$  66&  65 $\pm$  78&  7.6&h:SiII+FeII1260.5&Ly$\alpha$ \\
1261.355 $\pm$  0.034&  203 $\pm$  8&  $\lt 20$ & 144 $\pm$ 173& 16.9&h:SiII+FeII1260.5&\nodata \\
1261.585 $\pm$  0.034&  258 $\pm$  8&  $\lt 20$ &  43 $\pm$  51&  5.0&h:SiII+FeII1260.5&Ly$\alpha$ \\
1265.780 $\pm$  0.122& 12357 $\pm$  30&  89 $\pm$  39&  88 $\pm$  82& 10.6&Ly$\alpha$&\nodata \\
1266.474 $\pm$  0.206& 12529 $\pm$  51&  80 $\pm$  55&  44 $\pm$  53&  5.4&Ly$\alpha$&\nodata \\
1269.606 $\pm$  0.065& 13301 $\pm$  16&  83 $\pm$  22&  65 $\pm$  34&  8.3&Ly$\alpha$&\nodata \\
1287.438 $\pm$  0.031& 17698 $\pm$  7&  51 $\pm$  6& 167 $\pm$  31& 20.0&Ly$\alpha$&\nodata \\
1289.583 $\pm$  0.036& 18227 $\pm$  9&  38 $\pm$  9&  77 $\pm$  28&  8.9&Ly$\alpha$&\nodata \\
1291.754 $\pm$  0.067& 18763 $\pm$  16&  57 $\pm$  23&  42 $\pm$  31&  4.8&Ly$\alpha$&\nodata \\
\cutinhead{PG~1211+143}
1199.499 $\pm$  0.052&  -13 $\pm$  16&  43 $\pm$  4& 263 $\pm$  37& 22.3&g:NI1199.5&\nodata \\
1200.160 $\pm$  0.053&  -15 $\pm$  17&  46 $\pm$  5& 252 $\pm$  46& 21.7&g:NI1200.2&\nodata \\
1200.646 $\pm$  0.053&  -16 $\pm$  17&  41 $\pm$  5& 240 $\pm$  43& 20.9&g:NI1200.7&\nodata \\
1206.422 $\pm$  0.055&  -19 $\pm$  17&  50 $\pm$  6& 241 $\pm$  58& 21.5&h:SiIII1206.5&\nodata \\
1206.612 $\pm$  0.055&  28 $\pm$  17&  65 $\pm$  9& 239 $\pm$  75& 21.4&g:SiIII1206.5&\nodata \\
1207.212 $\pm$  0.055&  177 $\pm$  17&  $\lt 20$ &  45 $\pm$  28&  4.1&h:SiIII1206.5&\nodata \\
1207.627 $\pm$  0.055&  280 $\pm$  17&  25 $\pm$  18&  40 $\pm$  31&  3.6&h:SiIII1206.5&\nodata \\
1224.305 $\pm$  0.060&  2130 $\pm$  18&  97 $\pm$  3& 186 $\pm$  19& 19.1&Ly$\alpha$&\nodata \\
1235.718 $\pm$  0.057&  4944 $\pm$  17&  58 $\pm$  7& 189 $\pm$  46& 21.1&Ly$\alpha$&\nodata \\
1236.093 $\pm$  0.057&  5036 $\pm$  17&  34 $\pm$  6& 154 $\pm$  40& 17.6&Ly$\alpha$&\nodata \\
1239.146 $\pm$  0.057&  84 $\pm$  17&  30 $\pm$  14&  42 $\pm$  25&  4.8&h:NV1238.8&Ly$\alpha$ \\
1239.886 $\pm$  0.057&  -3 $\pm$  16&  37 $\pm$  16&  52 $\pm$  33&  5.9&g:MgII1239.9&\nodata \\
1242.492 $\pm$  0.062&  6615 $\pm$  18&  32 $\pm$  13&  89 $\pm$  54& 10.9&Ly$\alpha$&\nodata \\
1242.826 $\pm$  0.057&  5 $\pm$  17&  49 $\pm$  56&  28 $\pm$  33&  3.3&g:NV1242.8&\nodata \\
1244.064 $\pm$  0.055&  7002 $\pm$  17&  54 $\pm$  6& 150 $\pm$  30& 17.7&Ly$\alpha$&\nodata \\
1247.106 $\pm$  0.056&  7752 $\pm$  17&  75 $\pm$  6& 159 $\pm$  26& 25.7&Ly$\alpha$&\nodata \\
1250.578 $\pm$  0.062&  -5 $\pm$  16&  28 $\pm$  12& 119 $\pm$  73& 13.3&g:SII1250.6&\nodata \\
1253.714 $\pm$  0.057&  -21 $\pm$  17&  $\lt 20$ &  59 $\pm$  39&  6.5&h:SII1253.8&\nodata \\
1253.867 $\pm$  0.058&  13 $\pm$  17&  23 $\pm$  9& 124 $\pm$  54& 14.1&g:SII1253.8&\nodata \\
1259.444 $\pm$  0.058&  -13 $\pm$  17&  $\lt 20$ & 101 $\pm$  55& 11.7&h:SII1259.5&\nodata \\
1259.593 $\pm$  0.058&  22 $\pm$  17&  $\lt 20$ &  99 $\pm$  94& 11.2&g:SII1259.5&\nodata \\
1260.349 $\pm$  0.058&  -36 $\pm$  17&  34 $\pm$  15& 204 $\pm$ 209& 23.2&h:SiII+FeII1260.5&\nodata \\
1260.589 $\pm$  0.058&  21 $\pm$  17&  35 $\pm$  4& 238 $\pm$  65& 27.5&g:SiII+FeII1260.5&\nodata \\
1261.249 $\pm$  0.058&  178 $\pm$  17&  $\lt 20$ &  31 $\pm$  19&  3.7&h:SiII+FeII1260.5&Ly$\alpha$ \\
1268.428 $\pm$  0.058& 13010 $\pm$  17&  54 $\pm$  8& 216 $\pm$  56& 29.3&Ly$\alpha$&z:SiIII1206.5\\
1268.662 $\pm$  0.057& 13068 $\pm$  17&  31 $\pm$  7&  95 $\pm$  30& 13.0&Ly$\alpha$&z:SiIII1206.5\\
1270.625 $\pm$  0.056& 13552 $\pm$  17&  $\lt 20$ &  44 $\pm$  17&  5.9&Ly$\alpha$&\nodata \\
1277.318 $\pm$  0.058&  28 $\pm$  17&  28 $\pm$  12&  51 $\pm$  30&  6.4&g:CI1277.2&\nodata \\
1277.718 $\pm$  0.056& 15301 $\pm$  17&  55 $\pm$  5& 308 $\pm$  54& 40.5&Ly$\alpha$&\nodata \\
1278.221 $\pm$  0.057& 15426 $\pm$  17&  97 $\pm$  3& 691 $\pm$  25& 90.6&Ly$\alpha$&\nodata \\
1278.948 $\pm$  0.058& 15605 $\pm$  17&  42 $\pm$  5& 132 $\pm$  28& 17.2&Ly$\alpha$&\nodata \\
1281.838 $\pm$  0.058& 16318 $\pm$  17&  54 $\pm$  28&  51 $\pm$  49&  7.2&Ly$\alpha$&\nodata \\
1284.209 $\pm$  0.059& 16902 $\pm$  17&  61 $\pm$  19&  57 $\pm$  34&  8.3&z:SiIII1206.5&Ly$\alpha$ \\
1294.047 $\pm$  0.059& 19328 $\pm$  18&  74 $\pm$  3& 564 $\pm$  31& 92.4&Ly$\alpha$&\nodata \\
1294.612 $\pm$  0.059& 19468 $\pm$  18&  56 $\pm$  3& 249 $\pm$  26& 41.7&Ly$\alpha$&\nodata \\
1295.167 $\pm$  0.059& 19604 $\pm$  18&  60 $\pm$  12&  67 $\pm$  25& 11.4&Ly$\alpha$&\nodata \\
\cutinhead{PG~1351+640}
1198.910 $\pm$  0.039&  -147 $\pm$  14&  91 $\pm$  14& 283 $\pm$  86& 18.3&h:NI1199.5&\nodata \\
1199.405 $\pm$  0.040&  -24 $\pm$  14&  66 $\pm$  9& 450 $\pm$ 119& 28.7&g:NI1199.5&\nodata \\
1200.046 $\pm$  0.049&  -38 $\pm$  16&  70 $\pm$  18& 493 $\pm$ 261& 30.9&g:NI1200.2&\nodata \\
1200.547 $\pm$  0.055&  -38 $\pm$  17&  61 $\pm$  11& 405 $\pm$ 171& 25.0&g:NI1200.7&\nodata \\
1205.903 $\pm$  0.039&  -148 $\pm$  14&  45 $\pm$  6& 292 $\pm$  62& 19.0&h:SiIII1206.5&\nodata \\
1206.153 $\pm$  0.039&  -86 $\pm$  14&  20 $\pm$  11& 101 $\pm$  50&  6.4&h:SiIII1206.5&\nodata \\
1206.454 $\pm$  0.039&  -12 $\pm$  14&  73 $\pm$  6& 545 $\pm$  88& 33.2&g:SiIII1206.5&\nodata \\
1246.677 $\pm$  0.036&  7647 $\pm$  13&   $\lt 20$  &  34 $\pm$  15&  4.3&Ly$\alpha$&\nodata \\
1250.372 $\pm$  0.039&  -55 $\pm$  14&  $\lt 20$ &  80 $\pm$  37&  6.4&h:SII1250.6&\nodata \\
1250.570 $\pm$  0.040&  -7 $\pm$  13&  43 $\pm$  13& 109 $\pm$  54&  8.5&g:SII1250.6&\nodata \\
1253.582 $\pm$  0.037&  -52 $\pm$  13&  23 $\pm$  6& 156 $\pm$  40& 12.2&h:SII1253.8&\nodata \\
1253.810 $\pm$  0.040&  2 $\pm$  13&  $\lt 20$& 113 $\pm$  38&  8.8&g:SII1253.8&\nodata \\
1259.279 $\pm$  0.040&  -53 $\pm$  14&   $\lt 20$ &  48 $\pm$  47&  3.6&h:SII1259.5&\nodata \\
1259.559 $\pm$  0.040&  14 $\pm$  14&  88 $\pm$  16& 533 $\pm$ 222& 41.1&g:SII1259.5&\nodata \\
1260.019 $\pm$  0.040&  -114 $\pm$  14&  69 $\pm$  15& 389 $\pm$ 185& 30.3&h:SiII+FeII1260.5&\nodata \\
1260.269 $\pm$  0.040&  -55 $\pm$  14&  44 $\pm$  32&  95 $\pm$ 114&  7.4&h:SiII+FeII1260.5&\nodata \\
1260.509 $\pm$  0.040&  2 $\pm$  13&  59 $\pm$  7& 353 $\pm$  84& 27.7&g:SiII+FeII1260.5&\nodata \\
1276.878 $\pm$  0.041&  -755 $\pm$  14&  54 $\pm$  24&  50 $\pm$  39&  4.6&h:CI1280.1&Ly$\alpha$ \\
1277.298 $\pm$  0.041&  23 $\pm$  14&   $\lt 20$ &  35 $\pm$  22&  3.4&g:CI1277.2&\nodata \\
1280.278 $\pm$  0.041&  42 $\pm$  14&  43 $\pm$  24&  33 $\pm$  31&  3.5&g:CI1280.1&\nodata \\
1297.287 $\pm$  0.039& 20127 $\pm$  14&  $\lt 20$ &  42 $\pm$  18&  5.1&Ly$\alpha$&\nodata \\
\cutinhead{PKS~2005-489}
1197.140 $\pm$  0.053&  -15 $\pm$  13&  55 $\pm$  14&  47 $\pm$  22&  7.8&g:MnII1197.2&\nodata \\
1198.868 $\pm$  0.053&  -158 $\pm$  13&  41 $\pm$  33&  19 $\pm$  23&  3.3&h:NI1199.5&\nodata \\
1199.451 $\pm$  0.054&  -12 $\pm$  13&  58 $\pm$  6& 375 $\pm$  85& 62.8&g:NI1199.5&\nodata \\
1199.802 $\pm$  0.056&  -99 $\pm$  14&  36 $\pm$  16& 137 $\pm$ 118& 22.9&h:NI1200.2&h:NI1199.5 \\
1200.116 $\pm$  0.053&  -26 $\pm$  13&  37 $\pm$  9& 226 $\pm$  98& 37.7&g:NI1200.2&\nodata \\
1200.572 $\pm$  0.051&  -34 $\pm$  13&  75 $\pm$  6& 458 $\pm$  77& 75.8&g:NI1200.7&\nodata \\
1201.060 $\pm$  0.053&  87 $\pm$  13&  38 $\pm$  13&  40 $\pm$  25&  6.6&h:NI1200.7&\nodata \\
1206.254 $\pm$  0.053&  -61 $\pm$  13&  43 $\pm$  3& 221 $\pm$  43& 37.1&h:SiIII1206.5&\nodata \\
1206.554 $\pm$  0.053&  13 $\pm$  13&  57 $\pm$  12& 356 $\pm$ 126& 60.0&g:SiIII1206.5&\nodata \\
1206.818 $\pm$  0.051&  79 $\pm$  13&  45 $\pm$  9& 177 $\pm$  59& 30.0&h:SiIII1206.5&\nodata \\
1207.124 $\pm$  0.053&  155 $\pm$  13&  22 $\pm$  14&  35 $\pm$  26&  6.0&h:SiIII1206.5&\nodata \\
1226.831 $\pm$  0.061&  2752 $\pm$  15&  31 $\pm$  15&  24 $\pm$  15&  5.1&Ly$\alpha$&\nodata \\
1235.730 $\pm$  0.051&  4947 $\pm$  13&  59 $\pm$  3& 299 $\pm$  26& 68.1&Ly$\alpha$&\nodata \\
1236.192 $\pm$  0.051&  5061 $\pm$  12&  40 $\pm$  3& 281 $\pm$  21& 64.1&Ly$\alpha$&\nodata \\
1238.763 $\pm$  0.054&  -9 $\pm$  12&  70 $\pm$  10& 105 $\pm$  30& 23.5&g:NV1238.8&\nodata \\
1239.073 $\pm$  0.054&  66 $\pm$  13&  46 $\pm$  8&  87 $\pm$  26& 19.5&h:NV1238.8&\nodata \\
1239.673 $\pm$  0.054&  -61 $\pm$  13&  58 $\pm$  60&  19 $\pm$  23&  4.4&h:MgII1239.9&\nodata \\
1239.863 $\pm$  0.054&  -15 $\pm$  13&  $\lt 20$ &  16 $\pm$  19&  3.6&g:MgII1239.9&\nodata \\
1240.254 $\pm$  0.055&  86 $\pm$  13&  40 $\pm$  50&  16 $\pm$  19&  3.6&h:MgII1239.9&\nodata \\
1242.717 $\pm$  0.055&  -20 $\pm$  13&  45 $\pm$  50&  33 $\pm$  40&  7.4&g:NV1242.8&\nodata \\
1243.002 $\pm$  0.056&  49 $\pm$  14&  35 $\pm$  7&  56 $\pm$  25& 12.7&h:NV1242.8&\nodata \\
1246.445 $\pm$  0.055&  7589 $\pm$  13&  47 $\pm$  23&  14 $\pm$  11&  4.5&Ly$\alpha$&\nodata \\
1250.316 $\pm$  0.055&  -68 $\pm$  13&  $\lt 20$ &  20 $\pm$  15&  4.3&h:SII1250.6&\nodata \\
1250.564 $\pm$  0.052&  -5 $\pm$  13&  $\lt 20$ & 105 $\pm$  30& 22.9&g:SII1250.6&\nodata \\
1250.821 $\pm$  0.063&  53 $\pm$  15&  36 $\pm$  11&  71 $\pm$  37& 15.8&h:SII1250.6&\nodata \\
1253.603 $\pm$  0.055& -1404 $\pm$  13&  41 $\pm$  13&  40 $\pm$  21&  8.7&h:SII1259.5&\nodata \\
1253.798 $\pm$  0.052&  -24 $\pm$  12&  20 $\pm$  4& 131 $\pm$  29& 28.6&g:SII1253.8&\nodata \\
1254.088 $\pm$  0.055& -66 $\pm$  13&  36 $\pm$  7& 100 $\pm$  31& 21.6&h:SII1259.5&\nodata \\
1259.311 $\pm$  0.055&  -45 $\pm$  13&  28 $\pm$  44&  33 $\pm$  40&  7.4&h:SII1259.5&\nodata \\
1259.512 $\pm$  0.092&  3 $\pm$  12&  28 $\pm$  11& 178 $\pm$ 200& 40.2&g:SII1259.5&\nodata \\
1259.788 $\pm$  0.053&  64 $\pm$  13&  21 $\pm$  6& 104 $\pm$  34& 23.6&h:SII1259.5&\nodata \\
1260.310 $\pm$  0.055&  -45 $\pm$  13&  56 $\pm$  5& 417 $\pm$  79& 95.9&g:SiII+FeII1260.5&\nodata \\
1260.704 $\pm$  0.054&  49 $\pm$  13&  50 $\pm$  4& 365 $\pm$  72& 84.7&h:SiII+FeII1260.5&g:CI1260.7 \\
1261.131 $\pm$  0.055&  150 $\pm$  13&  $\lt 20$ &  29 $\pm$  17&  6.7&h:SiII+FeII1260.5&h:CI1260.7 \\
1266.740 $\pm$  0.065& 12594 $\pm$  16&  44 $\pm$  16&  22 $\pm$  13&  5.3&Ly$\alpha$&\nodata \\
1273.763 $\pm$  0.054& 14326 $\pm$  13&  39 $\pm$  7&  49 $\pm$  13& 11.8&Ly$\alpha$&\nodata \\
1277.114 $\pm$  0.064&  -20 $\pm$  15&  48 $\pm$  15&  32 $\pm$  17&  7.5&g:CI1277.2&\nodata \\
1277.572 $\pm$  0.057& 15265 $\pm$  14&  25 $\pm$  10&  27 $\pm$  12&  6.6&Ly$\alpha$&h:CII1277.2 \\
1285.845 $\pm$  0.053& 17306 $\pm$  13&  47 $\pm$  3& 294 $\pm$  15& 72.7&Ly$\alpha$&\nodata \\
1294.654 $\pm$  0.053& 19478 $\pm$  13&  37 $\pm$  3& 249 $\pm$  15& 63.3&Ly$\alpha$&\nodata \\
1295.191 $\pm$  0.057&  -107 $\pm$  13&  97 $\pm$  3&  30 $\pm$  19&  7.0&g:SI1295.7?&\nodata \\
\cutinhead{TON~S180}
1199.481 $\pm$  0.043&  -17 $\pm$  11&  52 $\pm$  5& 229 $\pm$  34& 21.5&g:NI1199.5&\nodata \\
1200.140 $\pm$  0.043&  -20 $\pm$  11&  18 $\pm$  5& 152 $\pm$  26& 14.1&g:NI1200.2&\nodata \\
1200.614 $\pm$  0.044&  -24 $\pm$  11&  45 $\pm$  5& 185 $\pm$  35& 17.0&g:NI1200.7&\nodata \\
1205.843 $\pm$  0.044&  -163 $\pm$  11&  53 $\pm$  5& 287 $\pm$  43& 26.1&h:SiIII1206.5&\nodata \\
1206.457 $\pm$  0.043&  -11 $\pm$  11&  57 $\pm$  3& 415 $\pm$  45& 38.2&g:SiIII1206.5&\nodata \\
1223.452 $\pm$  0.049&  1919 $\pm$  12&  35 $\pm$  11&  66 $\pm$  28&  6.5&Ly$\alpha$&\nodata \\
1226.921 $\pm$  0.051&  2774 $\pm$  12&  28 $\pm$  13&  49 $\pm$  26&  4.9&Ly$\alpha$&\nodata \\
1227.774 $\pm$  0.049&  2985 $\pm$  12&  21 $\pm$  13&  41 $\pm$  23&  4.3&Ly$\alpha$&\nodata \\
1237.982 $\pm$  0.046&  5502 $\pm$  11&  51 $\pm$  5& 268 $\pm$  54& 31.2&Ly$\alpha$&\nodata \\
1241.177 $\pm$  0.048&  6290 $\pm$  12&  28 $\pm$  10&  54 $\pm$  23&  6.2&Ly$\alpha$&\nodata \\
1244.125 $\pm$  0.045&  7017 $\pm$  11&  56 $\pm$  4& 222 $\pm$  29& 30.0&Ly$\alpha$&\nodata \\
1250.592 $\pm$  0.046&  2 $\pm$  11&  33 $\pm$  7&  87 $\pm$  25&  9.8&g:SII1250.6&\nodata \\
1253.591 $\pm$  0.048&  -50 $\pm$  12&  39 $\pm$  22&  41 $\pm$  36&  4.5&h:SII1253.8&\nodata \\
1253.798 $\pm$  0.045&  -3 $\pm$  11&  $\lt 20$ &  83 $\pm$  28&  9.1&g:SII1253.8&\nodata \\
1257.901 $\pm$  0.052& 10415 $\pm$  13&  30 $\pm$  13&  41 $\pm$  23&  5.0&Ly$\alpha$&\nodata \\
1259.506 $\pm$  0.045&  -3 $\pm$  11&  $\lt 20$ & 124 $\pm$  21& 14.9&g:SII1259.5&\nodata \\
1259.816 $\pm$  0.047&  -153 $\pm$  11&  $\lt 20$ &  30 $\pm$  17&  3.5&h:SiII+FeII1260.5&\nodata \\
1260.409 $\pm$  0.044&  -12 $\pm$  11&  51 $\pm$  3& 384 $\pm$  33& 45.1&g:SiII+FeII1260.5&\nodata \\
1268.027 $\pm$  0.049& 12912 $\pm$  12&  81 $\pm$  15& 107 $\pm$  39& 14.3&Ly$\alpha$&\nodata \\
1268.661 $\pm$  0.047& 13068 $\pm$  12&  52 $\pm$  5& 140 $\pm$  27& 18.7&Ly$\alpha$&\nodata \\
1270.473 $\pm$  0.048& 13515 $\pm$  12&  58 $\pm$  7& 140 $\pm$  29& 19.0&Ly$\alpha$&\nodata \\
1271.147 $\pm$  0.046& 13681 $\pm$  11&  54 $\pm$  4& 212 $\pm$  29& 28.1&Ly$\alpha$&\nodata \\
1290.376 $\pm$  0.048&  -149 $\pm$  11&  42 $\pm$  6&  74 $\pm$  20& 19.9&i:Ly$\alpha$&\nodata \\
1290.960 $\pm$  0.046&  -13 $\pm$  11&  $\lt 20$ &  39 $\pm$  14& 10.6&i:Ly$\alpha$&\nodata \\
1291.414 $\pm$  0.045&  92 $\pm$  11&  24 $\pm$  4& 107 $\pm$  14& 27.9&i:Ly$\alpha$&\nodata \\
\cutinhead{TON~1542}
1199.520 $\pm$  0.032&  5 $\pm$  8&  40 $\pm$  4& 244 $\pm$  41& 17.6&g:NI1199.5&\nodata \\
1200.172 $\pm$  0.032&  -7 $\pm$  8&  30 $\pm$  4& 237 $\pm$  40& 16.8&g:NI1200.2&\nodata \\
1200.671 $\pm$  0.033&  -7 $\pm$  8&  43 $\pm$  5& 270 $\pm$  48& 18.8&g:NI1200.7&\nodata \\
1206.287 $\pm$  0.037&  -53 $\pm$  9&  $\lt 20$&  69 $\pm$  82&  4.8&h:SiIII1206.5&\nodata \\
1206.479 $\pm$  0.086&  -5 $\pm$  8&  44 $\pm$  53& 290 $\pm$ 347& 20.5&g:SiIII1206.5&\nodata \\
1206.707 $\pm$  0.037&  52 $\pm$  9&  27 $\pm$  9& 124 $\pm$ 148&  8.9&h:SiIII1206.5&\nodata \\
1220.480 $\pm$  0.033&  1186 $\pm$  8&  47 $\pm$  5& 294 $\pm$  56& 19.7&Ly$\alpha$&\nodata \\
1223.355 $\pm$  0.033&  1895 $\pm$  8&  35 $\pm$  5& 216 $\pm$  42& 16.9&Ly$\alpha$&\nodata \\
1226.063 $\pm$  0.033&  2563 $\pm$  8&  49 $\pm$  5& 248 $\pm$  41& 19.6&Ly$\alpha$&\nodata \\
1238.882 $\pm$  0.057&  15 $\pm$  14&  46 $\pm$  19&  50 $\pm$  35&  4.6&g:NV1238.8&\nodata \\
1240.014 $\pm$  0.061&  22 $\pm$  15&  69 $\pm$  20&  76 $\pm$  42&  6.7&g:MgII1239.9&\nodata \\
1242.902 $\pm$  0.061&  24 $\pm$  15&  50 $\pm$  21&  50 $\pm$  36&  4.7&g:NV1242.8&\nodata \\
1250.607 $\pm$  0.035&  5 $\pm$  8&  $\lt 20$ &  80 $\pm$  29&  6.9&g:SII1250.6&\nodata \\
1253.832 $\pm$  0.034&  5 $\pm$  8&  $\lt 20$ & 124 $\pm$  31& 10.7&g:SII1253.8&\nodata \\
1259.540 $\pm$  0.041&  5 $\pm$  10&  32 $\pm$  9& 158 $\pm$  65& 13.7&g:SII1259.5&\nodata \\
1260.478 $\pm$  0.033&  -5 $\pm$  8&  57 $\pm$  3& 444 $\pm$  39& 40.3&g:SiII+FeII1260.5&\nodata \\
1282.406 $\pm$  0.034& 16458 $\pm$  8&  31 $\pm$  4& 105 $\pm$  20& 15.0&Ly$\alpha$&\nodata \\
1289.496 $\pm$  0.034&  -644 $\pm$  8&  $\lt 20$ &  54 $\pm$  11& 11.1&i:Ly$\alpha$&\nodata \\
\cutinhead{VII~ZW~118}
1197.440 $\pm$  0.029&  60 $\pm$  7&  97 $\pm$  3&  34 $\pm$  17&  3.5&h:MnII1197.2?&\nodata \\
1199.544 $\pm$  0.022&  -1 $\pm$  6&  42 $\pm$  3& 289 $\pm$  28& 30.8&g:NI1199.5&\nodata \\
1200.227 $\pm$  0.022&  2 $\pm$  6&  42 $\pm$  3& 271 $\pm$  32& 28.5&g:NI1200.2&\nodata \\
1200.712 $\pm$  0.022&  0 $\pm$  6&  33 $\pm$  4& 236 $\pm$  31& 24.6&g:NI1200.7&\nodata \\
1201.127 $\pm$  0.029&  107 $\pm$  7&  $\lt 20$ &  31 $\pm$  27&  3.2&h:NI1200.7&\nodata \\
1206.204 $\pm$  0.029&  -73 $\pm$  7&  66 $\pm$  8& 254 $\pm$  58& 27.0&h:SiIII1206.5&\nodata \\
1206.509 $\pm$  0.026&  2 $\pm$  6&  58 $\pm$  5& 344 $\pm$  58& 36.7&g:SiIII1206.5&\nodata \\
1222.648 $\pm$  0.039&  1721 $\pm$  10&  43 $\pm$  14&  54 $\pm$  29&  5.9&Ly$\alpha$&\nodata \\
1225.329 $\pm$  0.029&  2382 $\pm$  7&  50 $\pm$  16&  68 $\pm$  38&  7.8&Ly$\alpha$&\nodata \\
1225.646 $\pm$  0.023&  2460 $\pm$  6&  44 $\pm$  4& 267 $\pm$  35& 31.1&Ly$\alpha$&\nodata \\
1234.301 $\pm$  0.029&  4595 $\pm$  7&  20 $\pm$  13&  35 $\pm$  20&  4.6&Ly$\alpha$&\nodata \\
1234.701 $\pm$  0.029&  4693 $\pm$  7&  55 $\pm$  21&  45 $\pm$  31&  6.0&Ly$\alpha$&\nodata \\
1240.014 $\pm$  0.029&  21 $\pm$  7&  55 $\pm$  21&  43 $\pm$  29&  5.8&g:MgII1239.9&\nodata \\
1242.655 $\pm$  0.049&  -35 $\pm$  12&  44 $\pm$  19&  34 $\pm$  24&  4.7&g:NV1242.8&\nodata \enddata
\end{deluxetable}
\normalsize

\clearpage
\bibliographystyle{apj}

\end{document}